\documentclass[11pt]{JHEP}
\usepackage{amsmath}
\usepackage{amssymb}
\usepackage{amsmath,amsthm,epsfig,euscript,array,cancel}
\font\mybb=msbm10 at 10pt
\def\bb#1{\hbox{\mybb#1}}

\newcommand{\be}{\begin{equation}}
\newcommand{\ee}{\end{equation}}

\def\bea{\begin{eqnarray}}
\def\eea{\end{eqnarray}}
\preprint{
Nov. 2, 2017. V2: May 21, 2018. V3: Oct. 30, 2018
}
\renewcommand{\theequation}{\arabic{section}.\arabic{equation}}
\title{Spinor frame formalism for amplitudes and constrained superamplitudes of 10D SYM and 11D supergravity }

\author{Igor Bandos
\\ {\small\it Department of
Theoretical Physics, University of the Basque Country UPV/EHU, \\ P.O. Box 644, 48080 Bilbao, Spain} \\
{\small\it and IKERBASQUE, Basque Foundation for Science, 48011, Bilbao, Spain}
}

\date{11/04/2016-- 12/12/16--20/12/16--05/05/2017--17/08/2017--30/10/2017--04/04/2018--21/05/2018--30/07/2018--26/09/2018,  Printed \today }

\abstract{We show that 10D spinor helicity formalism can be understood as spinor moving frame approach to supersymmetric particles extended to the description of amplitudes. This allows us to develop the spinor helicity formalism for 11D supergravity and  a  new constrained superfield formalism for  10D SYM and 11D SUGRA amplitudes. We show how  the constrained on-shell superfields, one-particle counterparts of the superamplitudes, can be obtained by quantization of massless superparticle mechanics.
\\
We make some stages  towards the calculation of amplitudes of  10D SYM and 11D SUGRA  in this framework. In particular we have found supersymmetric Ward identities for constrained amplitudes and an especially  convenient gauge, fixed on the spinor frame variables corresponding to scattered particles, which promises to be  an extremely useful tool for further development of our approach.
\\
We also discuss a candidate for generalization of the BCFW recurrent relations for the constrained tree superamplitudes, indicate and discuss a problem of dependence of the expressions obtained with it on a deformation vector, which is not fixed uniquely in higher dimensional $D>4$  cases.
}


\begin{document}

\section{Introduction}

The seminal paper \cite{Witten:2003nn} initiated a wave of renewed interest to the applications of D=4 twistor approach \cite{Penrose:1967wn,Penrose:1972ia,Penrose:1986ca,Ferber:1977qx,Shirafuji:1983zd,Witten:1978xx} which resulted in development of new methods of computation of on-shell amplitudes and (in combination with other ideas \cite{Bern:2011qn})  to a significant progress in loop calculations
 \cite{Bern:2011qn,Drummond:2008vq,Drummond:2009fd,Eden:2011ku,Kallosh:2012yy,Elvang:2015rqa,ArkaniHamed:2017}.
Of these new methods let us mention the Britto-Cachazo-Feng-Witten (BCFW) approach \cite{Britto:2005fq}, which allowed  to obtain Britto-Cachazo-Feng (BCF) recursion relations for tree amplitudes \cite{Britto:2004ap} and their loop generalizations. It was further developed and generalized for superamplitudes of maximally supersymmetric 4D ${\cal N}=4$ super--Yang--Mills (SYM) and ${\cal N}=8$  supergravity (SUGRA) theories in  \cite{Bianchi:2008pu,ArkaniHamed:2008gz,Brandhuber:2008pf,Heslop:2016plj,Herrmann:2016qea} and refs. therein.

To be precise, the  original BCFW work \cite{Britto:2005fq} as well as \cite{Bianchi:2008pu,ArkaniHamed:2008gz,Brandhuber:2008pf,Heslop:2016plj,Herrmann:2016qea} used essentially the so-called ($D=4$) spinor helicity formalism\footnote{As it was shown in \cite{ArkaniHamed:2008gz}, the bosonic BCFW relation can be also proved without the use of spinor helicity variables; this fact does not diminish the significance of these  which were used quite extensively in \cite{ArkaniHamed:2008gz}.}, which is related but not identical to the above mentioned twistor approach \footnote{ Also a momentum twistor formalism, alternative (dual) to the standard twistor approach and related to dual superconformal symmetry \cite{Drummond:2008vq}, rather than to the standard conformal symmetry of ${\cal N}=4$ SYM, was developed in \cite{Hodges:2009hk,Mason:2009qx}.
}.
A D=10 dimensional generalization of the spinor helicity formalism was constructed by Caron-Huot and O'Connell in
\cite{CaronHuot:2010rj} and used in \cite{Boels:2012ie,Boels:2012zr,Wang:2015jna,Wang:2015aua} to study the 10D SYM and type IIB supergravity amplitudes (see \cite{Boels:2012zr} also  for application to type IIB string theory amplitudes). The authors of  \cite{CaronHuot:2010rj} also constructed a superfield generalization of 10D amplitudes, which we have called 'Clifford superamplitude' in \cite{Bandos:2016tsm}. However this formalism happens to be quite nonminimal and very complicated, so that the further use of 10D spinor helicity formalism was in the field of type IIB supergravity where an essential simplification can be reached with the use of the natural IIB complex structure.

In this paper we first show that the variables of the 10D spinor helicity formalism can be identified with spinor moving frame variables (Lorentz harmonics) used for the description of supersymmetric particle mechanics in \cite{Galperin:1991gk,Delduc:1991ir,Bandos:1996ju}\footnote{The identification of spinor helicity variables with Lorentz harmonics were noticed in \cite{Uvarov:2015rxa} and used there to construct  D=5 spinor helicity formalism. }. The spinor moving frame formalism was also developed for the case of  11D massless superparticle \cite{Galperin:1992pz,Bandos:2006nr,Bandos:2007mi,Bandos:2007wm}
\footnote{The spinor moving frame approach for D=4 and D=10 superstrings was proposed in \cite{BZ-str} and elaborated in \cite{Bandos:1992ze}, for 11D supermembrane in \cite{BZ-M2} and for the generic super-p-branes from the `standard brane scan' - in \cite{BZ-p}. The synthesis of spinor moving frame approach with the so-called STV (Sorokin-Tkach--Volkov) approach to superparticles and superstrings
\cite{Sorokin:1989zi,Delduc:1992fk,Delduc:1992fc}   (see \cite{Sorokin:1999jx} for the review and more references) resulted in the development of the superembedding approach to superstrings and super-p-branes
\cite{Bandos:1995zw}.  In particular, in the frame of this approach (also reviewed in \cite{Sorokin:1999jx})  the equations of motion of the M-theory 5-brane had been obtained in \cite{Howe:1996yn} some months before the covariant actions was constructed in \cite{Bandos:1997ui} and \cite{Aganagic:1997zq}.} and we have used it to develop the spinor helicity formalism for 11D supergravity amplitudes. This has been briefly reported in \cite{Bandos:2016tsm} and is elaborated here.

A key observation is that a suitable quantization of the D=4 ${\cal N}=4$ and  ${\cal N}=8$  superparticle models in its twistor--like  formulation \cite{Ferber:1977qx,Shirafuji:1983zd} results in  description of their quantum state spectrum by so-called  on-shell superfields. These are  chiral superfields in superspace with ${\cal N}$ complex fermionic and 2 complex (4 real) bosonic spinor coordinates, known as on--shell superspace, which  describe the fields of  the linearized  ${\cal N}=4$ SYM and ${\cal N}=8$ SUGRA multiplets (see e.g. \cite{Witten:2003nn,Drummond:2008vq}). The superamplitudes of  ${\cal N}=4$ SYM and ${\cal N}=8$  can be considered as multiparticle generalizations of these on-shell superfields.

Similarly, a suitable  quantization of the 10D and 11D massless superparticle models in their spinor moving frame formulations \cite{Bandos:1996ju,Bandos:2006nr,Bandos:2007mi,Bandos:2007wm} results in an on-shell superfield description of the linearized 10D SYM and 11D SUGRA supermultiplets which had been proposed in \cite{Galperin:1992pz} (as a generalization of the Penrose twistor transform \cite{Penrose:1967wn,Penrose:1972ia,Penrose:1986ca}). As, in distinction to the 4D case, these on-shell superfields obey some nontrivial differential equations with fermionic covariant derivatives, we call them {\it constrained on-shell superfields}.  Multiparticle generalizations
of these give us constrained superamplitudes which are much more economic than the Clifford superamplitudes  of \cite{CaronHuot:2010rj}. The development of such a {\it constrained superamplitude formalism} for  10D SYM and 11D SUGRA is one of the main aims of the present paper.

We describe the properties of  the constrained tree superamplitudes of 10D SYM and 11D SUGRA
and  make some stages necessary for practical calculations of amplitudes  in the frame of our spinor helicity formalism and constrained superamplitude approach. In particular we obtain the supersymmetric Ward identites and present a convenient gauge fixed on the spinor frame variables,
which can be considered as  Lorentz covariant counterpart of the light-cone gauge.

We have also presented a candidate for generalization of the BCFW recurrent relations for the constrained tree superamplitudes and studied its properties.
In this part the results are quite preliminary.
 The  use of our candidate BCFW relations  to calculate  4-point SYM amplitudes leads to expressions suffering an unwanted explicit dependence on a deformation vector, which is not fixed uniquely in the case of higher $D$.
We suggest  that this can be improved with the use of a complementary analytic superamplitude formalism and complex spinor frame variables of \cite{Bandos:2017zap}.

A part of the 11D results  was briefly reported in \cite{Bandos:2016tsm}. An alternative (although related)
analytic superamplitude  formalism has been proposed in \cite{Bandos:2017zap}. The use of this latter for further development of the constrained superamplitude formalism of the present  paper will be  the subject of future publication.


\subsection{Other superfield approaches to maximally supersymmetric QFTs}

We have to notice the existence of other superfield approaches to calculation of the amplitudes which are oriented on the use of the off-shell methods, actions and Feynman diagrams.

First of all there is the light-cone superspace approach used in particular to prove the perturbative finiteness of the ${\cal N}=4$ SYM theory in D=4 \cite{Mandelstam:1982cb,Brink:1982pd,Brink:1982wv}. It was also developed for superstring theory \cite{Green:1983hw}. Despite it was proved to be very useful, the lack of an explicit  Lorentz covariance was the reason to search for manifestly Lorentz covariant formalism.

The covariant superfield approach of \cite{Howe:1983sr}, which was  based on the off-shell supersymmetry and background superfield methods, allowed to obtain the non-renormalization theorems and analyse the structure of possible counterterms. It brought a number of impressive results, in particular an alternative proof of perturbative finiteness of ${\cal N}=4$ SYM \cite{Howe:1983sr} and of the absence of some loop divergences in ${\cal N}=8$ supergravity \cite{Bossard:2009sy}. The restriction of this method comes from the need in an off-shell superfield description of the supersymmetric theory, which is unknown in its complete form for the cases of most interesting higher ${\cal N}$ and higher dimensional supergravities and SYM theories \footnote{The result on finiteness of ${\cal N}=4$ SYM were obtained in \cite{Howe:1983sr}  using its description in terms of ${\cal N}=2$ superfields.}.

The analysis of the on-shell counterterms of ${\cal N}$-extended supergravity theories based on their on-shell superfield description was initiated in \cite{Howe:1980th,Kallosh:1980fi} and is still in the center of interest (see e.g. \cite{Kallosh:2012yy} and refs. therein).

The so-called pure spinor approach to superstring \cite{Berkovits:2000fe,Berkovits:2004px,Berkovits:2006vi,Berkovits:2017ldz} was also developed for supersymmetric field theories
\cite{Bjornsson:2010wu,Bjornsson:2010wm,Mafra:2014gja,Mafra:2015mja}.
In particular, a pure spinor formulation of the action for 11D supergravity was proposed in \cite{Cederwall:2010tn}, elaborated in \cite{Cederwall:2012es,Berkovits:2018gbq} and applied to analysis of UV divergences  in \cite{Karlsson:2014xva}.

This is a good place to notice by pass that the study of the relation of pure spinor description of superstring with the standard Green--Schwarz formulation resulted in proposing a number of versions of multidimensional twistor correspondences \cite{Berkovits:2004bw,Berkovits:2009by,Berkovits:2014aia} alternative to the Lorentz harmonic version of twistor-like approach used in this paper and {\it e.g.} in \cite{Bandos:1996ju,Bandos:2006nr}.

As we have already noticed, the main difference of the approach of this paper with the above mentioned superfield approaches to maximally supersymmetric quantum theories is that these use off-shell methods, Lagrangians and Feynman rules, while our study belongs to the line of the on-shell approaches to superamplitudes.
The experience gained in D=4 \cite{Elvang:2015rqa,ArkaniHamed:2017} suggests that the on-shell methods at the end might happen to be more practical in quantum calculations. Of course, we are at the beginning of the way and presently cannot show such impressive results of our approach as the ones obtained with the above methods.

\subsection{Outline of the paper}

The rest of the paper is organized as follows. In the next Sec. 2 we first  review the D=4 spinor helicity formalism (sec. 2.1) and BCFW recurrent relations for D=4 amplitudes (sec. 2.2), as well as the on-shell superspace, D=4 superamplitudes and their one--particle counterparts called 'on-shell superfields' (sec. 2.3). In sec. 2.4. we show how these  on-shell superfields can be obtained by quantization of ${\cal N}$-extended massless superparticle model.

This D=4 observation provides us with a guide in search for on-shell superamplitudes of higher dimensional SYM and SUGRA theories. Namely, first we have to find a suitable superparticle model and perform its quantization in a suitable representation thus arriving at on-shell superfields describing the superparticle quantum state vectors. The tree superamplitudes will be multiparticle counterparts of such on-shell superfields. In particular they  carry the same indices/charges/weights as on-shell superfields,  but multiplicated (one for each scattered particles), and obey the set of equations
which repeat the equations satisfied by on-shell superfield in variables corresponding to each of the scattered particles.

Spinor moving frame  formulation of D=10 and D=11 massless superparticle mechanics
(also known under the name `twistor-like Lorentz harmonic formulation') is reviewed in Sec. 3. Its quantization is described in Sec. 5 where it is shown how the constrained on-shell superfields and more complicated (and quite non-minimal) Clifford superfields appear in this way. But before that, in Sec. 4 we construct the
D=10 and D=11 spinor helicity formalism  by identifying the spinor helicity variables of scattered particles with the counterparts of spinor moving frame variables (Lorentz harmonics) of the massless superparticle mechanics.

In Sec. 4 we also show how the solutions of linearized equations of  motion of 10D SYM and 11D supergravity can written in terms of these spinor frame and related vector frame variables, and write supersymmetry transformations mixing the solutions of bosonic and fermionic equations. We also introduce there the spinor helicity representation of the amplitudes of 10D SYM and 11D SUGRA and obtain supersymmetric Ward identities for these amplitudes.

In Sec. 5, by quantizing 10D and 11D  superparticle models, we arrive at the on-shell superfield description of the linearized 10D SYM and 11D SUGRA which are generalized to the constrained superamplitudes in Sec. 6. The quantization, described in sec. 5.5.,  requires some  technical details on spinor moving frame variables and coordinate basis of Lorentz-harmonic superspace. These have been presented in secs. 5.1-5.4 and are not used in the forthcoming sections. The reader not interested in quantization procedure, nor in details on properties of  spinor moving frame in D=10 and D=11, might skip the first subsections and move directly to sub-subsection
\ref{On-shellSYM} using the previous sub-section \ref{On-shSSP=sec} just for  notation.


In Sec. 6 we develop the constrained superamplitude formalism for 10D SYM and 11D SUGRA and present a candidate for BCFW type recurrent relations for these superamplitudes. The explicit form of these candidate BCFW relations for 4-point 11D superamplitude  is discussed  in Sec. 7 where we also describe some  further development of spinor frame approach to amplitudes and superamplitudes. In particular, we discuss  there  the relation between spinor frames associated to different particles,  consequences of the momentum conservation, D=10, 11  supermomentum, and   a gauge fixing conditions for the auxiliary gauge symmetry. These conditions describe a Lorentz harmonic counterpart of the light cone gauge, which is fixed on spinor frame variables and promises to be especially convenient for the amplitude calculations.

Finally, in Sec. 8 we discuss the candidate BCFW relation for 10D four-point  superamplitudes, reduce them to the relation for 4--point amplitudes, present the explicit expressions for 3-point amplitudes  and use these to calculate  10D 4-point amplitudes with 4 and 2 fermionic legs.
We find that, unfortunately, the resulting expressions for four point amplitudes suffer an unwanted dependence on a deformation vector and suggest that this might be improved with the use of analytic superamplitude formalism and complex spinor frame of \cite{Bandos:2017zap}.

We conclude in Sec. 9.

In Appendix A we describe some aspects  of the Clifford superfield version of the 10D superamplitude formalism of \cite{CaronHuot:2010rj}.


\section{D=4 spinor helicity formalism, (super)twistors, (super)amplitudes and superparticle mechanics}

In this section we first review the D=4 spinor helicity formalism for gauge theory amplitudes, BCFW recurrent relations for these amplitudes, superamplitudes of maximally supersymmetric D=4 SYM and  SUGRA theory and their one-particle counterparts called on-shell superfields. Then we show how such on-shell superfield appears as a representation of quantum state vector in quantization of a suitable formulation of massless superparticle mechanics. This provides us with a guide in search for tree on-shell superamplitudes of 10D SYM and 11D SUGRA which will be the subject of secs. 5 and 6.

\subsection{Spinor helicity formalism for  D=4 (S)YM amplitudes}

In spinor helicity approach to D=4 scattered amplitudes of massless particles (see e.g. \cite{Drummond:2008vq,Elvang:2015rqa,ArkaniHamed:2017} and refs. therein)  the information on the (light-like) momentum $p_{\mu (i)}$  and on helicity of the $i$-th scattered particle is encoded in the complex 2-component bosonic spinor
$\lambda_{A(i)}$. The momentum is given by the product  of this and its complex conjugate $\bar{\lambda}{}_{\dot{A}(i)}= (\lambda_{A(i)})^*$:
\begin{eqnarray}\label{p=ll=4D}
p_{{A\dot{A}(i)}}:= p_{\mu (i)} \sigma^\mu_{{A\dot{A}}}=  2\lambda{_{{A(i)}  }}\bar{\lambda}{}_{{\dot{A}(i)}}   \quad \Leftrightarrow \quad p_{\mu(i)}= \lambda_{(i)} \sigma_\mu \bar{\lambda}_{(i)},  \quad \begin{cases} A=1,2\; , \quad \dot{A}=1,2\; , \cr   \mu=0,...,3\;  , \end{cases}
\end{eqnarray}
where $\sigma^\mu_{{A\dot{A}}}$ are relativistic Pauli matrices.  The polarization vectors of $i$-th scattering spin 1  particles ('gluons') of negative and positive helicity can be written as
\begin{eqnarray}\label{polar=}
\varepsilon^{(+)}_{{A\dot{A}(i)}}= \frac {\lambda{{_{A(i)}  }}\bar{\mu}{}_{{\dot{A}}} }{[\bar{\lambda}_{(i)}\bar{\mu}]} \; ,  \qquad \varepsilon^{(-)}_{{A\dot{A}(i)}}= \frac {\mu_{A}\bar{\lambda}{}_{{\dot{A}(i)}}} {<\mu\lambda_{(i)}>}
\; ,   \qquad  \end{eqnarray}
where $\bar{\mu}{}_{{\dot{A}}}= (\mu_{A})^*$ is a (constant) reference spinor,
 $<\mu\lambda_{(i)}>:= \epsilon_{AB} \mu^A\lambda^B_{(i)}$, ${}[\bar{\mu}\bar{\lambda}_{(i)}]=
\epsilon_{{\dot{A}\dot{B}}} \bar{\mu}^{{\dot{A}}}\bar{\lambda}^{\dot{B}}{(i)}$, and \begin{eqnarray}\label{epsAB=} \epsilon_{{AB}} =\left(\begin{matrix} 0 & -1\cr 1 & 0 \end{matrix}\right) = -\epsilon^{{AB}}\; , \qquad  \epsilon_{{\dot{A}\dot{B}}}  =\left(\begin{matrix} 0 & -1\cr 1 & 0 \end{matrix}\right) =  -\epsilon^{{\dot{A}\dot{B}}}\;   \qquad
\end{eqnarray}
are unit antisymmetric spin tensors used to lower and to rise the Weyl spinor indices,
$$ \lambda^A= \epsilon^{AB}\lambda_B\; , \qquad \lambda_A= \epsilon_{AB}\lambda^B\; , \qquad
 \bar{\lambda}_{{\dot{A}}}=\epsilon_{{\dot{A}\dot{B}}}\bar{\lambda}^{{\dot{B}}}\; .
$$
Below we will use even shorter notation for the contraction of the `momentum spinors' or {\it spinor helicity variables},
\begin{eqnarray}\label{lilj=}
 <ij> \equiv  <\lambda_{(i)}\lambda_{(j)}>:= \lambda^{A}_{(i)}\lambda_{A(j)}= \epsilon_{AB} \lambda^A_{(i)}\lambda^B_{(j)}\; , \qquad \nonumber \\ {} [ij]\equiv
{}[\bar{\lambda}_{(i)}\bar{\lambda}_{(j)}]= \bar{\lambda}^{\dot{A}}_{(i)}\bar{\lambda}{}_{_{\dot{A}(j)}} =
\epsilon_{{\dot{A}\dot{B}}} \bar{\lambda}^{{\dot{A}}}_{(i)}\bar{\lambda}^{{\dot{B}}}_{(j)}\;  . \qquad \end{eqnarray}
One easily observes that $<ij>=-<ji>$,
${}[ij]=-[ji]$ (as $\epsilon_{{\dot{A}\dot{B}}}=-\epsilon_{{\dot{B}\dot{A}}}$ (\ref{epsAB=})) and, hence,  $<ii>=0=[ii]$.
 Then the light-likeness of the momentum (\ref{p=ll=4D}) and its orthogonality to polarization vectors  (\ref{polar=}),  $p^{\mu}_{(i)}\varepsilon^{(\pm )}_{{\mu(i)}} =0$, follow from the identity
 \begin{eqnarray}\label{sisi=2ee}
 \sigma^\mu{}_{{A\dot{A}}}\sigma_{{\mu B\dot{B}}}=2\epsilon_{{AB}}\epsilon_{{{\dot{A}\dot{B}}}}\;  . \qquad \end{eqnarray}

The scattering amplitudes ${\cal A}(1,...,n):= {\cal A}(p_{(1)},\varepsilon_{(1)}; ...,p_{(n)},\varepsilon_{(n)})$
 are independent of the choice of $\mu$ in (\ref{polar=}), so that
\begin{eqnarray}\label{cAn=cAnl}
{\cal A}(1,..,n):= {\cal A}(p_{(1)},\varepsilon_{(1)}; ...;p_{(n)},\varepsilon_{(n)}) =
{\cal A}(\lambda_{(1)}, \bar{\lambda}_{(1)}; \ldots ;\lambda_{(n)}, \bar{\lambda}_{(n)})
\;  .\qquad \end{eqnarray}
Furthermore, as a function of bosonic spinors, the  amplitude should obey the helicity constraints,
\begin{eqnarray}\label{hhiA=hiA}
\hat{h}_{(i)} {\cal A}(1,...,n) = h_i {\cal A}(1,...,n)\; ,\qquad
\end{eqnarray}
where $h_i$ is the helicity of the state, $h_i=\pm 1$ in the case of gluons, and $\hat{h}_{(i)} $ is helicity operator which has the form
\begin{eqnarray}
 \label{Ui=b}
\hat{h}_{(i)}:= \frac 1 2 \lambda^A_{(i)} \frac \partial  {\partial \lambda^A_{(i)} }  -
\frac 1 2 \bar{\lambda}{}^{{\dot{A}}}_{(i)} \frac \partial  {\partial \bar{\lambda}{}^{{\dot{A}}}_{(i)}} \, .  \qquad \end{eqnarray}
Thus $n$-particle amplitude is also characterized by $n$ helicities. In the case of $n$ gluons these are $\pm 1$ and the amplitude carries $n$ sign indices, \begin{eqnarray}
 \label{cAMHV:=4D+-}{\cal A}(1,...,n) ={\cal A}^{-...-...+...+}(1,...,n)\; .
 \end{eqnarray}

It can be shown that the amplitudes with all and with all but one gluons of the same helicity vanish, i.e. $ {\cal A}^{+...+}(1,...,n)=0$,  $ {\cal A}^{-+...+}(1,...,n)=0$, so that the  simplest {\it maximal helicity violation (MHV)} amplitude is  \begin{eqnarray}\label{cAMHV:=4D++-} {\cal A}^{MHV}(1,...,n)= {\cal A}^{+...+-_{_i}+...+-_{_j}+...+}(1,...,n) \; . \qquad \end{eqnarray} This can be expressed in a simple way \cite{Parke:1986gb}  in terms of the contractions (\ref{lilj=})  of the left-handed  bosonic spinors corresponding to different external particles,
\begin{eqnarray}
 \label{cA=MHV=4D}
 \delta^4\big( \sum\limits_i p_{a(i)}\big)
 {\cal A}^{MHV}(1,...,n) = \delta^4\left( \sum\limits_i \lambda_{A(i)}  \bar{\lambda}_{\dot{A}(i)}\right) \frac{  <ij>^4  }{<12> ... <(n-1)n> <n1>} \; .
 \end{eqnarray}
In (\ref{cA=MHV=4D})  $i$-th and $j$-th particles are assumed to be of negative helicity ($-1$) (as explicitly written in  (\ref{cAMHV:=4D++-})).

\subsection{BCFW recurrent relations}

The BCFW recursion relations for $n$-point (S)YM amplitudes are formulated with the use of the on-shell amplitudes depending on {\it deformed} momentum spinors of, say, 1-st and n-th of the scattered particles. The BCFW deformation rule reads
 \begin{eqnarray}\label{BCFWln=4D}
  & \lambda^{A}_{(n)}\mapsto \widehat{\lambda^{A}_{(n)}} = \lambda^{A}_{(n)} + z \lambda^{A}_{(1)} , \qquad  & \qquad \bar{\lambda}{}^{\dot A}_{(n)}\mapsto \widehat{\bar{\lambda}{}^{\dot A}_{(n)}}=\bar{\lambda}{}^{\dot A}_{(n)} , \qquad
  \\
  \label{BCFWl1=4D}
 & \lambda^{A}_{(1)}\mapsto \widehat{\lambda^{A}_{(1)}} = \lambda^{A}_{(1)}\; , \qquad
 & \bar{\lambda}{}^{\dot A}_{(1)} \mapsto  \widehat{\bar{\lambda}{}^{\dot A}_{(1)}}=\bar{\lambda}{}^{\dot A}_{(1)} -z \bar{\lambda}{}^{\dot A}_{(n)}  , \qquad  \end{eqnarray}
where $z$ is a complex number. {\it Notice that $\widehat{\bar{\lambda}{}^{\dot A}_{(1,n)}}$ differs from complex conjugate of $\widehat{{\lambda}{}^{ A}_{(1,n)}}$}, so that the deformed momenta  of  1-st and n-th particle
  \begin{eqnarray}\label{BCFWp=4D}
  p_{(n)}^a \mapsto \widehat{p{}_{(n)}^a}(z)=  p_{(n)}^a   + z q^a\; , \qquad
   p_{(1)}^a \mapsto \widehat{p{}_{(1)}^a}(z) =  p_{(1)}^a - z \bar{q}^a\; \qquad
  \end{eqnarray}
are complex. It is important that they remain light-like   \begin{eqnarray}\label{pnapna=0=}
\widehat{p_{(n)}^a}\, \widehat{p_{(n)a}}=0 \; , \qquad \widehat{p_{(1)}^a}\, \widehat{p_{(1)a}}=0 \; \qquad
  \end{eqnarray}
and hence can be used as arguments of an on-shell amplitude.
In (\ref{BCFWp=4D}) the vector $q^a$ is constructed as bilinear of the 1-st left and n-th right bosonic spinors,
  \begin{eqnarray}\label{qa=4D}
  q^{A\dot{A}} = q^a\tilde{\sigma}_a^{A\dot{A}}  = \lambda^{A}_{(1)} \bar{\lambda}^{\dot{A}}_{(n)} \; .   \qquad \end{eqnarray}
As a result, it is complex, light-like and orthogonal both to 1-st and n-th momentum,
 \begin{eqnarray}\label{qapa=0=}
  q^aq_a=0 \; , \qquad p_{(n)}^aq_a=0 \; , \qquad p_{(1)}^aq_a=0 \; . \qquad
  \end{eqnarray}
  One can easily check the lightlikeness of complex deformed momenta  (\ref{pnapna=0=}) using just
(\ref{qapa=0=}) and (\ref{BCFWp=4D}).

The BCFW recurrent relations for tree amplitudes of D=4 gluons read \cite{Britto:2005fq,ArkaniHamed:2008gz}
\begin{eqnarray}\label{cA=rrBCFW}
&{\cal A}^{(n)}& ({p_1},p_2 ,\ldots ;{p_n})= \qquad \\ \nonumber
&& = \sum_h \sum\limits_{l=2}^n
{\cal A}^{(l+1)}_{h}(\widehat{p_1}(z_l);p_2; \ldots ;{p_l};\widehat{P_{\Sigma_l}}(z_l)) \, \frac {1}{(P_{\Sigma_l})^2}\, {\cal A}^{(n-l+1)}_{-h}(-\widehat{P_{\Sigma_l}}(z_l), p_{l+1};\ldots ;\widehat{p_n}(z_l))
\; ,
\end{eqnarray}
where $h$ is the helicity of intermediate state with momentum $\widehat{P_{\Sigma_l}}(z_l)$.
Notice that, for shortness, we included the momentum conservation delta functions inside the amplitudes
({\it cf.} (\ref{cA=MHV=4D})). In the denominator of (\ref{cA=rrBCFW})  we find the (minus) sum of the first $l$ undeformed momenta,
\begin{eqnarray}\label{kSl=4d}
{P_{\Sigma_l}^a}=- \sum\limits_{m=1}^l  {p_m^a}\; , \qquad
\end{eqnarray}
while the arguments contain the sum of corresponding deformed momenta
\begin{eqnarray}\label{kSlz=4d}
\widehat{P_{\Sigma_l}^a}(z)=- \sum\limits_{m=1}^l  \widehat{p_m^a}(z)
\;  \qquad
\end{eqnarray}
 with the specific $l$-dependent value of the complex parameter $z$,
\begin{eqnarray} \label{zl:=4d}
 z_l\equiv  z_{_{P_{_{\Sigma_l}}}}(q)= \frac {P_{\Sigma_l}^a P_{\Sigma_l a}} {2P_{\Sigma_l}^b q_b}\; .
\end{eqnarray}
This values is designed in such a way that the sum of deformed momenta becomes light-like $(\widehat{P_{\Sigma_l}^a} (z_l))^2=0$, and, hence, the {\it r.h.s.} of (\ref{cA=rrBCFW}) contains on-shell amplitudes, although some of the light-like momenta they depend on are  complex.

Schematically one can write (\ref{cA=rrBCFW}) in the form \cite{Brandhuber:2008pf}
\begin{eqnarray}\label{cA=BCFW=sch}
&& {\cal A}=  \sum\limits_{{\hat{P}}} \sum_h
\widehat{{\cal A}}_{h}(z_{_{{P}}}) \, \frac {1}{{{P}}^2}\, \widehat{{\cal A}}_{-h}(z_{_{{P}}})
\; ,
\end{eqnarray}
in which in denominator  ${P}$ is a sum of a subset of momenta,  as in (\ref{kSl=4d}) above,
$z_{_{{P}}}$ is related to $P$ by (\ref{zl:=4d}),
and $\sum\limits_{{\hat{P}}} $ represents symbolically the sum   over the division of the set of particle momenta on two subsets (the sum over such divisions is implicit in the  expression (\ref{cA=rrBCFW}) above).

\subsection{ ${\cal N}=4$  SYM and ${\cal N}=8$ SUGRA superamplitudes and generalized BCFW }

 One can also collect the data of all the n-particle amplitudes of the fields of SYM supermultiplet by considering a superfield amplitude, called superamplitude, depending, besides  the set of $n$ bosonic spinors
 $\lambda^{^A}_{(i)}= (\bar{\lambda}{}^{^{\dot{A}}}_{(i)})^*$, also on the set of  $n$ complex  fermionic coordinates
 $\eta_{(i)}^q$ carrying the index $q=1,...,4$ of the fundamental representation of $SU(4)$:
  \begin{eqnarray}\label{cAn=cAnlf}
{\cal A}(1; ...;n) = {\cal A}(\lambda_{(1)}, \bar{\lambda}_{(1)}, \eta_{(1)}; ...;\lambda_{(n)}, \bar{\lambda}_{(n)},  \eta_{(n)})
\;  . \qquad \end{eqnarray}

The fact staying beyond the above superfield representation for SYM amplitudes is that the unconstrained superfield depending on the above bosonic spinor and Grassmann variable $\eta^q$, but not on its complex conjugate $\bar{\eta}_{q}=(\eta^q)^*$,
 \begin{eqnarray}\label{Phi=3,4}
 \Phi (\lambda ,\bar{\lambda}, \eta^q) = f^{(+)} +  \eta^q \chi_q + \frac 1 2 \eta^q \eta^p s_{pq}+
 \frac 1 {3!} \eta^q \eta^p\eta^r \epsilon_{rpqs}\bar{\chi}^s + \frac 1 {4!} \eta^q \eta^p\eta^r\eta^s  \epsilon_{rpqs}  f^{(-)}
\;   \qquad \end{eqnarray}
describes the on-shell degrees of freedom  of the ${\cal N}=4$ SYM multiplet provided it has a super-helicity  +1, i.e. if it obeys
 \begin{eqnarray}\label{UPhi=1}
 \hat{h}  \Phi (\lambda ,\bar{\lambda}, \eta^q) &=&
 \Phi (\lambda ,\bar{\lambda}, \eta^q) \; , \qquad q=1,...,4\; ,
 \end{eqnarray}
 with
 \begin{eqnarray}
 \label{U=}
 && 2\hat{h}:=
 \lambda^A \frac \partial  {\partial \lambda^A }-
 \bar{\lambda}{}^{^{\dot{A}}} \frac \partial  {\partial \bar{\lambda}{}^{^{\dot{A}}}} +
 \eta^q \frac \partial  {\partial  \eta^q} \; .  \qquad \end{eqnarray}

In this sense one calls the superfield (\ref{Phi=3,4}) and the superspace with coordinates
  $(\lambda ,\bar{\lambda}, \eta^q)$ it is defined on  the
{\it on-shell superfield} and {\it on-shell superspace}, respectively.
The component fields in (\ref{Phi=3,4}) describe photons (or gluons) of helicity  $+1$ ($ f^{(+)}=f^{(+)}(\lambda ,\bar{\lambda})$) and $-1$ ($ f^{(-)}=f^{(-)}(\lambda ,\bar{\lambda})$),
photinos (or gluinos) of helicity $+1/2$ ($\chi_q=\chi_q (\lambda ,\bar{\lambda})$) and $-1/2$ ($\bar{\chi}^q =\bar{\chi}^q (\lambda ,\bar{\lambda})$), and 6 scalars of the maximal ${\cal N}=4$ SYM multiplet
($s_{pq}=s_{[pq]} (\lambda ,\bar{\lambda})$).

The superfield amplitudes (superamplitudes)  (\ref{cAn=cAnlf}) of ${\cal N}=4$ SYM theory, which can be considered as multiparticle counterparts of the superfields $\Phi (\lambda ,\bar{\lambda}, \eta^q)$,  obey $n$ superhelicity constraints (with the same eigenvalue +1)
 \begin{eqnarray}\label{UcAn=cAn}
\hat{h}_{(i)} {\cal A}(\{\lambda_{(i)}, \bar{\lambda}_{(i)}, \eta^q_{(i)}\}) = {\cal A}(\{\lambda_{(i)}, \bar{\lambda}_{(i)}, \eta_{(i)}^q\})\; ,\qquad q=1,..., 4\; , \qquad i=1,..,n\; , \quad \\ \label{Ui=}
2\hat{h}_{(i)}:=
 \lambda^A_{(i)} \frac \partial  {\partial \lambda^A_{(i)} }-
 \bar{\lambda}{}^{^{\dot{A}}}_{(i)} \frac \partial  {\partial \bar{\lambda}{}^{^{\dot{A}}}_{(i)}} +
 \eta_i^q \frac \partial  {\partial  \eta_i^q}  \, .   \qquad \end{eqnarray}

Similarly, the ${\cal N}=8$ supergravity multiplet is described by chiral superfield on an  ${\cal N}=8$ on-shell superspace with 2 complex bosonic and 8 complex fermionic coordinates $(\lambda^A \, ,
 \bar{\lambda}{}^{{\dot{A}}} \, ,
 \eta^q)$, $q=1,\ldots, 8$, which has superhelicity +2, i.e. obeys
\begin{eqnarray}\label{UPhi=2}
 \hat{h}  \Phi (\lambda ,\bar{\lambda}, \eta^q) &=&
 2\Phi (\lambda ,\bar{\lambda}, \eta^q) \; , \qquad q=1,...,8\; ,
 \end{eqnarray}
with $\hat{h}  $ defined by (\ref{U=}) with  $q=1,...,8$.  The $n$-particle tree superamplitude of ${\cal N}=8$ supergravity can be described as a function in a direct product of $n$ copies of such an ${\cal N}=8$ on-shell superspace, depending holomorphically on the fermionic coordinates and carrying superhelicity +2 on  each of the set of  2 complex bosonic and 8 complex fermionic coordinates $(\lambda^A_{(i)} \, ,
 \bar{\lambda}{}^{^{\dot{A}}}_{(i)} \, ,
 \eta^q_{(i)})$ corresponding to
 external legs,
  \begin{eqnarray}\label{UcAn=2cAn}
  \hat{h}_{(i)} {\cal A}(\{\lambda_{(i)}, \bar{\lambda}_{(i)}, \eta^q_{(i)}\}) = 2{\cal A}(\{\lambda_{(i)}, \bar{\lambda}_{(i)}, \eta_{(i)}^q\})\; ,\qquad q=1,..., 8\, . \qquad \end{eqnarray}

To write the superfield generalization of the BCFW relations (\ref{cA=rrBCFW}) for tree superamplitudes of
  ${\cal N}=8$ supergravity and  ${\cal N}=4$ SYM \cite{ArkaniHamed:2008gz,Brandhuber:2008pf}, one should supplements the deformations of the bosonic spinors (\ref{BCFWln=4D}),  (\ref{BCFWl1=4D}) by the deformation of the fermionic variables $\eta^q_i=(\bar{\eta}_{q \, i})^*$ corresponding to 1-st and $n$-th particles,
  \begin{eqnarray}\label{BCFWeta=4D}
  \eta_{(n)}^q \mapsto \widehat{\eta_{(n)}^q }(z)= \eta_{(n)}^q    + z \eta_{(1)}^q \; , \qquad
  \eta_{(1)}^q \mapsto \widehat{\eta_{(1)}^q }(z)= \eta_{(1)}^q   \; . \qquad
  \end{eqnarray}
 Notice that, actually, the fermionic variable of the 1-st particle has not been changed \cite{ArkaniHamed:2008gz,Brandhuber:2008pf}.

The schematic form of the generalization of the BCFW recurrent relations for ${\cal N}=4$ SYM superamplitudes reads
 \begin{eqnarray}\label{cA-Sf=BCFW=sch}
&& {\cal A}=  \sum\limits_{\hat{P}} \int d^4\eta_{\hat{P}}
{\cal A}_{L}(z_{_{\hat{P}}}) \, \frac {1}{{{P}}^2}\, {\cal A}_{R}(z_{_{\hat{P}}})
\; .
\end{eqnarray}
A more explicit form can be easily restored by comparison with   (\ref{cA=BCFW=sch}) and (\ref{cA=rrBCFW}), also taking into account the deformation of not only bosonic spinors (\ref{BCFWln=4D}),  (\ref{BCFWl1=4D}) but also fermionic variables,
(\ref{BCFWeta=4D}). A new element is the Grassmann integration over the fermionic variable corresponding to the intermediate states,  $\eta_{\hat{P}}$.

The BCFW recurrent relations allow to construct all higher points on-shell superamplitudes from the basic 3-point superamplitudes. One of the aim of our work, the first stages of which are the subject of the present paper, was to find the generalization of these relations for the case of  superamplitudes of 11D supergravity and 10D SYM. In secs. 6-8 we will present and study a candidate for such a generalization (see \cite{Bandos:2016tsm} for brief description of D=11 case). As we show, these however suffers an unwanted dependence on the deformation vector which is not fixed uniquely in higher dimensional cases. Thus the bottom line of the results of present paper will be  the properties of the constrained on-shell superamplitudes of 10D SYM and 11D SUGRA, the counterparts of (\ref{cAn=cAnlf}),  and the equations they satisfy, the counterparts of (\ref{UcAn=cAn}) and (\ref{UcAn=2cAn}).

\subsection{Massless superparticle mechanics and on-shell superfields}

\label{on-shell4D}

The on-shell superfields describing the states of maximal D=4 SYM and maximal supergravity  can be obtained by quantization of the massless ${\cal N}=4$ and ${\cal N}=8$ superparticle mechanics defined by the following Ferber-Shirafuji action \cite{Ferber:1977qx,Shirafuji:1983zd}
  \begin{eqnarray}\label{S-FS=X}
  S_{4D}&=& \int d\tau \lambda_{A} \bar{\lambda}{}_{{\dot{A}}} (\partial_\tau X^{{A\dot{A}}} - i
  \partial_\tau \Theta_q^{{A}}\, \bar{\Theta}{} ^{{\dot{A}q}}  + i
  \Theta_q^{{A}}\,  \partial_\tau\bar{\Theta}{} ^{{\dot{A}q}} )   , \quad q=1,..., {\cal N} \; . \qquad \end{eqnarray}
We review this well known procedure here as a guide in our search for the on-shell superfield description of 10D SYM and 11D supergravity multiplets
which then will be generalized to  10D and 11D tree superamplitudes.

\subsubsection*{\ref{on-shell4D}.1. Twistor form of the massless superparticle action}

Just by moving the derivative with the use of Leibnitz rules, we can write the action (\ref{S-FS=X}) in the form
 \begin{eqnarray}\label{S-FS=mu}
  S_{4D}&=& \int d\tau (\bar{\lambda}{}_{{\dot{A}}}  \partial_\tau \mu ^{{\dot{A}}}-
  \mu^A  \partial_\tau
  \lambda_{_A}  -2i \partial_\tau\eta_q \, \bar{\eta}{} ^q) = \qquad \nonumber \\   &=& \int d\tau ( \partial_\tau \lambda^{A} \, \mu_{A} - \partial_\tau \mu_{{\dot{A}}}   \bar{\lambda}^{{\dot{A}}} -2i \partial_\tau\eta_q \, \bar{\eta}{} ^q)  , \quad q=1,..., {\cal N} \; , \qquad
  \end{eqnarray}
where
 \begin{eqnarray}\label{mu=}
 \mu ^{{\dot{A}}} &=&  {\lambda}{}_{{{A}}} X^{{A\dot{A}}}_L   \; , \qquad  X^{{A\dot{A}}}_L = X^{{A\dot{A}}}+i
 \Theta_q^{{A}}\, \bar{\Theta}{} ^{{\dot{A}q}}\, =(X^{{A\dot{A}}}_R)^* , \qquad
 \\ \label{eta=} \eta_q &=& \Theta_q^{{A}} \lambda_{_A} = (\bar{\eta}{} ^q)^* \; .
 \end{eqnarray}
 Eqs. (\ref{mu=}) and (\ref{eta=}) are called Penrose incidence relations. They define super-spacetime  as a surface in the supertwistor superspace with complex coordinates  $( \lambda_{A},  \mu ^{{\dot{A}}} , \eta_q)$
 \cite{Penrose:1967wn,Penrose:1972ia,Ferber:1977qx}\footnote{To be more precise, (\ref{mu=}) and (\ref{eta=}) define the super-generalization of the space of light-like geodesics in super-spacetime as a surface in the space of supertwistors. This statement is tantamount to the observation that
 (\ref{mu=}) and (\ref{eta=}) are invariant under the transformations $$
 \delta  X^{A\dot{A}} = t \lambda^A \bar{\lambda}^{\dot{A}} + i \lambda^A {\kappa}_{q} \bar{\Theta}^{\dot{A}q} - i \Theta^A_q \bar{\kappa}{}^q \bar{\lambda}^{\dot{A}} \; , \qquad  \delta  \Theta^A_q =  {\kappa}_{q}  \lambda^A \; ,   \qquad  \delta  \bar{\Theta}^{\dot{A}q} = \bar{\kappa}{}^q \bar{\lambda}^{\dot{A}} \; \qquad  $$
  with an arbitrary real bosonic parameter $t$ and ${\cal N}$ arbitrary fermionic parameters $\kappa_q$.
  The fermionic transformations can be associated with the $\kappa$--symmetry  of the massless superparticle action (\ref{S-FS=X}) which was observed a bit latter in \cite{Siegel:1983hh} (see earlier  \cite{de Azcarraga:1982dw} for the $\kappa$--symmetry of massive ${\cal N}=2$ superparticle).}
    and describe the general solution of the 'helicity  constraint'
 \begin{eqnarray}\label{h=}
2 h=
  \bar{\mu}{}^A
  \lambda_{A} -\bar{\lambda}{}_{{\dot{A}}}  \mu ^{{\dot{A}}} +2i \eta_q \, \bar{\eta}{} ^q \approx 0\; .
 \end{eqnarray}

\subsubsection*{\ref{on-shell4D}.2.Hamiltonian mechanics in twistor formulation }

In the generic case the  canonical momenta are defined as derivative of the Lagrangian with respect to velocities,
 \begin{eqnarray}\label{cP=}
 {\cal P}_M = (-)^M \frac {\partial L}  {\partial (\partial_\tau Y^M)}\; , \qquad
 \end{eqnarray}
 where $$(-)^M:=(-)^{\varepsilon ( Y^M)}= \begin{cases}+ 1\, for \, bosonic \; Y^M \cr -1\, for \, fermionic \; Y^M \; . \end{cases}$$
The canonical Poisson brackets are
 \begin{eqnarray}\label{BPB:=}
 [ \quad  ,\quad \}_{_{P.B.}} =     \frac \partial {\partial   Y^M}\;  \quad \frac \partial {\partial  {\cal P}_M} \quad - (-)^M \frac \partial {\partial  {\cal P}_M} \quad \frac \partial {\partial   Y^M}\quad  \; , \end{eqnarray}
so that $[ {\cal P}_M , {\cal P}_N \}_{_{P.B.}}=0=[ Y^M , Y^N\}_{_{P.B.}}$ and

\begin{eqnarray}\label{PY=PB}
[ Y^N,  {\cal P}_M \}_{_{P.B.}}=  \delta_M{}^N= - (-)^{MN} [ {\cal P}_M , Y^N\}_{_{P.B.}}
\; .
 \end{eqnarray}
 Canonical Hamiltonian ${\cal H}$ is defined by Legandre transform of the Lagrangian $L$
\begin{eqnarray}\label{H=VP-L}
{\cal H}= (-)^M \partial_\tau Y^M {\cal P}_M- L\; ,
 \end{eqnarray} and the Hamiltonian equations of motion have the form
$\partial_\tau Y^M = [ Y^M, {\cal H}]_{_{P.B.}}$.

In our dynamical system described by $S_{4D}=\int d\tau L_{4D}$ of (\ref{S-FS=mu}), the definition
(\ref{cP=}) results in the (second class) constraints which identify the bosonic spinor $\mu$ with the momentum of the bosonic spinor $\lambda$, and the momentum of the fermionic coordinate $\eta_q$ with its complex conjugate $\bar{\eta}^q$, so that the basic nonvanishing Poisson brackets (actually, these are Dirac brackets) read
\begin{eqnarray}\label{PB=b}
&& {}[\mu_A , \lambda^B]_{_{P.B.}}= - \delta_A{}^B \; , \qquad
[\mu_{_{\dot{A}}} ,  \bar{\lambda}{}^{^{\dot{B}}}]_{_{P.B.}}= - \delta_{_{\dot{A}}} {}^{^{\dot{B}}}  \quad (= [\bar{\lambda}{}_{_{\dot{A}}} , \mu ^{^{\dot{B}}}]_{_{P.B.}})\; , \qquad \\
\label{PB=f}
  && \{ \eta_q, \bar{\eta}^p\}_{_{P.B.}}= - \frac i 2 \delta_q{}^p \; .
 \end{eqnarray}
The only constraint (after getting read of the second class constraints by passing to Dirac brackets (\ref{PB=b}), (\ref{PB=f})) is (\ref{h=}). It generates  (on the Dirac brackets) U(1) gauge symmetry characteristic for the $D=4$ (super)twistor approach.

\subsubsection*{\ref{on-shell4D}.3. Quantization and appearance of the on-shell superfields }

Quantization implies the replacement of canonical variables by operators and Poisson brackets by
graded  commutators (i.e. commutators or anti-commutators)
$${} [... , ...\} _{_{P.B.}}\mapsto \frac 1 {i\hbar} [... , ...\} \;  . $$ In the coordinate representation  $$ \mu_A \mapsto -i \hbar \frac {\partial}{\partial \lambda^A}\, , \qquad
\mu_{_{\dot{A}}} \mapsto  -i\hbar  \frac {\partial}{\partial \bar{\lambda}{}^{^{\dot{B}}}}\;  \qquad
\bar{\eta}^p \mapsto  \frac \hbar  2 \frac {\partial}{\partial {\eta}_p}\; , $$  (below we will set $\hbar=1$) and the 'wavefunction' (or 'classical field') depends holomorphically on the fermionic variable,  {\it i.e.} it is given by an on-shell superfield (\ref{Phi=3,4}). The constraints (\ref{h=}) should be imposed on the wave function in its operator form (\ref{U=}), $h\mapsto -i\hbar\hat{h}$.
 Actually, the operator ordering can produce an arbitrary constant contribution so that the quantum constraints for the wavefunction of massless superparticle in ${\cal N}$-extended $D=4$ on-shell superspace reads
\begin{eqnarray}\label{h-s=0}
(\hat{h}-s)\Phi=0\; , \qquad 2s\in {\bb Z}\; .
 \end{eqnarray} The fact that $s$ should be quantized in half-integer unit of $\hbar$ follows from the requirement that $\Phi$ is a single--valued function of complex variables $\lambda_A$, while the choice
$s=1$ for ${\cal N}=4$ and  $s=2$ for ${\cal N}=8$ is made from 'physical reasons' (of absence in the multiplet under consideration of gravity in the former and of higher spin fields in the latter case).

The equation (\ref{PB=b}) for the wavefunction of superparticle suggests  that the scattering amplitudes obey
Eqs. (\ref{UcAn=cAn}) and (\ref{UcAn=2cAn}) in the case of ${\cal N}=4$ SYM ($s$=1)  and ${\cal N}=8$ supergravity ($s=2$), respectively.

\subsubsection*{\ref{on-shell4D}.4. From D=4 to D=10 and D=11}

In this paper our main interest is in the properties of tree  amplitudes and superamplitudes of the D=10 SYM and D=11 SUGRA theories. Our way to arrive at the appropriate spinor helicity and on-shell superfield formalism will pass through quantization of  the D=10 and D=11 counterparts of the Ferber-Shirafuji action (\ref{S-FS=X}) \cite{Bandos:1996ju,Bandos:2007mi} which we will describe in the next section.
Instead of unconstrained 4D Weyl spinors $\lambda_A$, $\bar{\lambda}_{\dot{A}}$ these D=10 and D=11 twistor-like actions use the spinor moving frame variables, also called Lorentz harmonics, parametrizing the celestial sphere $S^{(D-2)}$ as a coset
of D-dimensional Lorentz group $SO(1,D-1)$ over its Borel subgroup $[SO(1,1)\otimes SO(D-2)]\subset \!\!\!\!\!\! \times K_{D-2}$ \cite{Galperin:1991gk,Delduc:1991ir,Galperin:1992pz}.

The counterparts of these $\frac {SO(1,D-1)}{[SO(1,1)\otimes SO(D-2)]\subset \!\!\!\! \times K_{D-2}}$ Lorentz harmonics, spinor frame variables associated to each of the scattered particles,  will be used below to construct spinor helicity formalism for amplitudes of $D=10$ SYM  and $D=11$ SUGRA.

In the case of $D=10$ and $D= 11$ superparticle, the spinor {\it moving} frame variables can be considered as additional coordinate of target superspace which then can be called {\it  Lorentz harmonic superspace}\footnote{See \cite{Galperin:1984av,Galperin:1984bu,Galperin:2001uw} for the concept of harmonic superspace.}. Their  presence allows to perform a  change of variables similar to passing to the twistor variables in (\ref{S-FS=mu}), which can be seen now  as change of coordinates of this from
 'central' basis, given by  the standard bosonic vector and fermionic spinor coordinates of D=10 and D=11 superspaces plus the above mentioned spinor harmonics,  to a different, so-called analytical coordinate basis
 \footnote{The quantization of superparticle in the analytical basis of Lorentz harmonic superspace was discussed for the first time in \cite{Sokatchev:1985tc,Sokatchev:1987nk} using the vector moving frame variables (vector harmonics) only. The formulation of D=4 Ferber--Shirafuji superparticle in terms of corresponding spinor moving frame variables (spinor harmonics) can be found in
\cite{Bandos:1990ji}. There the massless superparticle  was considered both in the central and in the analytic coordinate basis of D=4 Lorentz harmonic superspace.
}. The quantization of superparticle in this analytical basis will provide us with the on-shell superfields describing the degrees of freedom of 10D SYM and 11D SUGRA. The tree superamplitudes will be constructed as multiparticle generalizations of these on-shell superfields.

\section{Superparticle in ten and eleven dimensions. Spinor moving frame
formulation}

\subsection{Brink-Schwarz action for massless superparticle in D=10 and D=11}
The Brink-Schwarz superparticle action can be defined in any dimensions
\begin{eqnarray}\label{S-BS=}
S_{BS}[X,p,\theta]= \int d\tau  \left( p_a \left({\partial}_\tau X^a -i {\partial}_\tau \theta \Gamma^a \theta \right)
-{e\over 2}  p^ap_a\right) \; .    \qquad
\end{eqnarray}
Here $X^a=X^a(\tau)$ and  $\theta^\alpha=\theta^\alpha(\tau)$ are bosonic and fermionic coordinate functions, $p_a= p_a(\tau)$ is auxiliary momentum variable, $e(\tau)$ is an einbein field playing the role of  Lagrange multiplier, and ${\partial}_\tau\theta \Gamma^a \theta = {\partial}_\tau\theta^{\alpha} \Gamma^a_{\alpha\beta} \theta^\beta$.

In D=11 case $a=0,1,...,9,10$,  $\alpha=1,...,32$, the real fermionic $\theta^\alpha$ are transformed as   Majorana spinor of $SO(1,10)$  and $\Gamma^a_{\alpha\beta}=\Gamma^a_{\beta\alpha}= \Gamma^a_{\alpha}{}^{\gamma}C_{\gamma\beta}$ are the products of 11D Dirac matrices $\Gamma^a_{\alpha}{}^{\gamma}=- (\Gamma^a_{\alpha}{}^{\gamma})^*$ obeying the Clifford algebra, $\Gamma^a\Gamma^b+\Gamma^b\Gamma^a = 2\eta^{ab}{\bb I}_{32\times 32}$, and of 11D charge conjugation matrix $C_{\gamma\beta}= - C_{\beta\gamma}=- (C_{\gamma\beta})^*$. We will also use
$\tilde{\Gamma}^{a \; \alpha\beta}= \tilde{\Gamma}^{a \; \alpha\beta}= C^{\alpha\gamma} {\Gamma}^{a \;\beta}_\gamma$.

In D=10 case the fermionic $\theta^\alpha$ is real  Majorana-Weyl (MW) spinor of $SO(1,9)$,  $\alpha=1,...,16$, and $\Gamma^a_{\alpha\beta}= \sigma^a_{\alpha\beta}=\sigma^a_{\beta\alpha}$ obey $\sigma^a\tilde{\sigma}^b +\sigma^b\tilde{\sigma}^a=
2\eta^{ab}{\bb I}_{16\times 16}$ in product with their upper-indices  counterpart
$\tilde{\Gamma}^{a \; \alpha\beta}= \tilde{\sigma}^{a \; \alpha\beta}=\tilde{\sigma}^{a \;\beta\alpha}$. There is no covariant way of rising or lowering Mayorana-Weyl spinor indices in D=10 as these correspond to different chiralities (in other words,  the charge conjugation matrix does not exist in D=10 MW spinor representation).

In any dimension D the  action (\ref{S-BS=})  is invariant under rigid supersymmetry
\begin{eqnarray}\label{susyXP}
\delta_\varepsilon X^a= i \theta\Gamma^a\varepsilon  \; , \qquad
\delta_\varepsilon \theta^\alpha =\varepsilon^{\alpha} \; , \qquad
\delta_\varepsilon p_a=0 \; ,  \qquad \delta_\varepsilon e=0 \; , \qquad
\;
\end{eqnarray}
and under local fermionic $kappa$--symmetry \cite{Siegel:1983hh} (see also earlier \cite{de Azcarraga:1982dw})
\begin{eqnarray}\label{kappa=red}
\delta_\kappa X^a= -{i} \theta\Gamma^a\delta_\kappa \theta \;  \; , \qquad
\delta_\kappa \theta^\alpha = p_a\tilde{\Gamma}^{a\alpha\beta}\kappa_\beta \; , \qquad
\delta_\kappa  p_a = 0\; , \qquad  \delta_\kappa e= -4i {\partial}_\tau \theta\kappa   \; . \qquad
\;
\end{eqnarray}

\bigskip

\subsection{Moving frame formulation of massless superparticle mechanics}

In this section we follow closely the discussion in \cite{Bandos:2007mi}.
The variation of (\ref{S-BS=}) with respect to the Lagrange multiplier $e(\tau)$ results in the mass--shell condition
\begin{eqnarray}\label{p2=0}
p_ap^a=0\, . \qquad
\end{eqnarray} This is algebraic equation and thus, knowing its solution we can substitute it back to the action. Hence, formally we can  write the action (\ref{S-BS=}) in the equivalent form
\begin{eqnarray}\label{S'-BS=} S'_{BS} = \int d\tau  \left({\partial}_\tau X^a -i {\partial}_\tau \theta \Gamma^a \theta \right)\, p_a\vert_{p^2=0}\, . \qquad
\end{eqnarray}
 A particular solution of the constraint (\ref{p2=0})  is given by
\begin{eqnarray}\label{p=sol} p_a= \rho \, (1,0,...,0,-1)\,  \qquad
\end{eqnarray} with arbitrary $ \rho (\tau) $  describing the energy of the massless particle.
Any other  solution can  be obtained from this
 by performing  a $\tau$-dependent $SO(1,D-1)$  Lorentz transformation
\begin{eqnarray}\label{uab=in}
 u_a^{(b)}(\tau ) \; \in \; SO(1,D-1)\,  \qquad \Leftrightarrow \qquad u_a^{(b)}u^{a(c)}= \eta^{(b)(c)} =
 diag (+1,-1,...,-1)\; .
\end{eqnarray}
Hence, the general solution of the mass-shell constraint  reads
\begin{eqnarray}\label{Pa=Up}
p_a(\tau ) = u_a^{(b)}
p_{(b)}=
\rho (\tau ) (u_a^{0}- u_a^{(D-1)})=: \rho^{\#} (\tau ) u_a^{=}(\tau )\, ,
\end{eqnarray}
where we have supplied the energy variable $\rho$ of (\ref{p=sol}) by the indices
$^{\#}=^{++}$ which indicate the invariance of the solution (\ref{Pa=Up}) under arbitrary scaling
of $u_a^{=}$ compensated by an opposite scaling of $\rho^{\#}$. As we will see below this scaling symmetry may be associated with $SO(1,1)$ subgroup of Lorentz group $SO(1,D-1)$.

By construction, the vector $u_a^{=}(\tau )= (u_a^{0}- u_a^{(D-1)})$ is light-like.
Substituting this general solution into the Brink--Schwarz action we arrive at
\begin{eqnarray}\label{S=mf}
S = \int d\tau \rho^\# u_a^{=}\left({\partial}_\tau X^a -i{\partial}_\tau \theta \Gamma^a \theta \right)
\; ,    \qquad   u_a^{=}u^{a=}=0\; .
\end{eqnarray}
This action describes  {\it moving frame} formulation of massless superparticle, as it includes a light-like vector (\ref{Pa=Up}) which we have introduced as a difference of two columns of the Lorentz group valued matrix (\ref{uab=in}). It is convenient to consider this matrix to be an element of proper Lorentz group simply connected to unity $SO^\uparrow (1,D-1)$ (see \cite{Galperin:1991gk,Delduc:1991ir}), and to write this matrix  explicitly in terms of light-like vector $u_a^{=}$  and its complementary $u_a^{\#}(\tau )= (u_a^{0}+ u_a^{(D-1)})$,
\begin{eqnarray}\label{Uab=in}
 u_a^{(b)}(\tau ) = \left( {1\over 2}\left( u_a^{=}+u_a^{\#}
 \right), \; u_a^i \, , {1\over 2}\left( u_a^{\#}-u_a^{=}
 \right)\right)\; \in \; SO^\uparrow (1,D-1)\,   . \qquad
\end{eqnarray}

At this stage it might seem that we just have substituted a simple action (\ref{S'-BS=}) by an almost identical but a bit more complicated (\ref{S=mf}) without any benefit. The benefit however is present (although still hidden) as the clear geometrical and group theoretical meaning of moving frame will allow us to extract a square root of (\ref{Uab=in}), which gives us the {\it spinor moving frame} providing the 10D and 11D counterparts of spinor helicity variables. We will see that, when the moving frame variables are understood as constructed from the spinor moving fame variables, the action (\ref{S=mf}) becomes a D-dimensional counterpart of the D=4 Ferber-Shirafuji action (\ref{S-FS=X}).

But before introducing the spinor moving frame, let us discuss some technical details about moving frame variables.

\subsection{Properties of moving frame variables (vector harmonics)}

The statement (\ref{Uab=in}) is equivalent to preservation of the flat Lorentzian metric under similarity transformations with the moving frame matrices $U=|| u_a^{(b)}||$, i.e. to $U^T\eta U=\eta$ and $U\eta U^T=\eta$. These equations can be equivalently written as $u_a^{(b)}u^{a(c)}= \eta^{(b)(c)}$ (see (\ref{uab=in})), which  splits into
\begin{eqnarray}\label{uu=0}
&& u_a^{=}u^{a=}=0 \; , \quad \\ \label{uu=2}
 && u_a^{\#}u^{a\#}=0 \; , \quad  u_a^{=}u^{a\#}=2 \; , \quad \\ \label{uui=0} && u_a^{I}u^{a=}=0 =u_a^{I}u^{a\#} \; , \quad  u_a^{I}u^{aJ}=-\delta^{IJ} \; , \qquad
\end{eqnarray}
and as
\begin{eqnarray}\label{I=UU}
\delta_a{}^b= {1\over 2}u_a^{=}u^{b\#}+ {1\over 2}u_a^{\#}u^{b=}-  u_a^{I}u^{bI} \; , \qquad
\end{eqnarray}
respectively.

The set of constrained variables (\ref{Uab=in}) considered as worldline  fields, $u_a^{=}=u_a^{=}(\tau)$, etc., describe a moving frame attached to every point of the worldsheet in such a way that canonical momentum of $X^a(\tau)$ field is proportional to $u_a^=(\tau)$ (as it can be seen from (\ref{Pa=Up})). Hence these fields can be called moving frame fields or {\it moving frame variables}.

On the other side, one can treat the model (\ref{S=mf}) as a model in superspace enlarged by additional bosonic directions parametrizing $SO(1,D-1)$ group or its coset. Such an enlarged  superspace, first introduced in \cite{Sokatchev:1985tc} is called {\it Lorentz
harmonic superspace}\footnote{In \cite{Sokatchev:1985tc,Sokatchev:1987nk} the name 'light-cone superspace' was used. The term 'harmonic superspace' was introduced in the seminal papers
\cite{Galperin:1984av,Galperin:1984bu} where the off-shell description of ${\cal N}=2,3$ SYM theory and
${\cal N}=2$ supergravity in terms of unconstrained superfields have been constructed; see also
\cite{Galperin:2001uw}.}. $u_a^=$, $u_a^\#$ and $u_a^I$ constrained by (\ref{uu=0})--(\ref{I=UU}) (or equivalently by the condition (\ref{Uab=in}) on $u_a^{(b)}$ matrix collecting them) can be considered as coordinates of the 'internal' sector of this superspace and can be called {\it Lorentz harmonic variables}. Of course these name can be applied as well to the worldline fields, $u_a^{(b)}(\tau)$, etc., so that the names Lorentz harmonics and moving frame variables can be used as synonyms.

Notice also that the splitting (\ref{Uab=in}) of the SO(1,D-1) valued matrix $u_a^{(b)}(\tau)$ is invariant under the subgroup $SO(1,1)\times SO(D-2)$ of the Lorentz group $SO(1,D-1)$ so that in a model which is gauge invariant under local $SO(1,1)\times SO(D-2)$ transformations, the  fields $u_a^{(b)}=(u_a^=, u_a^\# ,u_a^I)$ can be identified as homogeneous coordinates of the coset
$\frac {SO(1,D-1)}{SO(1,1)\times SO(D-2)}$,
\begin{eqnarray}\label{U=G-H=str}
\{ u_a^=, u_a^\# ,u_a^I\} \; = \;  \frac {SO(1,D-1)}{SO(1,1)\times SO(D-2)} \; . \qquad
\end{eqnarray}
The (spinor) moving frame variables used to describe (super)string model in its ''twistor-like Lorentz harmonic formulation'' of \cite{BZ-str,Bandos:1992ze} parametrize this non-compact coset.

The superparticle model is more economic: of all $u_a^{(b)}(\tau) \in SO(1,D-1)$ it contains only one light-like vector $u_a^{=}(\tau)$. Hence, if we treat this light-like vector as a part of moving frame matrix  $u_a^{(b)}(\tau)$, this model is gauge invariant, besides $SO(1,1)\times SO(D-2)$, under such $SO(1,D-1)$ transformations which mixes the complementary light--like vector  $u_a^{\#}(\tau)$ with spacelike vectors $u_a^{I}(\tau)$. These transformation form an Abelian subgroup $K_{D-2}$ of the Lorentz group $SO(1,D-1)$. The complete set of the gauge symmetries in the (spinor) moving frame sector of our superparticle model is thus a semidirect product $[SO(1,1)\times SO(D-2)]\subset\!\!\!\!\!\!\times K_{D-2}$ which is the so-called Borel subgroup of $SO(1,D-1)$. The latter fact implies that the coset $\frac {SO(1,D-1)} {[SO(1,1)\times SO(D-2)]\subset\!\!\!\!\times K_{D-2}}$ is compact; it is isomorphic to the sphere ${\bb S}^{D-2}$ which (when the massless (super)particle model is considered) can be identified as celestial sphere of a D-dimensional observer \cite{Galperin:1991gk,Delduc:1991ir}. Resuming,
\begin{eqnarray}\label{U=G-H}
\{ u_a^=, u_a^\# ,u_a^I\} \; = \;  \frac {SO(1,D-1)}{[SO(1,1)\times SO(D-2)]\subset\!\!\!\!\!\!\times K_{D-2}} = {\bb S}^{D-2}\; , \qquad
\end{eqnarray}
or, more schematically,
\begin{eqnarray}\label{u=G-H}
\{ u_a^=\} \; = \;  \frac {SO(1,D-1)}{[SO(1,1)\times SO(D-2)]\subset\!\!\!\!\!\!\times K_{D-2}} = {\bb S}^{D-2}\; . \qquad
\end{eqnarray}

Again, the above observations result in nontrivial consequences only after 'extracting the square roots' of the moving frame vectors, i.e. after reformulating this in terms of spinor moving frame variables, which are of interest for us as basic elements of D-dimensional spinor helicity formalism. Let us describe these in a notation especially suitable for  D=10 and D=11 dimensional cases.

\subsection{Spinor moving frame variables (spinor harmonics)}

The basic fact allowing for  extracting a square root of the moving frame matrix is the Lorentz invariance of D-dimensional Dirac (or Pauli)  matrices $\Gamma^a$ and also of
D-dimensional charge conjugation matrix $C_{\alpha\beta}$ if such exists in minimal D-dimensional spinor representation.
The conditions of these  Lorentz invariance can be written, in particular, for the moving fame matrix
(\ref{Uab=in}), (\ref{uab=in}),
\begin{eqnarray}\label{VGVt=G} V\Gamma_b V^T =  u_b^{(a)} {\Gamma}_{(a)}\, , \qquad V^T \tilde{\Gamma}^{(a)}  V = \tilde{\Gamma}^{b} u_b^{(a)}\;
 \, , \qquad \\ \label{VCVt=C}   VCV^T=C \; , \qquad if\quad  C\; exists\; for\; given\; D\; .  \qquad
\end{eqnarray}
These equations define the {\it spinor moving frame matrix} $V$ which  takes values in the fundamental representation of the doubly covering group $Spin(1,D-1)$ of the Lorentz group $SO(1,D-1)$, $V \in Spin(1,D-1)$. It is $n\times n$ matrix, where $n$ is dimension of a minimal spinor representation of $D$-dimensional Lorentz group: $n=16$ for D=10 and $n=32$ for D=11.

The $SO(1,1)\times SO(D-2)$ invariant splitting (\ref{uab=in}), $u_b^{(a)} =(u_b^{=}, u_b^{\#}, u_b^{I})$, is reflected by splitting the spinor moving frame matrix on two rectangular blocks,
\begin{eqnarray}\label{harmV=D}
V_{\alpha}^{(\beta)}= \left(\begin{matrix} v_{\alpha \dot{q}}^{\; +} , & v_{\alpha q}^{\; -}
  \end{matrix}\right) \in Spin(1,D-1)\; , \qquad
 \;  \qquad
\end{eqnarray}
The columns of two blocks of these matrices are enumerated by indices of (the same or different) representations
of $SO(D-2)$ subgroup of $SO(1,D-1)$. In particular, in the case of D=10, where the minimal MW spinor representation
is 16-dimensional,  these are c- and s-spinor indices of $SO(8)$,
\begin{eqnarray}\label{q=D10}
 D=10:\qquad \alpha=1,...,16\; , \quad \dot{q}=1,...,8 \; , \quad q=1,...,8
 \;  , \qquad
\end{eqnarray}
while in $D=11$ these are two copies of the same real (Majorana) spinor representation of the $SO(9)$ group,
\begin{eqnarray}\label{q=D11}
 D=11:\qquad \alpha=1,...,32\; , \quad q= \dot{q}=1,...,16\; , \quad   v_{\alpha \dot{q}}^{\; +}\equiv v_{\alpha {q}}^{\; +}\; . \quad
\end{eqnarray}
The sign indices $^\pm$ of two blocks, $v_{\alpha \dot{q}}^{\; +}$ and $v_{\alpha q}^{\; -}$, of the spinor moving frame matrix (\ref{harmV=D}) indicate their scaling properties with respect to the $SO(1,1)$ transformations.

Generically, working with spinor moving frame variables one cannot avoid the use of
the inverse of the spinor moving frame matrix
\begin{eqnarray}\label{harmV-1=D}
V_{(\beta)}^{\;\;\; \alpha}= \left(\begin{matrix}  v^{+\alpha}_{{q}}
 \cr  v^{-\alpha}_{_{\dot{q}}} \end{matrix} \right) \in Spin(1,D-1)
 \;  \qquad
\end{eqnarray}
the blocks of which obey  $V_{\alpha}^{(\beta)}V_{(\beta)}{}^\gamma :=  v_{\alpha}^{-\dot{q}} v^{+\gamma}_{\dot{q}}
+ v^{-\alpha}_q v^{-\gamma}_q
=
\delta_{\alpha}^\gamma$ and
\begin{eqnarray}\label{v-qv+p=}
&
v^{+\alpha}_{{q}} v_{\alpha {p}}^{\; -} = \delta_{qp}
 \; ,  \qquad & v^{+\alpha}_{{q}}  v_{\alpha \dot{p}}^{\; +}=0\;  , \qquad
 \nonumber  \\
 &
 v^{-\alpha}_{\dot{q}}   v_{\alpha q}^{\; -}=0\;  , \qquad & v^{-\alpha}_{\dot{q}}   v_{\alpha p}^{\; +}=\delta_{\dot{q}\dot{p}}\;  . \qquad
\end{eqnarray}

In D=11 the elements of the inverse spinor moving frame matrix can be constructed from the elements of  (\ref{harmV=D}) with the use of charge conjugation matrix  \begin{eqnarray}
\label{V-1=CV-A} D=11\; : \qquad  v_{q}^{\pm \alpha}= \pm  i C^{\alpha\beta}v^{\; \pm}_{\beta q}\, .
\qquad
 \end{eqnarray}
In D=10 the charge conjugation matrix does not exist and we define the elements of (\ref{harmV-1=D}) by the constraints (\ref{v-qv+p=}).

For our dynamical system an especially important relations between the vector and spinor moving frame variables (vector and spinor Lorentz harmonics) (\ref{VGVt=G}) are
\begin{eqnarray}\label{u==v-v-}
 && u_a^= \Gamma^a_{\alpha\beta}= 2v_{\alpha q}{}^- v_{\beta q}{}^-  \; , \qquad
 v^-_{{q}} \tilde{\Gamma}_{a}v^-_{{p}}= u_a^= \delta_{{q}{p}}, \qquad  \qquad \nonumber \\  && u_a^= \tilde{\Gamma}{}^{a\, \alpha\beta}= 2v^{-\alpha}_{\dot q}   v^{-\beta}_{\dot q} \; , \qquad  v^-_{\dot{q}} {\Gamma}_{a}v^-_{\dot{p}}= u_a^= \delta_{\dot{q}\dot{p}}
 \; . \qquad
\end{eqnarray}
The equations in the first line of (\ref{u==v-v-}) allow to state that $v_{\alpha q}^{\; -}$ is the square root of the light--like moving frame vector $u_a^=$ in the same sense as in D=4 the bosonic spinor $\lambda_A$ can be called square root of the light-like 4--momentum due to Eq. (\ref{p=ll=4D}). The equations in the second line state  the same for the block $v^{-\alpha}_{\dot q}$ of the inverse spinor moving frame matrix (\ref{harmV-1=D}). This fact is nontrivial for D=10, while for D=11 $v^{-\alpha}_{\dot q}\equiv v^{-\alpha}_{q}$ is expressed through
 $v_{\alpha q}^{\; -}$ by (\ref{V-1=CV-A}) and the equations in the second line of
(\ref{u==v-v-}) follow from the first line.

Similar  relations with other  moving frame vectors involve the complementary harmonic variables $v_{\alpha {q}}^{\; + }$ and $v_{\dot{q}}^{+ {\alpha}}$:
\begin{eqnarray}
\label{v+v+=u++}
& v_{\dot{q}}^+ \tilde{\Gamma}_{ {a}} v_{\dot{p}}^+ = \; u_{ {a}}^{\# } \delta_{\dot{q}\dot{p}}\; , & \qquad 2 v_{{\alpha}\dot{q}}{}^{+}v_{{\beta}\dot{q}}{}^{+}= {\Gamma}^{ {a}}_{ {\alpha} {\beta}} u_{ {a}}^{\# }\; , \qquad \\
\label{v+v+=u++1}
& v_{{q}}^+ {\Gamma}_{ {a}} v_{{p}}^+ = \; u_{ {a}}^{\# } \delta_{{q}{p}}\; , & \qquad 2 v_{{q}}^{+ {\alpha}}v_{{q}}^{+}{}^{ {\beta}}= \tilde{\Gamma}^{ {a} {\alpha} {\beta}} u_{ {a}}^{\# }\; , \qquad
\\ \label{uIs=v+v-}
& v_{{q}}^- \tilde {\Gamma}_{ {a}} v_{\dot{p}}^+=u_{ {a}}^{I} \gamma^I_{q\dot{p}}\; , &\qquad
 2 v_{( {\alpha}|{q} }{}^- \gamma^I_{q\dot{q}}v_{|{\beta})\dot{q}}{}^{+}= {\Gamma}^{a}_{\alpha\beta} u_{ {a}}^{I}\; , \quad\\ \label{uIs=v-v+}
&
 v_{\dot q}^- {\Gamma}_{ {a}} v_{{p}}^+ = - u_{ {a}}^{I} \gamma^I_{p\dot{q}}\; , &\qquad
 2 v_{\dot q}^{-( {\alpha}}\gamma^I_{q\dot{q}}v_{{q}}^{+}{}^{ {\beta})}=-  \tilde{\Gamma}^{ {a} {\alpha} {\beta}} u_{ {a}}^{I}\; . \quad
\end{eqnarray}
Here, for D=10 $\gamma^I_{p\dot{q}}=:\tilde{\gamma}^I_{\dot{q}p}$ are Klebsh-Gordan coefficients of
SO(8) group, $q,p=1,..., 8$ are  s-spinor (8s) indices, $\dot{q},\dot{p}=1,...,8$ are c-spinor (8c) indices  and I=1,.., 8 is SO(8) vector  index (8v-index); all three representations are 8 dimensional  in this case. For D=11 $q,p\equiv \dot{q}, \dot{p}=1,...,16$ are spinor indices of SO(9) and
$\gamma^I_{qp}=\gamma^I_{pq}$ are SO(9) gamma matrices.
For completeness, let us also repeat here that in D=10 $\Gamma^a_{\alpha\beta}=\sigma^a_{\alpha\beta}$,
 $\tilde{\Gamma}{}_a^{\alpha\beta}=\tilde{\sigma}{}_a^{\alpha\beta}$ are generalized 10D Pauli
matrices while for D=11 $ \tilde{\Gamma}{}_a^{\alpha\beta}=
 C^{\alpha\gamma}{\Gamma}{}_{a\gamma \delta}C^{\delta\beta} $.

\subsection{Spinor moving frame formulation of the massless superparticle action}

Using (\ref{u==v-v-}), we can write (\ref{S=mf}) in the form
\begin{eqnarray}\label{S=smf}
S =  \int d\tau  \rho^\#
\left(u_a^{=}{\partial}_\tau X^a -2 i {\partial}_\tau \theta^\alpha v_{\alpha q}^{\; -} \, \theta^\beta v_{\beta {q}}^{\; -}  \right)
\; ,    \qquad   \begin{cases} u_a^{=}u^{a=}=0\; , \cr u_a^= \Gamma^a_{\alpha\beta}= 2v_{\alpha q}^{\; -} v_{\beta  q}^{\; -}  \; , \cr
 v^-_{{q}} \tilde{\Gamma}_{a}v^-_{{p}}= u_a^= \delta_{{q}{p}}. \end{cases}  \;
\end{eqnarray}
This counterpart of the D=4 Ferber-Schirafuji action is the basis of the spinor moving frame formulation of massless D--dimensional superparticle (also called twistor-like Lorentz harmonic formulation) proposed and investigated in   \cite{Bandos:1996ju} for D=10 and in \cite{Bandos:2006nr,Bandos:2007mi,Bandos:2007wm} for 11D cases.

In distinction to the Brink-Schwarz formulation (\ref{S-BS=}), the  $\kappa$--symmetry of the spinor moving frame  action (\ref{S=smf}) is irreducible,
\begin{eqnarray}\label{kappa=irr}
\delta_\kappa X^a= -{i} \theta\sigma^av_q^- \; \kappa^{+q} \; , \qquad
\delta_\kappa \theta^\alpha =\kappa^{+q} v_q^{-\alpha} \; , \qquad
\delta_\kappa  v_q^{-\alpha} = 0= \delta_\kappa  v_{\alpha \dot{q}}{}^- \; . \qquad
\;
\end{eqnarray}
But moreover, as we will see
below, in section \ref{secAnalit}, we can perform such a change of variables after which this local fermionic symmetry is gauge fixed automatically: the action in this basis contains only variables which are inert under $\kappa$--symmetry.

However,  before passing to this issue, for which we will need to consider differential calculus on the space of spinor moving frame (Lorentz harmonic) variables, we would like to show that the D=10 spinor helicity variables  of Caron-Huot and O'Connell are actually spinor frame variables, and present the basic  equations of this D=10 spinor helicity formalism and of its  D=11 generalization.

\section{Spinorial frames, D=10 spinor helicity formalism of Caron-Huot and O'Connell, and its D=11 generalization}
\label{SpHel}

In this section we show that spinor helicity formalism for D=10 amplitudes proposed by Caron--Huot and O'Connell in \cite{CaronHuot:2010rj} can be treated as spinor frame approach to the amplitudes, and use this observation to construct D=11 generalization of this formalism.

It is worth stressing that in this paper, following the line of
\cite{Britto:2005fq,ArkaniHamed:2008gz,CaronHuot:2010rj}  we will consider the on-shell amplitudes only and do not use neither (full field theory) Lagrangian nor Feynman diagrams in their derivation. Such  pure on-shell approaches to the amplitudes (the set of which also incudes the unitary cut technique \cite{Bern:2011qn}) revived
key ideas of the S-matrix program \cite{Eden:1966dnq} aiming to restore the amplitudes mainly from the kinematics, symmetries and locality principle. For maximally supersymmetric D=4 theories such approaches have been quite successful resulting in a progress which cannot be reached with Feynman diagrams, see e.g. \cite{Elvang:2015rqa,ArkaniHamed:2017} and refs therein. For higher dimensional cases, especially for D=10 and D=11, the on-shell (super)amplitude methods  still need to be further elaborated and new ideas and new guides for the construction of the amplitudes are welcome. The aim of this paper is to contribute in such a development.

Particularly, in secs. 5,6 we use the statement that tree superamplitudes are multiparticle counterparts of the wavefunctions obtained in a suitable quantization of superparticle mechanics to find the small group index structure and transformation properties of  the superamplitudes as well as the equations which obey these superamplitudes. In this section we discuss the tree amplitudes which appear as leading components of these superamplitudes and obtain
the supersymmetric Ward identities relating several amplitudes. To introduce the amplitudes independently of superamplitudes, we consider them as multiparticle counterparts of the solution of the linearized equations of motion for the fields of 10D SYM and 11D SUGRA supermultiplets obtained with the use of spinor moving frame. This way also makes manifest how the supersymmetry transformations act on the amplitudes which, together with the evident fact that the amplitude for the process involving an odd number of fermions (counting all the incoming and outcoming particles) should vanish, gives us the above mentioned Ward identities.

Although we are not ready to reproduce the amplitudes just from the solution of  the Ward identities (as suggests the S-matrix program), below (in sec. 8.2.) we will check that 3-point amplitudes with two fermionic legs suggested by the form of SYM (and superstring) vertices found in \cite{Green:1981xx,Schwarz:1982jn} do obey these Word identities, and also use these to restore the form of  3-point amplitude with all three bosonic legs \cite{Green:1981xx,Schwarz:1982jn}.

\subsection{D=10 and D=11
spinor moving frame as spinor helicity variables}

To arrive at D=10 spinor helicity formalism \cite{CaronHuot:2010rj} used in \cite{Boels:2012ie,Wang:2015jna,Wang:2015aua} to study the 10D SYM and type IIB supergravity amplitudes, and at its D=11 generalization briefly presented  in \cite{Bandos:2016tsm}, let us consider a vector frame $(u_{a i}^=, u_{a i}^\#, u_{a i}^I)$ attached  to a light-like  10D (11D) momentum $k_{ai}$,  $k_{ai}k^a_{i}=0$,  of i-th of scattered particles in such a way that ({\it cf.} (\ref{Pa=Up}))
\begin{eqnarray}\label{kia=Up}
k_{i}^a =  \rho^{\#}_{i} u_{i}^{a=}\, . \qquad
\end{eqnarray}
We  allow $ \rho^{\#}_{i}\;$ to be negative, associating such with
in-states of the scattering, while positive $ \rho^{\#}_{i}\;$ are associated to out-states. This corresponds to the usual convention on that all the momenta in the amplitude  are considered as, say, outgoing  but the
 incoming particles have negative energy $k_{i}^0$.

The other vectors of the $i$-th frame
\begin{eqnarray}\label{harmUi=}
u_{a\, i}^{(b)}= (u_{a i}^=, u_{a i}^\#, u_{a i}^I)\; \in \; SO(1,D-1)  \qquad
\end{eqnarray}
are not fixed by any additional conditions except for that they form an orthogonal and normalized frame with $u_{a i}^=$ ({\it i.e.} obey  (\ref{uu=2}), (\ref{uui=0})). Thus the transformations mixing
$u_{a i}^I$'s among themselves ($SO(D-2)$) and with $ u_{a i}^\#$ ($K_{D-2}$) can be considered as a kind of gauge symmetry transformations acting on the frame. In the same manner one can treat the scaling transformations of $u_{a i}^=$: supplemented by opposite scaling of $\rho^{\#}_{(i)}$. These leave invariant the momentum (\ref{kia=Up}) and, when supplemented by the opposite scaling of $u_{a i}^\#$,
can be identified with  $SO(1,1)\subset SO(1,D-1)$ transformations
leaving  invariant (\ref{uu=2}) and, hence, the splitting  (\ref{harmUi=}).

Thus the complete set of transformations which can be used as identification relations on the class  of frames defined by the only condition (\ref{kia=Up}) (set of gauge symmetries relating equivalent vector frames) form the $[SO(1,1)\otimes SO(D-2)]\subset\!\!\!\!\!\! \times K_{D-2}$ subgroup of $SO(1,D-1)$. The frames obeying (\ref{kia=Up}) with some light-like momentum span the coset
$\frac {SO(1,D-1)} {[SO(1,1)\otimes SO(D-2)]\subset\!\!\!\! \times K_{D-2}}$ isomorphic to the celestial sphere of a D-dimensional observer. We would like to express this fact by writing ({\it cf.} (\ref{u=G-H}))
\begin{eqnarray}\label{u=i=G-H}
\{ u_{a\, i}^{=}\}  \; \in   \; \frac {SO(1,D-1)} {[SO(1,1)\otimes SO(D-2)]\subset\!\!\!\!\!\!\times  K_{D-2}} \; .  \qquad
\end{eqnarray}
Notice that, in distinction to (\ref{u=G-H}), here we have used   $ \in $ symbol (rather then $=$), as this  is more appropriate when we are speaking about a definite scattering process.


In D=10 case we introduce the corresponding spinor  frame matrix
\begin{eqnarray}\label{harmVi=D}
V_{\alpha \, i}^{(\beta)}= \left(\begin{matrix} v_{\alpha \dot{q}\, i}^{\; +} , & v_{\alpha q\, i }^{\; -}
  \end{matrix}\right) \in Spin(1,D-1)\;  \qquad
 \;  \qquad
\end{eqnarray}
and its inverse
\begin{eqnarray}\label{harmV-1i=D}
V_{(\beta)\, i}^{\;\;\; \alpha}= \left(\begin{matrix}  v^{+\alpha}_{{q}\, i}
 \cr  v^{-\alpha}_{{\dot{q}}\, i} \end{matrix} \right) \in Spin(1,D-1)
 \;  \qquad
\end{eqnarray}
related among themselves by (\ref{v-qv+p=}) and
 to the above  frame vectors by Eqs. (\ref{u==v-v-})--(\ref{uIs=v+v-}). Eqs. (\ref{u==v-v-}) imply
\begin{eqnarray}\label{k=pv-v-}
D=10\; : \qquad &&
k_{a i}
\Gamma^a_{\alpha\beta}= 2\rho^{\#}_{i} v_{\alpha q i}^{\; - } v_{\beta q i}^{\; - } \; , \qquad
 \rho^{\#} v^-_{\dot{q} i} \tilde{\Gamma}_{a}v^-_{\dot{p}i }= k_{ai} \delta_{{q}{p}}, \qquad  \qquad \nonumber \\  && k_{ai} \tilde{\Gamma}{}^{a\, \alpha\beta}= 2\rho^{\#}_{i}v^{-\alpha}_{\dot{q}i }   v^{-\beta}_{\dot{q}i} \; , \qquad \rho^{\#}  v^-_{\dot{q}i} {\Gamma}_{a}v^-_{\dot{p}i}=  k_{ai}  \delta_{\dot{q}\dot{p}}
 \; . \qquad
\end{eqnarray}
 Contracting suitable equations (\ref{k=pv-v-})   with $v^{-\beta}_{qi}$ and $v_{\alpha \dot{q}i}{}^-$ and using (\ref{v-qv+p=}) we easily find that these obey the massless Dirac equations (Weyl equations)
\begin{eqnarray}\label{Dirac=10D}
k_{a i}\Gamma^a_{\alpha\beta}  v^{-\beta}_{\dot{q}i}=0\; , \qquad  k_{ai} \tilde{\Gamma}{}^{a\, \alpha\beta}v_{\beta qi}^{\; -} =0
\; . \qquad
\end{eqnarray}
Thus they  can be identified with D=10 spinor helicity variables of \cite{CaronHuot:2010rj}
\begin{eqnarray}\label{shel=l=}
\lambda_{\alpha q}= \sqrt{\rho^{\#}} v_{\alpha  q}^{\; -}  \qquad
\end{eqnarray}
and, as we will see in a moment, with its D=11 generalization.

The counterparts of the basic relations   of the corresponding spinor helicity formalism for D=11 supergravity can be extracted from    \cite{Bandos:2007wm}. In this case we can also adapt 11D vector frame to a light--like  momentum according to (\ref{kia=Up}); then the spinor moving frame variables/sinorial harmonics obey the counterpart of the constraints (\ref{k=pv-v-})
\begin{eqnarray}\label{k=pv-v-11}
D=11\; : \qquad &&
k_{a i}
\Gamma^a_{\alpha\beta}= 2\rho^{\#}_{i} v_{\alpha q i}^{\; -} v_{\beta qi}^{\; -}  \; , \qquad
 \rho^{\#} v^-_{{q}i} \tilde{\Gamma}_{a}v^-_{{p}i}= k_{ai} \delta_{{q}{p}}\; , \qquad
\end{eqnarray}
with $ \alpha, \beta =1,...,32$, $q,p=1,..,16$ and $a=0,1,...,10$.
Notice that in this case the counterparts of the equations in the second line of (\ref{k=pv-v-}),
\begin{eqnarray}\label{k=pv-v-11-1}
D=11\; : \qquad && k_{ai} \tilde{\Gamma}{}^{a\, \alpha\beta}= 2\rho^{\#}_{i}v^{-\alpha}_{qi}   v^{-\beta}_{qi}\; , \qquad  \rho^{\#}  v^-_{{q}i} {\Gamma}_{a}v^-_{{p}i}=  k_{ai}  \delta_{{q}{p}}\; , \qquad
\end{eqnarray}
are equivalent to (\ref{k=pv-v-11}). Again, one can identify spinor harmonics with the solutions of massless Dirac equation
\begin{eqnarray}\label{Dirac}
 k_{ai} \tilde{\Gamma}{}^{a\, \alpha\beta}v_{\beta  q i}^{\; -} =0\;  \qquad \Leftrightarrow \qquad k_{a  i}\Gamma^a_{\alpha\beta}  v^{-\beta}_{qi}=0 \; . \qquad
\end{eqnarray}
Hence, the 11D generalization of the 10D spinor helicity variables is given by (\ref{shel=l=})
with $\alpha=1,...,32$ and $q=1,...,16$.

As many equations of 10D and 11D spinor frame/spinor helicity formalism differs formally by just replacing dotted indices $\dot{q}, \dot{p}$ by undotted  $q,p$, and by changing the range of the values of the indices,  below, when this cannot lead to confusion,  we will not duplicate the equations writing  them separately for D=10 and D=11 cases, but rather write single equation assuming that for D=11 case
$\dot{q}=q$, etc.

Polarization spinor of the D=10 and D=11 fermionic fields can be associated with the element of the inverse D=10 and D=11 spinor frame matrix,
\begin{eqnarray}\label{spol=l=} \lambda^{\alpha}_{\dot{q}i}= \sqrt{\rho^{\#}} v^{-\alpha}_{\dot{q}i}\; . \qquad
\end{eqnarray}
For $D=11$, $\;\dot q =q$ and Eq. (\ref{spol=l=}),
$\lambda^{\alpha}_{{q}i}= \sqrt{\rho^{\#}} v^{-\alpha}_{{q}i}$, is equivalent to (\ref{shel=l=}) due to (\ref{V-1=CV-A}).

Notice that $v_{\alpha \dot{q} i}^{\; +}$ and $ v^{+\alpha}_{{q}i}$ are not included explicitly in the definition of momentum and polarization variables and in this sense can be treated as constrained reference spinors restricted by the requirement to form, together with $v_{\alpha {q} i}^{\; -} $ and $ v^{-\alpha}_{\dot{q}i}$, the $Spin(1,D-1)$ valued  matrix (\ref{harmVi=D}) and its inverse (\ref{harmV-1i=D}).
The freedom in the definition of $v_{\alpha \dot{q} i}^{\; +}$ includes the $K_{D-2}$ transformations
\begin{eqnarray}\label{K8i}
v_{\alpha \dot{q} i}^{\; +}\mapsto v_{\alpha \dot{q} i}^{\; +} + \frac 1 2 K_i^{\# I} v_{\alpha {p} i}^{\; -} \gamma^I_{p\dot{q}}\; , \qquad v_{\alpha {q} i}^{\; -}\mapsto v_{\alpha {q} i}^{\; -}
\qquad
\end{eqnarray} as well as the natural action of $Spin(D-2)$ on $q,p$ and $\dot{q},\dot{p}$ indices and scaling, supplemented by an opposite scaling of $v_{\alpha {q} i}^{\; -} $. With the same line of argument as presented for the vector frame when arriving at (\ref{u=i=G-H}),
we arrive at
\begin{eqnarray}\label{v-qi=G-H}
\{ v_{\alpha q i}^{\; -} \}  \; \in  \; \frac {Spin(1,D-1)} {[SO(1,1)\otimes Spin(D-2)]\subset\!\!\!\!\!\!\times  K_{D-2}} = \; {\bb S}^{D-2}\; .  \qquad
\end{eqnarray}
The simplest application of our spinor frame form of the spinor helicity formalism is to write the solution of the momentum representation of the linearized equations of motion of 10D SYM and 11D SUGRA multiplets.

\subsection{D=10 SYM multiplet in spinor helicity formalism  }

As, according to (\ref{spol=l=}),  polarization of a 10D spinor field can be described by $8$ constrained spinors from the  inverse spinor  frame matrix (\ref{harmV-1i=D}),
$v^{-\alpha}_{\dot q }$ which obey (\ref{Dirac=10D}), the general  solution of linearized massless Dirac equation reads
\begin{eqnarray}\label{chi10D=vpsi}
D=10:\quad \chi^\alpha=v^{-\alpha}_{\dot q } \psi_{\dot q }\; , \qquad \alpha=1,...,16\; , \qquad \dot{q}=1,...,8
\; ,  \qquad
\end{eqnarray}
and  the superpartner of the gauge field is characterized by a fermionic SO(8) c-spinor field  $\psi_{\dot q }$.

Polarization vector of a gauge field can be identified with the spacelike vector $u_a^I$ of the frame (\ref{harmUi=}) adapted to the light-like momentum $k_a$  through (\ref{kia=Up}) ({\it cf.} \cite{CaronHuot:2010rj}). Then   the basic solutions of the linearized YM equations can be written as   $F_{ab}{}^I= k_{[a}u_{b]}{}^I$
and  the  general solution
\begin{eqnarray}\label{Fab=kuwI}
D=10:\quad F_{ab} = k_{[a}u_{b]}{}^I\, w^I \; , \qquad k_{a}=\rho^{\#} u_a^=\; , \qquad
a=0,1,...,9\; , \qquad I=1,...,8 \; , \qquad
\end{eqnarray}
 is characterized by an SO(8) vector $w^I$.

When the formalism is applied to external particles of scattering amplitudes, coefficients in the above  expressions for bosonic and fermionic fields, the bosonic  $w^I$ and fermionic  $\psi_q$, can be taken to be dependent on the on-shell momentum $k_{a}$ ; in our formalism this implies the dependence  on the constrained spinors $v_{\alpha q }^{\; -}$ (\ref{v-qi=G-H}) and density  $\rho^{\#}$   (see (\ref{k=pv-v-})). Alternatively, we can replace $\rho^{\#}$ by  its conjugate coordinate and consider the field $w^I=w^I(x^=, v_{q}^{ -})$ and
$\psi_q=\psi_q(x^=, v_{q}^{ -})$ on the nine-dimensional space ${\bb R}\otimes {\bb S}^{8}$.  We will see below that supersymmetry acts on these 9d fields as
\begin{eqnarray}\label{susy=8+8a}
\delta_{\epsilon} w^I(x^=, v_{q}^{ -}) = 2i \epsilon^{-{q}} \gamma^I_{q\dot{q}} \psi_{\dot q }(x^=, v_{q}^{ -})\; \; , \qquad \delta_{\epsilon}\psi_{\dot q }(x^=, v_{q}^{ -}) = \epsilon^{-{q}}  \gamma^I_{q\dot{q}}  \;  \partial_{=}  w^I(x^=, v_{q}^{ -}) \; , \qquad
\end{eqnarray}
where the fermionic $SO(8)$ s-spinors $\epsilon^{-{q}}$ are expressed through the constant fermionic spinor parameter of rigid 10D ${\cal N}=1$ supersymmetry, $\epsilon^\alpha$,  by
\begin{eqnarray}\label{susy8=16v}
\epsilon^{-{q}}  =\epsilon^\alpha v_{\alpha q}^{\; -}  \; .  \qquad
\end{eqnarray}

In D=4 the field describing physical degrees of freedom of the maximal SYM multiplet,
appear in the decomposition of an on-shell superfield on fermionic coordinate, (\ref{Phi=3,4}).
The natural question is whether such an on-shell superfield may exist in the case of D=10 SYM. As we will see below, the answer on this question is affirmative.

\subsection{11D SUGRA in spinor helicity formalism}

In D=11 it is convenient to begin with the solution of the linearized equations for the on-shell field strength $F_{abcd}$ of the 3-form  gauge field. By analogy with the above described 10D gauge field strength, the solution   can be expressed in terms of 11D $u_a^I$,
$F_{abcd}^{\; IJK}=k_{[a}u_{b}{}^I u_{c}{}^J u_{d]}{}^K$, so that the generic linearized field strength
\begin{eqnarray}\label{FabcdIJK=}
D=11:\quad F_{abcd}=k_{[a}u_{b}{}^I u_{c}{}^J u_{d]}{}^K\, A_{IJK} \; , \qquad a=0,1,...,10\; , \qquad  I=1,...,9\, ,
\end{eqnarray}
is expressed in terms of an antisymmetric SO(9) tensor $A_{IJK}$. Its superpartners,
$\gamma$--traceless $\Psi_{I{q}}$ and symmetric traceless $h_{IJ}$, are used to make a decomposition of linearized 11D graviton and gravitino fields,
\begin{eqnarray}\label{Psiaal=}
D=11: \quad && \psi_{ab}^\alpha = k_{[a} u_{b]}^I  v^{-\alpha}_q \Psi_{Iq}
\; , \qquad \gamma^I_{qp}  \Psi_{Ip}=0\; , \qquad  \\ \label{hab=hIJ} &&
h_{ab}= u_{(a}^I  u_{b)}^J h_{IJ} \; , \qquad h_{II}=0\; . \qquad
\end{eqnarray}
The linearised on-shell Riemann tensor reads
\begin{eqnarray}\label{Rabcd=kukuh}
R_{ab}{}^{cd}\; = k_{[a} u_{b]}^{\, I} k^{[c} u^{d]J} h_{IJ}\; . \qquad
\end{eqnarray}
It is easy to check that this obeys the characteristic identities
$R_{[ab\, c]}{}^d=0$, $R_{ab\, cd}=R_{cd\, ab}$ as well as the linearized Einstein  equations  $R_{ab}{}^{cb}=0$.

Again, it is natural to expect that the above  fields appear as independent components of a constrained $d=1$, ${\cal N}=16$ on-shell superfield. As we will show below, this is indeed the case.

\subsection{10D amplitudes in spinor helicity formalism, supersymmetry and Ward identities}

\label{10D=WId}

The tree amplitudes of 10D SYM  should depend on the light-like momentum and polarization vectors or spinors of scattered particles and can carry some nontrivial representations of a small group, in particular of its $SO(8)$ subgroup. In the light of the above discussion, this implies that the n-point tree  amplitudes, a D=10 counterpart of (\ref{cAn=cAnl}),  should depend on  $n$ sets of
spinor moving frame variables $v_{\alpha q (i)}^{\; -}$ and $n$ 'energies' $\rho^\#_{(i)}$, corresponding to  each of the scattered particles; it can also carry the SO(8) polarization indices for each of the scattered particles. The  superamplitudes, counterparts  of 4D (\ref{cAn=cAnlf}),   shall also depend on fermionic variables, but in this section we would like to discuss  briefly a 'component' (rather than superfield) approach to SYM amplitudes.

To understand the structure and properties of the supersymmetry transformations of the amplitudes,
it is convenient to consider 16-parametric rigid supersymmetry  as a superposition of $n$ 8-parametric SUSY  transformations defined such that $i$-th  'item supersymmetries' with parameters $\epsilon^{-q}_{i}$
acts on the variables and/or indices corresponding to $i$-th scattered particle.
The true supersymmetry transformations will involve all the item supersymmetries with parameters expressed in terms of the constant fermionic spinor of rigid supersymmetry and $i$-th spinor frame variables by
\begin{eqnarray}\label{el-q=}
\epsilon_i^{- {q}}= \epsilon^{\alpha} v_{\alpha {q} i}^{\; -}  \; . \qquad
\end{eqnarray}

The structure of the supersymmetry transformation relating on-shell fields of the SYM supermultiplet, (\ref{susy=8+8a}), suggests that the supermultiplet with respect to $i-th$ 'item supersymmetry' is formed by two amplitudes which differs by only one of their $n$ polarization indices, namely by the index corresponding to $i$-th scattered particle, and that this index can be either of {\bf 8v} or of {\bf 8c} representation
\footnote{The third, {\bf 8s} representation is singled out by that, in our notation, its indeex is carfried by the spinor frame variables $v_{\alpha qi}^{\; -}$ related to the light-like momentum $k_{ai}$ of $i$-th particle by (\ref{k=pv-v-}).},
\begin{eqnarray}\label{S-multA}
{\cal A}^{(n)}_{... I_i...}(k_1,..., k_i,...,k_n)=
{\cal A}^{(n)}_{... I_i...}(\{\rho^{\#}_{j}, \, v^-_{{q}j}\}  )  \qquad     \nonumber \\  and \nonumber \qquad  \\ {\cal A}^{(n)}_{... \dot{q}_l...}(k_1,..., k_{i},...,k_n)=
{\cal A}^{(n)}_{... \dot{q}_i...}(\{\rho^{\#}_{j}\}; \{v^-_{{q}j}\}  ) \; .
\end{eqnarray}
Of course, for any particular combinations of the indices denoted by multidots, either
${\cal A}^{(n)}_{... I_i...}$ or ${\cal A}^{(n)}_{... \dot{q}_i...}$ vanishes by the fermionic number preservation (see below for more details).
Nevertheless it is instructive first to write the generic  $i-th$ SUSY transformations with Grassmann parameter $\epsilon_{i}^{-q}$ as they are suggested  by transformations (\ref{susy=8+8a}) of the on-shell SYM fields  $(w^I, \psi_{\dot{q}})$:
\begin{eqnarray}\label{susy=8A+8A} && \delta_{\epsilon_i}{\cal A}^{(n)}_{... I_i...}(...,k_i,...)
=    2 i\; (-)^{\Sigma_i}\,  \epsilon_i^{-{q}} \gamma^{I_i}_{q\dot{q}_i} \;   {\cal A}^{(n)}_{... \dot{q}_i...} (...,k_l,...) \; ,
  \qquad \nonumber \\   && \delta_{\epsilon_i}{\cal A}^{(n)}_{... \dot{q}_i...}(...,k_i,...)
=  i (-)^{\Sigma_i} \rho^{\#}_{i} \epsilon_i^{-{q}} \gamma^{I_i}_{{q}\dot{q}_i} \; {\cal A}^{(n)}_{... I_i...} (...,k_i,...) \; , \qquad
\end{eqnarray}
 The sign factor  $ (-)^{\Sigma_i} $ in (\ref{susy=8A+8A}) will be specified below (see also Eq. (\ref{Sl=10D})) \footnote{Here for shortness we have written the argument $k_i$ instead of $\rho^\#_i, v_{\alpha q i}^{\; -}$. Notice also  that $\partial_{=}$ in the second equation of (\ref{susy=8+8a}) is replaced by $ i\rho^{\#}_{i}$ in (\ref{susy=8A+8A})  (momentum versus coordinate representation).}.
The complete expressions  (\ref{susy=8A+8A}) are also valid for the case of coefficients in the decomposition of superamplitudes on  fermionic coordinate corresponding to one of $n$ scattered particles.

To write the complete form of the rigid supersymmetry transformations of the amplitudes, it is convenient to introduce a cumulative index
\begin{eqnarray}
\label{Qi=10D}
Q_i=(I_i, \dot{q}_i)\; ,
\qquad I_i= 1,...,8\; , \qquad \dot{q}_i=1,...,8\;
\end{eqnarray}
allowing to describe all the amplitudes  (\ref{S-multA}) by  the universal expression
\begin{eqnarray}\label{cAQ1-Qn=10D}
{\cal A}^{(n)}_{Q_1...Q_n}(\rho^\#_{1}, v^{\; -}_{\alpha q\, 1}; ...;\rho^\#_{n}, v^{\; -}_{\alpha q\, n})=:{\cal A}^{(n)}_{... Q_i...}( ... ; \rho^\#_{i}, v^{\; -}_{\alpha q\, i} ; ...)\; . \qquad
\end{eqnarray}
Formally, the rigid 10D supersymmetry acts on the amplitudes by
\begin{eqnarray}\label{susyAmp=10D}
&& \delta_\epsilon {\cal A}^{(n)}_{Q_1...Q_n}(\rho^\#_{1}, v^{\; -}_{\alpha q\, 1}; ...;\rho^\#_{n}, v^{\; -}_{\alpha q\, n})= \qquad \nonumber \\ && \qquad = \epsilon^\alpha \, \sum\limits_{i=1}^n
(-)^{\Sigma_i} v_{\alpha q_i}^{\; -} \Delta_{q_iQ_iQ^\prime_i} {\cal A}^{(n)}_{Q_1...Q^\prime_i...Q_n}(\rho^\#_{1}, v^{\; -}_{\alpha q\, 1}; ...;\rho^\#_{n}, v^{\; -}_{\alpha q\, n})    \; , \qquad
\end{eqnarray}
where $\Delta_{q_iQ_iQ^\prime_i}=\Delta_{q_iQ_iQ^\prime_i}(\rho^\#_i)$ are expressed through the $SO(8)$ Klebsh-Gordan coefficients $\gamma^{I}_{q\dot{q}} $ by
\begin{eqnarray}
\label{gqQQ'=10D}
\Delta_{q_iQ_iQ^\prime_i} = i\gamma^{I_i}_{q_i\dot{q}_i} \left( 2 \delta_{Q_i}{}^{I_i} \delta _{Q^\prime_i\dot{q}_i}+ \rho^\#_i\delta _{Q_i\dot{q}_i} \delta_{Q^\prime_i}{}^{I_i} \right)\;  \qquad
\end{eqnarray}
and ${\Sigma_i}$ counts the number of fermionic indices $\dot{q}_j$ among $Q_j$'s with $1\leq j< i$.

Now let us recall that the preservation of the fermionic number requires vanishing of  all the amplitudes describing scattering  of an odd number of fermions (when counting both incoming and outgoing particle). Then their supersymmetry transformations should also vanish. This implies, firstly, that all the (potentially) nonvanishing amplitudes, describing the scattering of an even number of fermionic and some number of bosonic particles, are supersymmetric invariant
\begin{eqnarray}\label{Amp10D=susy}
&& \delta_\epsilon {\cal A}^{(n)}_{Q_1...Q_n}(\rho^\#_{1}, v^{\; -}_{\alpha q\, 1}; ...;\rho^\#_{n}, v^{\; -}_{\alpha q\, n})=0\;   \qquad
\end{eqnarray}
and, secondly, that they obey the  following supersymmetric Ward identities
\footnote{Of course, Eqs.   (\ref{susyWard=10D})  are nontrivial only
when the amplitudes carry the even number of fermionic indices, but we do not feel necessary to stress this in the formulae as  otherwise  they are  trivially satisfied.}
\begin{eqnarray}\label{susyWard=10D}
&&   \, \sum\limits_{i=1}^n
(-)^{\Sigma_i} v_{\alpha q_i}^{\; -} \Delta_{q_iQ_iQ^\prime_i} {\cal A}^{(n)}_{Q_1...Q^\prime_i...Q_n}(\rho^\#_{1}, v^{\; -}_{\alpha q\, 1}; ...;\rho^\#_{i}, v^{\; -}_{\alpha q\, i}; ...;\rho^\#_{n}, v^{\; -}_{\alpha q\, n})   =0  \; . \qquad
\end{eqnarray}

As an example, let us write the explicit form of the Ward identities for 3-point amplitudes:
\begin{eqnarray}\label{Ward10D=A3fff}
&&   \rho^\#_{1} v_{\alpha q1}^{\; -} \gamma^{I_1}_{q \dot{q}_1} {\cal A}^{(3)}_{I_1\dot{q}_2\dot{q}_3}- \rho^\#_{2} v_{\alpha q2}^{\; -} \gamma^{I_2}_{q \dot{q}_2} {\cal A}^{(3)}_{\dot{q}_1I_2\dot{q}_3}+ \rho^\#_{3} v_{\alpha q3}^{\; -} \gamma^{I_3}_{q \dot{q}_3} {\cal A}^{(3)}_{\dot{q}_1\dot{q}_2I_3}=0 \; , \qquad \\
\label{Ward10D=A3I1}
&&   2 v_{\alpha q1}^{\; -} \gamma^{I_1}_{q \dot{q}_1} {\cal A}^{(3)}_{\dot{q}_1I_2\dot{q}_3}+  2 v_{\alpha q2}^{\; -} \gamma^{I_2}_{q \dot{q}_2} {\cal A}^{(3)}_{I_1\dot{q}_2\dot{q}_3}+ \rho^\#_{3} v_{\alpha q3}^{\; -} \gamma^{I_3}_{q \dot{q}_3} {\cal A}^{(3)}_{I_1I_2I_3}=0 \; . \qquad
\end{eqnarray}
These could be used to find the structure of the  3-point amplitudes of 10D SYM (see \cite{Schwarz:1982jn}). Below we will present a specific parametrization of spinor frames (special gauge fixed on spinor frame variables) which is especially useful for addressing such type of problems.

But before we would like to develop a constrained superamplitude formalism, based on superfield generalization of the above amplitudes. These superamplitudes can be considered as multi-particle generalizations of the constrained on-shell superfield description of the linearized SYM multiplet, which, as we will show, can be obtained from superparticle quantization.

We conclude this section by a brief description of the 11D amplitudes and the  supersymmetric Ward identities for them. Their similarity with 10D counterparts allows to reduce this description to specification of the cumulative indices and $\Delta$-symbols in
(\ref{cAQ1-Qn=10D}),  (\ref{gqQQ'=10D}), (\ref{Amp10D=susy}) and  (\ref{susyWard=10D}).

\subsection{Supersymmetric Ward identities for 11D SUGRA amplitudes }

The superamplitudes of 11D supergravity can be described by formula (\ref{cAQ1-Qn=10D}) with 11D spinor frame variables $v_{\alpha q i}^{\; -}$ and cumulative indices
\begin{eqnarray}
\label{Qi=11D}
Q=\big( [IJK],\,  ((IJ)),\,  I{q}\big)\; ,
\qquad I,J,K= 1,...,8,9\; , \qquad {q}=1,...,16\; ,
\end{eqnarray}
including antisymmetric combination of $SO(9)$ vector indices $[IJK]$, symmetric traceless combination of two $SO(9)$ vector indices $((IJ))$ [to lighten equation we will sometimes write this with one set of brackets, as   $(IJ)$] and $\gamma$-traceless combination of the vector and spinor indices of $SO(9)$, $I{q}$.

Then the supersymmetric Ward identities for these amplitudes will have the form of Eq. (\ref{susyWard=10D}) with the following components of  $\Delta_{q_iQ_iQ^\prime_i}=\Delta_{q_iQ_iQ^\prime_i}(\rho^\#_i)$:
\begin{eqnarray}
\label{gqQQ'=11D}
&& \Delta_{q\; [IJK] \; Lp } = i\delta^{L[I} \gamma^{JK]}_{qp} \; , \qquad
\Delta_{q\; ((IJ)) \; Kp } = i\delta^{K((I} \gamma^{J))}_{qp} \; , \qquad \nonumber \\
&& \Delta_{q\; Jp \; ((KL)) } = 2i\rho^{\#}_i \delta^{K((I} \gamma^{J))}_{qp} \; , \qquad
\Delta_{q\; Jp \; [KLM] } =
\frac {i} {18} \rho^\#_{(i)}  \left(\gamma^{JKLM}_{qp}+ 6 \delta ^{J[K}\gamma^{LM]}_{qp}  \right)\;  . \qquad
\end{eqnarray}

\section{D=10 and D=11 on-shell superfields from quantization of massless superparticle  in analytical basis of Lorentz harmonic superspace}

In this section we show how the above descriptions of linearized 10D SYM and 11D SUGRA supermultiplet appear, in their superfield form, in quantization of D=10 and D=11 massless superparticle models. The reader not interested in quantization procedure may omit the first four subsections and pass directly to subsection  \ref{On-shellSYM}, using subsection \ref{On-shSSP=sec} just for notation.

\subsection{Changing variables in  D=10,11 massless superparticle action }

As we discussed in sec. \ref{on-shell4D}, 4D on-shell superfields can be obtained from quantization of  D=4 massless superparticle mechanics reformulated in terms of bosonic spinor and their conjugate variables.
The original Ferber-Shirafuji action (\ref{S-FS=X}) can be written in terms of these variables, (\ref{S-FS=mu}), by using the Leibnitz rule to move $\partial_\tau$ derivatives.

Let us try to do the same in the analogous spinor moving frame action for D=10 and D=11 massless superparticle, which is given in (\ref{S=smf}) with the range of indices described in (\ref{q=D10}) and (\ref{q=D11}), respectively.
One can easily arrive at
\begin{eqnarray}\label{S=mf=v2}
S = \int d\tau \rho^\# \left({\partial}_\tau X^{=}  -2i {\partial}_\tau\theta_{ {q}}{}^- \, \theta_{{q}}{}^-  - {\partial}_\tau  v_{\alpha {q}}^{\; -} \, \left(\ldots  \right)^{\alpha {q}} \right),  \quad \end{eqnarray}
where
\begin{eqnarray}\label{X--=}
X^= = X^a u_a^=, \qquad \theta_{ {q}}{}^- = \theta^\alpha  \, v_{\alpha {q}}^{\; -}    , \qquad  \end{eqnarray}
which can be associated with coordinates of the so-called {\it analytical coordinate basis of Lorentz harmonic superspace}, which we describe in the next sec. \ref{secAnalit},
and multidots denote $\frac 1 4 X^a\tilde{\Gamma}_a^{\alpha\beta} v_{\beta {q}}{}^-   -i \theta^\alpha \theta^\beta v_{\beta {q}}{}^-$.
Actually this expression can also be rewritten in terms of coordinates of the analytical basis of the Lorentz harmonic superspace. However,  to perform this in a brief and clear manner, we need first to clarify the structure of the   derivative of spinor moving frame variable, ${\partial}_\tau  v_{\alpha {q}}^{\; -}$ entering the third term in the integrand of (\ref{S=mf=v2}), or more generally of the differential of this variable $dv_{\alpha {q}}^{\; -}$.
In sec. \ref{secDuDv} we  will show how to express these in term of Cartan forms of the Lorenz group.

\subsection{Analitical basis of Lorenz harmonic superspace and its invariant sub-superspaces}
\label{secAnalit}

As we have already mentioned, the  spinor moving frame formulation of massless superparticle  can be considered as a dynamical system in an enlarged superspace called Lorentz harmonic superspace  with the bosonic body given by the direct product of D-dimensional Minkowski space and $2(D-2)$-dimensional coset of the D-dimensional Lorentz group, ${\bb R}^{1,(D-1)}\times {SO(1,D-1)\over {SO(1,1)\otimes SO(D-2)}}$.

The set of coordinates of Lorentz harmonic superspace includes, in addition to the usual bosonic D-vector and fermionic spinor coordinates, also Lorentz harmonics (spinor moving frame variables) $v_{\alpha  q}^{\; -}$, $ v_{\alpha \dot  q}^{\; +}$, 'parametrizing' (as a kind of homogeneous coordinates)  a non-compact coset ${SO(1,D-1)\over {SO(1,1)\otimes SO(D-2)}}$,
\begin{eqnarray}\label{cZ=10D-11D}
 {\cal Z}^{\underline{{\cal M}}}=( X^a, \theta^\alpha , v_{\alpha  q}^{\; -},  v_{\alpha \dot  q}^{\; +})
 \;  \qquad
\end{eqnarray}
(the ranges of values of indices for D=10 and D=11 cases are given in (\ref{q=D10}) and (\ref{q=D11}), respectively).
Notice that these Lorentz harmonic variables  are also appropriate for the description of (super)string \cite{BZ-str,Bandos:1992ze}\footnote{
The spinor  moving frame formulations of  super-p-branes \cite{BZ-str,BZ-p} use the Lorentz harmonics  parametrizing the  noncompact cosets ${SO(1,D-1)\over SO(1,p)\otimes SO(D-p-1)}$ with an appropriate values of D; e.g. for superstring we can consider $D=3,4,6,10$. The spinor moving frame formulation of superparticle \cite{Bandos:1996ju,Bandos:2006nr,Bandos:2007mi,Bandos:2007wm} uses
$v_{\alpha q}^{\; -}$ variables  parametrizing, modulo gauge symmetries, a compact coset \cite{Galperin:1991gk,Delduc:1991ir,Galperin:1992pz}. However,
when such a formulation is treated as superparticle in Lorentz harmonic superspace, it is convenient to have also $v_{\alpha \dot  q}^{\; +}$, the counterpart of reference spinor, as a superspace coordinate (see \cite{Bandos:1990ji} for D=4 model).}.

The coordinate basis  (\ref{cZ=10D-11D}) is called 'central basis'. Besides this we  can define an analytical basis  with coordinates ({\it cf.} \cite{Sokatchev:1985tc,Sokatchev:1987nk} and, in a more general perspective, \cite{Galperin:1984av,Galperin:1984bu,Galperin:2001uw})
\begin{eqnarray}\label{cZA=10D}
&& {\cal Z}_A^{\underline{{\cal M}}}=( X^\#, X^=, X^I, \theta^+_{\dot{q}} ,\theta^-_q ,
v_{\alpha q}^{\; -}\, , \, v_{\alpha \dot  q}^{\; +}) \; , \qquad \\ \label{cZA=X10D}
&& X^=:=  X^au_a^=\; , \quad X^\#:= X^au_a^\#\; , \quad X^I:= X^au_a^I+{i} \theta^-_{{q}}\gamma^I_{q\dot{q}}\theta^+_{\dot{q}}    \; , \qquad  \\ \label{cZA=th10D} && \theta^+_{\dot{q}} = \theta^\alpha  v_{\alpha \dot q}{}^+\, , \quad \theta^-_q= \theta^\alpha  v_{\alpha q}{}^-
 \; . \qquad
\end{eqnarray}


Notice the nontrivial fermionic bilinear contributions in the definition of $X^I$ in (\ref{cZA=X10D}). They are  designed in such a way, that the superspace  supersymmetry transformations
\begin{eqnarray}\label{susyXth}
\delta_\varepsilon X^a= i \theta\Gamma^a\varepsilon  \; , \qquad
\delta_\varepsilon \theta^\alpha =\varepsilon^{\alpha} \;  \qquad
\end{eqnarray}
 can be closed on smaller set of coordinates
\begin{eqnarray}\label{czA=10D11D}
\tilde{{\zeta}}^{\tilde{{\cal M}}}=( X^=, X^I, \theta^-_{{q}},
v_{\alpha q}^{\; -}\, , \, v_{\alpha \dot  q}^{\; +})  \; . \qquad
\end{eqnarray}
These
parametrize an invariant sub-superspace of the Lorentz-harmonic superspace which is called {\it analytic (sub)superspace}. The supersymmetry transformations of the analytic superspace coordinates (\ref{czA=10D11D}) read
\begin{eqnarray}\label{susyAn}
\delta_\varepsilon X^== 2i\theta^-_{{q}}\varepsilon^{-{q}} \; , \qquad \delta_\varepsilon X^I= 2i\theta^-_{{q}}{\gamma}^I_{q\dot{p}}\varepsilon^{+\dot{p}}\; , \qquad \delta_\varepsilon \theta^-_{{q}}= \varepsilon^{-{q}}
 \; , \qquad
\delta_\varepsilon v_{\alpha\dot{q}}{}^-=0 = \delta_\varepsilon v_{\alpha {q}}{}^+ \; ,   \qquad
\;
\end{eqnarray}
where ({\it cf.} (\ref{susy8=16v}))
\begin{eqnarray}\label{e-=ev-}
\varepsilon^{-{q}}= \varepsilon^{\alpha}v_{\alpha {q}}{}^-\; , \qquad \varepsilon^{+\dot{q}}= \varepsilon^{\alpha}v_{\alpha \dot{q}}{}^+
\; .  \qquad
\end{eqnarray}

Actually
the smaller superspace with coordinates
\begin{eqnarray}\label{czA=10D-11D}
 {\zeta}^{{{\cal M}}}=( X^=, \theta^-_{{q}},
v_{\alpha  q}^{\; -}\, , \, v_{\alpha \dot q}^{\; +}) \;   \qquad
\end{eqnarray}
is  also invariant under (\ref{susyAn}),
\begin{eqnarray}\label{susyAn-Min}
\delta_\varepsilon X^== i\theta^-_{{q}}\varepsilon^{-{q}} \; , \qquad \delta_\varepsilon \theta^-_{{q}}= \varepsilon^{-{q}}
 \; , \qquad
\delta_\varepsilon v_{\alpha {q}}{}^-=0 = \delta_\varepsilon v_{\alpha \dot{q}}{}^+ \; .   \qquad
\;
\end{eqnarray}
One  can observe however  that only a half of the  supersymmetries acts efficiently on this {\it \underline{minimal} analytic (sub)superspace}.

Below we will see that the minimal superspaces (\ref{czA=10D-11D}) are sufficient  to provide the arena for on-shell superfields describing 10D SYM and linearized 11D supergravity multiplets. Then, $n$-point  superamplitudes will be defined on the direct product of $n$ copies of minimal analytic superspaces   (\ref{czA=10D-11D}).  In contrast, to write  the massless superparticle actions the coordinates $X^I$ are also needed, so that the superparticle can be considered as a particle  in the 'non-minimal' analytic superspace (\ref{czA=10D11D}).

Another important remark is related to the fact that the massless superparticle action (\ref{S=smf}) contains the coordinate functions $v_{\alpha  q}^{\; -}$, but not $v_{\alpha \dot q}^{\; +}$. As it  is also  invariant under  $SO(1,1)\otimes SO(D-2)$ symmetry, using this as identification relation on the set of
$v_{\alpha  q}^{\; -}$, we concluded, following \cite{Galperin:1991gk,Delduc:1991ir},  that these constrained variables parametrize the compact coset
isomorphic to the celestial sphere (see (\ref{v-qi=G-H}) and the descussion above it).

This statement can be reformulated by considering the superparticle action as apparently depending on  the complete set of Lorentz harmonics $v_{\alpha  q}^{\; -}\, , \, v_{\alpha \dot q}^{\; +}$ parametrizing the non-compact coset
$ {SO(1,D-1)\over SO(1,1)\otimes SO(D-2)}$, {\it i.e.} as an action of superparticle in Lorentz harmonic superspace (\ref{cZ=10D-11D}),
which possesses  the gauge symmetry
under $K_{D-2}$\footnote{See \cite{Bandos:1990ji} for D=4 case and  also  earlier \cite{Sokatchev:1985tc,Sokatchev:1987nk} where the 'light-cone superspace' with additional vector frame coordinates $u_a^=, u_a^\#, u_a^I$ was introduced.
One more equivalent form of the same statement is that the action depends on
  $v_{\alpha  q}^{\; -}\, , \, v_{\alpha \dot q}^{\; +}$ which are constrained
by (\ref{harmV=D}), and has the gauge symmetry under
$[SO(1,1)\otimes SO(D-2)]\subset\!\!\!\!\!\!\times K_{D-2}$.}.
 The role of $K_{D-2}$ symmetry, which completes $SO(1,1)\otimes SO(D-2)$ till Borel subgroup of $SO(1,D-1)$, is to make 'unphysical' (pure gauge) the complementary element of Spin(1,D-1) valued matrix, $v_{\alpha \dot{q}}^{\; +}$ ($v_{\alpha  q}^{\; +}$ in 11D case). Such a point of view
    is reflected by stating that homogeneous coordinates of the coset
  $ {SO(1,D-1)\over [SO(1,1)\otimes SO(D-2)]\subset\!\!\!\!\times K_{D-2}}={\bb S}^{D-2}$ (i.e. of celestial sphere) are given by
 $(v_{\alpha  q}^{\; -}, v_{\alpha \dot{q}}^{\; +})$, i.e. by writing
 \begin{eqnarray}
\label{v-v+=G-H}  \{ (v_{\alpha  q}^{\; -}, v_{\alpha \dot{q}}^{\; +})\}={Spin(1,D-1)\over [SO(1,1)\otimes Spin(D-2)]\subset\!\!\!\!\!\!\times K_{D-2}}={\bb S}^{D-2} \; ,
\qquad
\end{eqnarray} instead of (\ref{v-qi=G-H}) and
 \begin{eqnarray}
\label{u-u+uI=G-H} \{ (u_a^=, u_a^\#, u_a^I)\} ={SO(1,D-1)\over [SO(1,1)\otimes SO(D-2)]\subset\!\!\!\!\!\!\times K_{D-2}}={\bb S}^{D-2} \; ,
\qquad
\end{eqnarray}
instead of (\ref{U=G-H}).

This point of view is actually preferable for our discussion in this Section.

\subsection{Cartan forms and derivatives of the spinor moving frame variables}
\label{secDuDv}

To understand the structure of the last term in the analytical basis form of the massless superparticle action (\ref{S=mf=v2}), we need to describe the structure of derivative of the spinor moving frame variables. This is not apparent as these variables are strongly constrained by Eqs. (\ref{u==v-v-})--(\ref{uIs=v-v+}) the set of  which is equivalent to (\ref{harmV=D}). However, actually  the problem can be easily solved by using clear group theoretical meaning   of the moving frame and spinor moving frame variables. Indeed, as these  can be understood as elements of Lorentz $SO(1,D-1)$ ($Spin(1,D-1)$) group valued matrix, their  differential (and variation)  should belong to the space (co-)tangent to the Lorentz group which  is isomorphic (actually dual) to the Lie  algebra $so(1,D-1)$. A basis of this space is provided by the $SO(1,D-1)$ Cartan forms. We briefly describe these in the next subsections referring to  \cite{Bandos:2007mi,Bandos:2007wm} for  more details.

\subsubsection{Cartan forms and vector harmonics}
The derivatives of the  moving frame variables (vector harmonics) are expressed in terms of $SO(1,D-1)$ Cartan forms. Thier set  can be split on subsets of $(D-2)$ 1--forms $\Omega^{= I}:=u_a^{=}du^{aI}$  which provide a covariant basis of the space cotangent to the coset ${SO(1,D-1)\over [SO(1,1)\otimes SO(D-2)]\subset\!\!\!\!\times  K_{D-2}}={\bb S}^{D-2}$,  of $(D-2)$ forms $\Omega^{\# I}:=u_a^{\#}du^{aI}$ dual to $K_{D-2}$ generators, and of the forms  $\Omega^{(0)}:={1\over 4}u_a^{=}du^{a\#}$ and $\Omega^{IJ}:=u_a^{I}du^{aJ}$ which have the properties of connection under $SO(1,1)$ and $SO(D-2)$ gauge transformations.
This definition of Cartan forms can be encoded in  expressions for the $SO(1,1)\otimes SO(D-2)$ covariant derivatives of the moving frame variables
\begin{eqnarray}
\label{Du--}  Du^{=}_a &:=& du^{=}_a + 2u^{=}_a \Omega^{(0)} = u^{I}_a \Omega^{=I}
\; , \qquad \\ \label{Du++}  Du^{\#}_a &:=&d u^{\#}_a - 2u^{\#}_a \Omega^{(0)} =
u^{I}_a \Omega^{\# I} \; , \qquad \\ \label{Dui}  Du^{I}_a &:=& du^{I}_a + u^{J}_a
\Omega^{JI} = {1\over 2} u^{\#}_a \Omega^{= I} + {1\over 2} u^{=}_a \Omega^{\# I} \; .
\qquad
\end{eqnarray}
These expressions automatically take into account the constrained nature of the vector harmonics, {\it i.e.} they guaranty the preservation of the set of constraints (\ref{uu=0}), (\ref{uu=2}) and  (\ref{uui=0}) (equivalent to  (\ref{Uab=in})) under the action of differential $d$.

The selfconsistency conditions for (\ref{Du--}), (\ref{Du++}) and (\ref{Dui}) are given by   Ricci identities $$DDu^{=}_a = 2u^{=}_a d\Omega^{(0)}, \qquad DDu^{\#}_a =- 2u^{\#}_a d\Omega^{(0)}\; , \qquad
DDu^{I}_a =u^{J}_a
G^{JI}$$ which are equivalent to the Maurer-Cartan equations of the $SO(1,D-1)$ group
\begin{eqnarray}
\label{DOm--}  D\Omega^{=I}&:=& d\Omega^{=I} + 2 \Omega^{=I}\wedge  \Omega^{(0)} + \Omega^{IJ}\wedge  \Omega^{=J}=0 \; , \qquad \\
\label{DOm++}  D\Omega^{\#I}&:= & d\Omega^{\#I} - 2 \Omega^{\#I}\wedge  \Omega^{(0)}+ \Omega^{IJ}\wedge  \Omega^{\# J}=0 \; , \qquad \\
\label{dOm0}   && d\Omega^{(0)}= {1\over 4}\Omega^{=I}\wedge \Omega^{\#I}\; , \qquad
\\
\label{dOmIJ}   G^{IJ}&:= & d\Omega^{IJ}+\Omega^{IK}\wedge  \Omega^{KJ}= -
\Omega^{=[I}\wedge \Omega^{\# J]}\; . \qquad
\end{eqnarray}

\subsubsection{Derivatives of spinorial harmonics}

$Spin (1,D-1)$, the double covering of the Lorentz group $SO(1,D-1)$, is locally isomorphic to it.
Hence the tangent space to  $Spin (1,D-1)$ is isomorphic to tangent space to  $SO(1,D-1)$.
Hence the covariant derivatives of D=10 and D=11 spinor harmonics
 are also expressed in terms of the above  Cartan forms. One finds
\begin{eqnarray}
\label{Dv-dq=} &  Dv_{\alpha {q}}^{\; -} := dv_{\alpha {q}}^{\; -}+ \Omega^{(0)} v_{\alpha {q}}^{\; -} + {1\over 4} \Omega^{IJ}
v_{\alpha {p}}^{\; -}\gamma_{{p}{q}}^{IJ} = {1\over 2} \Omega^{=I}  \gamma_{q\dot{q}}^{I} v_{\alpha \dot{q}}^{\;+} \; , \qquad
\\
 \label{Dv+q=} & Dv_{\alpha \dot{q}}^{\;+}   :=  dv_{\alpha \dot{q}}^{\;+}    -
\Omega^{(0)} v_{\alpha \dot{q}}^{\;+}   + {1\over 4} \Omega^{IJ}  v_{\alpha \dot{p}}^{\;+}  \tilde{\gamma}_{\dot{p}\dot{q}}^{IJ} =  {1\over 2}
v_{\alpha {q}}^{\; -}  \Omega^{\# I} \gamma_{q\dot{q}}^{I}\; . \qquad
\end{eqnarray}
For the components of the inverse spinor moving frame matrix we find
\begin{eqnarray}
\label{Dv-1-q} &  Dv_{\dot{q}}^{-\alpha} := dv_{\dot{q}}^{-\alpha} + \Omega^{(0)} v_{\dot{q}}^{-\alpha} +
{1\over 4} \Omega^{IJ} \tilde{\gamma}_{\dot{q}\dot{p}}^{IJ} v_{\dot{p}}^{-\alpha} = - {1\over 2} \Omega^{=I}
 v_{{q}}^{+\alpha} \gamma_{q\dot{q}}^{I}\; , \qquad \\
\label{Dv-1+q} &  Dv_{\dot{q}}^{+\alpha} := dv_{\dot{q}}^{+\alpha} - \Omega^{(0)} v_{\dot{q}}^{+\alpha} +
{1\over 4} \Omega^{IJ} v_{\dot{p}}^{+\alpha} \gamma_{\dot{p}\dot{q}}^{IJ} = - {1\over 2} \Omega^{\#I }
v_p^{-\alpha} \gamma_{p\dot{q}}^{I}\; . \qquad
\end{eqnarray}
Notice that Eqs. (\ref{Dv-dq=}) can be used to find the derivatives or variations of  the spinor helicity variable (\ref{shel=l=}),
while (\ref{Dv-1-q}) gives the derivative/admissible variation of the   polarization spinors  (\ref{spol=l=}).

The above equations have been written for D=10 Lorentz harmonics, while
 the corresponding $D=11$ relation can be reproduced by identifying dotted and undotted indices in (\ref{Dv-dq=}) --- (\ref{Dv-1+q}) and assuming that $I,J=1,...,9$, $\; p,q=1,...,16$.
In particular the SO(8) Klebsh-Gordan coefficients $\gamma_{p\dot{q}}^{I}$ in 11D case are replaced by $16\times 16$ nine dimensional  gamma matrices   $\gamma_{p{q}}^{I}=\gamma_{qp}^{I}$.

\subsection{Massless superparticle action in the analytical basis of Lorentz harmonic superspace}

 Eqs. (\ref{Dv-dq=}) allows us to specify the last term in D=10 and D=11 superstring action (\ref{S=mf=v2}): after  simple algebra using  (\ref{u==v-v-})--(\ref{uIs=v-v+}), we arrive at
\begin{eqnarray}\label{S=smf=An}
S =\int d\tau {\cal L}  = \int d\tau  \rho^\#(\tau) \left(D_\tau X^{=} -2i D_\tau \theta^-_{{q}}\;\theta^-_{{q}} - X^I \Omega^{=I}_{\tau} \right)
\; .    \qquad
\end{eqnarray}
Here $\Omega^{=I}_{\tau}$ is the pull--back of the Cartan form  $\Omega^{=I}=u^{=a}du_a^I=-u_a^Idu^{=a}$
(see (\ref{Du--}) and  (\ref{Dui}))  to the worldline divided by $d\tau$, generically
\begin{eqnarray}\label{Om=dtOmt}
\Omega^{=I}=d\tau \Omega^{=I}_{\tau}, \qquad \Omega^{(0)}=d\tau \Omega^{(0)}_{\tau}, \qquad  \Omega^{IJ}=d\tau \Omega^{IJ}_{\tau}, \qquad
\end{eqnarray}
and
\begin{eqnarray}\label{DX==}
D_\tau X^{=}=\partial_\tau X^{=} + 2 \Omega^{(0)}_\tau X^{=}\; , \qquad
D_\tau \theta^-_{{q}}= \partial_\tau \theta^-_{{q}}
+ \Omega^{(0)}_{\tau}\theta^-_{{q}}  + {1\over 4} \Omega^{IJ}_{\tau}
\theta^-_{{p}}\gamma_{{p}{q}}^{IJ}\;
\end{eqnarray}
 ({\it cf.}  (\ref{Du--}) and (\ref{Dv-dq=})).

By construction, the action (\ref{S=smf=An})  is invariant under the rigid supersymmetry (\ref{susyAn}),  (\ref{e-=ev-}),
\begin{eqnarray}\label{susyAn10-11}
\delta_\varepsilon X^== 2i\theta^-_{{q}}\varepsilon^{-{q}}  , \qquad \delta_\varepsilon X^I= 2i\theta^-_{{q}}{\gamma}^I_{q\dot{q}}\varepsilon^{+\dot{q}} , \qquad \delta_\varepsilon \theta^-_{{q}}= \varepsilon^{-{q}}
 , \qquad
\delta_\varepsilon \Omega^{=I}_{\tau}=0\; , \qquad \delta_\varepsilon \rho^\# =0  . \qquad
\end{eqnarray}
As far as the $\kappa$--symmetry (\ref{kappa=irr}) is concerned, it is easy to check that all the fields in the action (\ref{S=smf=An}) are just invariant under it,
\begin{eqnarray}\label{kappaAn}
\delta_\kappa X^== 0 \; , \qquad \delta_\kappa  X^I= 0 \; , \qquad \delta_\kappa  \theta^-_{{q}}= 0
 \; , \qquad
\delta_\kappa  \Omega^{=I}_{\bar{z}}=0  \; , \qquad \delta_\kappa  \rho^\# =0  \;   .
\end{eqnarray}
In this sense the $\kappa$--invariance of the analytic basis form of the massless superparticle   action is trivial.
Let us recall that this  property is also characteristic for the pure twistor form of the  Ferber-Schirafuji action, (\ref{S-FS=mu}), the quantization of which results in the on-shell superfields of maximal 4D SYM and SUGRA theories.

\subsection{Massless D=10 and D=11 superparticle quantization in the analytical basis}

\label{11Dsuperparticle}
The quantization of massless 11D superparticle in analytical basis of Lorentz harmonic superspace has been considered in \cite{Bandos:2007wm} and the 10D case can be carried out in a similar manner (see also \cite{Sokatchev:1985tc,Sokatchev:1987nk} where only vector harmonics were used). So we refer to \cite{Bandos:2007wm}  for the discussion on the structure of bosonic constraints and   their resolution in classical and quantum theory, and concentrate here on the quantization of the fermionic variables (omitting the details when these can be found   in  \cite{Bandos:2007wm}).
Notice only that the worldline field $\rho^\#(\tau)$ becomes identified with the momentum $P_=$ of the $X^=$ coordinate function, $\rho^\#-P_=\approx 0$, so that after quantization it can be replaced by $-i\hbar \partial_== -i\hbar \frac \partial {\partial x^=}$ (below, for shortness, we set $\hbar=1$).
The fermionic constraints which follow from the action (\ref{S=smf=An}),
\begin{eqnarray}\label{d+q}
d^+_q=-\pi^+_q + 2i\rho^\# \theta^-_q\approx 0\; , \qquad  \pi^+_q =-\frac {\partial L} {\partial \dot{\theta}{}^-_q}
\; ,
\end{eqnarray}
obey the Poisson bracket relations ($\{ \theta^-_q, \pi^+_p\}_{_{P.B.}} = \delta_{qp}$)
\begin{eqnarray}\label{d+qd+q}
\{ d^+_q, d^+_p\}_{_{P.B.}} =-4i\rho^\# \delta_{qp}\;  \qquad
\end{eqnarray}
and hence are  second class constraints in Dirac's classification \cite{Dirac:1963}.

After the quantization, as a consequences of the bosonic first class constraints and the (explicitly resolved) bosonic second class constraints,  the 'wavefunction' of the superparticle depends on one bosonic coordinate $x^=$ (or on its 'momentum' $\rho^\#$), on a set of homogeneous coordinates of  ${\bb S}^{D-2}$ given  by Lorentz harmonics $v_{\alpha q}^{\; -}$ (see (\ref{v-qi=G-H})), and on a set of fermionic  variables. The type of these latter depends on the way of quantization chosen for the fermionic coordinate function $\theta^-_q$ obeying the second class constraints (\ref{d+q}) with characteristic Poisson brackets (\ref{d+qd+q}).

There is no SO(8) (SO(9)) covariant way to solve the second class constraints (\ref{d+q}) explicitly. If one solves them implicitly by passing to Dirac brackets \cite{Dirac:1963}, one finds that $\theta^-_q$ obey
\begin{eqnarray}\label{th-th-=DB}
\{ \theta^-_q , \theta^-_p\}_{_{D.B.}} = - \frac {i}{4\rho^{\#}}\delta_{qp}\; . \qquad
\end{eqnarray}
This implies that after quantization
\begin{eqnarray}\label{th=G}
\hat{\theta}^-_q= \frac {1} {\sqrt{2\rho^\# } }\; \mathfrak{C}_{q } \; , \qquad
\end{eqnarray}
where $\mathfrak{C}_{q } $  obey the Clifford algebra
\begin{eqnarray}\label{GG=I} {} \{ \mathfrak{C}_{q} \, , \mathfrak{C}_{p}  \} = 2   \delta_{qp} I \; , \qquad \begin{cases}q=1,..,8\; for\; D=10 \\ q=1,..,16\;  for \; D=11. \end{cases}
\end{eqnarray}

One might want to try a possibility to consider the wavefunction of the superparticle to be a Clifford superfield
$\Phi (x^=, v_{\alpha q}^{\; -}, \mathfrak{C}_{q })$. Similar approach to  D=10 superamplitudes, which  implies their dependence on $n$ sets of Clifford variables $\mathfrak{C}_{q(i) }$, was developed in
\cite{CaronHuot:2010rj} (we briefly describe it in  Appendix \ref{CliffSF}).
However, as unconstrained  $\Phi (x^=, v_{\alpha q}^{\; -}, \mathfrak{C}_{q })$ contains the terms up to 8-th degree in $\mathfrak{C}_{q }$ in D=10  and up to 16-th degree in the case of D=11, the Clifford superfield approach  does not look economic and the corresponding description of amplitudes seems to be reducible  (see also discussion in the Conclusion of recent \cite{Bandos:2017zap}). A more economic constrained superfield description originates in an alternative quantization of the fermionic degrees of freedom of the D=10 and D=11 massless superparticle which we are going to  describe now.

The idea of this alternative quantization can be followed back  to the first studies of  spinning particle mechanics  \cite{Berezin:1976eg,Brink:1976uf} in which the fermionic superpartners $\psi^\mu (\tau)$ of the bosonic coordinate function
$x^\mu(\tau)$ obeys the Dirac brackets  $\{ \psi^\mu ,\psi^\nu\}_{D.B.}= -2i  \eta^{\mu\nu}$. After quantization these produce the Clifford algebra commutation relation  $\{ \hat{\psi}{}^\mu ,\hat{\psi}{}^\nu\}=2  \eta^{\mu\nu}$ for the fermionic operator. Then the  standard way to arrive at Dirac (or Weyl) equation for the wavefunction of the spinning particle passes through using the representation of the fermionic operators by Dirac matrices, $\hat{\psi}{}^\mu \mapsto \Gamma^\mu{}_\alpha{}^\beta$, and allowing the wavefunction to carry the corresponding spinorial index.

Similar method was used in \cite{Green:1999by} in light-cone quantization of D=11 superparticle and in  \cite{Bandos:2007wm} for covariant quantization and the study of  hidden symmetries of this. There we identified  the quantum version of the fermionic second class constraints with Clifford algebra valued element  $\mathfrak{C}_{q }$, \begin{eqnarray}\label{d+q=Clq} {}  \hat{d}^+_q\;=\; \sqrt{2\rho^\#}\;\mathfrak{C}_{q }
\end{eqnarray}and represented  $\mathfrak{C}_{q }$ with $q=1,..,16$   by  sixteen  $256\times 256$ gamma  matrices acting on
the 'wavefunction' of the 11D superparticle which belongs to  256 component Majorana spinor of $SO(16)$ group. This is split on two 128 component Mayorana-Weyl spinors, one describing the on-shell fermionic gravitino degrees of freedom $\Psi_{Iq}$ and other ({ 128}={84}+{44}) describing the bosonic on-shell degrees of freedom. Under
SO(9) subgroup of SO(16) this latter can be decomposed on antisymmetric tensor $A_{IJK}$ ({\bf 84}) and symmetric traceless
$h_{IJ}$ ({\bf 44}). Choosing the block-anti-diagonal representation of the d=16 gamma matrices, one finds that
the action of the Gamma matrix on Majorana spinor mixes the two Majorana Weyl spinors,
\begin{eqnarray}\label{Cliff=16+16=}
\mathfrak{C}_{{q}} A_{IJK}= 3\gamma_{qp[IJ} \Psi_{K]p}\; , \qquad
\mathfrak{C}_{{q}} h_{IJ}= 2 \gamma_{qp(I} \Psi_{J)p}\; , \qquad
 \nonumber \\
 \mathfrak{C}_{{q}} \Psi_{Ip} = \gamma^J_{qp} h_{JI} + {1\over 3!} \left(\gamma^{IJKL}_{qp}+ 6\delta^{I[J}\gamma^{KL]}_{qp} \right)A_{JKL} \; . \qquad
\end{eqnarray}

One can perform a similar quantization of the D=10 superparticle. In it
$\mathfrak{C}_{q }$ with $q=1,..,8$  are represented by eight $16\times 16$ gamma  matrices of
$SO(8)$ and the wavefunction is a 16 component Majorana spinor of  $SO(8)$ which can be split on
two 8-dimensional representations. The triality of the  $Spin(8)$ group implies the equivalence of its two spinor and one vector representations,  8s, 8c and 8v, so that we can decompose the Majorana spinor on, say,
8c and 8v  and identify these as a fermionic c-spinor $ \psi_{\dot q }$ and a bosonic vector $w^I$. The action of the Majorana representation of gamma matrices on the Majorana spinor wavefunction is then described by  (\ref{susy=8+8a}) (in which the $\rho^\#$ multiplier can be removed by a proper redefinition of the fields),
\begin{eqnarray}\label{Cliff=8+8=}
\mathfrak{C}_{{q}} {\psi}_{\dot q } =   i \gamma^I_{q\dot{q}}  \;{w}^I \; , \qquad \mathfrak{C}_{{q}} {w}^I =   \gamma^I_{q\dot{q}}  {\psi}_{\dot q } \; \; , \qquad
\end{eqnarray}

Superfields collecting the above  supermultiplets of fields will be the on-shell superfields of D=11 SUGRA and  D=10 SYM  supermultiplets, the counterpart of D=4 superfield (\ref{Phi=3,4}).
To obtain such superfields in superparticle quantization, we use
a new (to the best of our knowledge) manner of quantization of the system with second class constraints, in which the state vector is represented by multi-component field(s) and  the quantum second class constraints act on it as differential operators mixing the components of the state vector rather than annihilating them.

Namely, when quantizing the dynamical system described by the action (\ref{S=smf=An}),  instead of passing to Dirac brackets (\ref{th-th-=DB}),
we allow the quantum counterparts  $\hat{d}^+_q$ of the second class constraints  $d^+_q$ to obey the Clifford-like algebra (as follows from straightforward quantization of (\ref{d+qd+q})), realize them as differential operators in an appropriate on-shell superspace,  $\hat{d}^+_q=D^+_q$,   introduce an appropriate set of superfields in  this on-shell superspace
 (these can be guessed from (\ref{Cliff=8+8=}) and (\ref{Cliff=16+16=}) for D=10 and D=11)
 and
assume that the differential operator  $D^+_q$ acts on this set of superfields in the way which is essentially determined by
 (\ref{Cliff=8+8=}) and (\ref{Cliff=16+16=}).

The result of such a quantization can be called constrained on-shell superfield formalism.
Interestingly enough, such on-shell superfield formalism
had been  proposed in  \cite{Galperin:1992pz} without any discussion  of a superparticle model.

\subsubsection{ On-shell superspaces for 10D SYM and 11D SUGRA }

\label{On-shSSP=sec}

The D=10 and D=11 on-shell superfields are defined on the superspace with bosonic coordinates  $x^=$ and $v_{\alpha q}^{\; -}$, and fermionic  $\theta^-_q$ coordinates,
\begin{eqnarray}\label{On-shellSSP}
\Sigma^{(D-1|n_D)} : \qquad &&  \{ (x^= , \theta^-_q, v_{\alpha q}^{\; -})\} \; ,  \qquad \{ v_{\alpha q}^{\; -}\} = {\bb S}^{D-2}\; , \qquad \\ \nonumber &&
q=1,..., n_D\; , \qquad \alpha =1,..., 2n_D\; , \qquad
n_D= \begin{cases} \; 8\qquad for\; D=10 \cr  16\qquad  for\; D=11 \; . \end{cases}
\qquad \nonumber
\end{eqnarray}
which, hence, are called D=10 and D=11 on-shell superspaces. The generic superfields on these superspaces contain too many component fields so that D=10 and D=11 on-shell  superfields on $\Sigma^{(D-1|n_D)}$ cannot be unconstrained, {\it i.e.} they should obey some superfield equations.

To arrive at these equations when quantizing 10D and 11D massless superparticle, we begin by realizing
the quantum version of the constraint $\hat{d}^+_q$ as differential operator on
(\ref{On-shellSSP})
\begin{eqnarray}\label{D+dq:=}
D^+_{{q}}={\partial}^+_{{q}} + 2i  \theta^-_{{q}}\partial_{=} \; , \qquad \partial_{=}
:={\partial\over \partial x^=}\; , \qquad \partial^+_{{q}}
:={\partial\over \partial \theta^-_{{q}}}\; .
\end{eqnarray}
Then, as the fermionic constraint is of the second class, i.e.
the anticommutator of two such constraints is nonzero,
\begin{eqnarray}\label{D+qD+p=I}
\{ D^+_{{q}},  D^+_{{p}}\} = 4i \delta_{qp}\partial_=\,  ,
\end{eqnarray}
we cannot assume that the action of $ D^+_{{q}}$ annihilates the state vector, {\it i.e.} we should expect the presence of a nonvanishing {\it r.h.s.} in the equation reflecting the existence of the second class constraint by defining the action of  $ D^+_{{q}}$  on the state vector (wavefunction)\footnote{An analytic superfield formalism of \cite{Bandos:2017zap} is based on  homogeneous equations, but requires breaking of $SO(D-2)$ symmetry down to $SO(D-4)$ by splitting $D^+_{{q}}$ on two sets of complex  operators $D^A$ and $\bar{D}_A$ with $A=1,2,3,4$. }. This can be done consistently if the wavefunction carries certain indices of $SO(D-2)$ group.

To find the index structure of the wavefunctions and the {\it r.h.s.}-s of the above mentioned equations, it is convenient to notice that (\ref{D+qD+p=I}) differs from the Clifford algebra by the presence of $\partial_=$ in the r.h.s.
Then the  superfield equations which should be imposed on the on-shell superfield to reduce their field content to the fields describing 10D SYM and 11D supergravity multiplets are suggested by the action of the formal Clifford algebra elements on the fields of these multiplets: (\ref{Cliff=8+8=})
and (\ref{Cliff=16+16=}).

\subsubsection{ On-shell superfields and superfield equations  of 10D SYM }

\label{On-shellSYM}
The basic D=10 equations  \cite{Galperin:1992pz} are imposed on the fermionic superfield
 $\Psi_{\dot{q}}=\Psi_{\dot{q}}(x^=, \theta^-_{\dot{q}}, v_{\alpha \dot{q}}{}^-)$ carrying $c$-spinor index of SO(8). They read
\begin{eqnarray}\label{D+Psi=gV}
D=10\; : \qquad D^+_{{q}}\Psi_{\dot{q}} =\gamma^I_{q\dot{q}}\, \partial_= W^I\; , \qquad
q= 1,...,8\; , \quad \dot{q}=1,..,8 \; , \quad I =1,..,8 \; .
\end{eqnarray}
The superfield $W^I$ is defined by eq. (\ref{D+Psi=gV}) itself which also implies that it obeys\footnote{
To be more precise, Eq. (\ref{D+Psi=gV}) determines the form of $V^I=\partial_= W^I$. Just this superfield was used in  \cite{Galperin:1992pz}. }
\begin{eqnarray}\label{D+W=gdPsi}
D^+_{{q}}W^I= 2i \gamma^I_{q\dot{q}}\Psi_{\dot{q}}
\; . \qquad
\end{eqnarray}
This equation shows that there are no other independent components in the constrained on-shell superfield
$\Psi_{\dot{q}}$. Indeed, one can find that its decomposition in the powers of $\theta^-_{\dot{q}}$ reads
\begin{eqnarray}\label{Psi=psi+}
& \Psi_{\dot{q}} (x^=, v^-_q; \theta^-_{{q}})= \psi_{\dot{q}}(x^=,v^-_q) + \theta^-_{{q}}   \gamma^I_{q\dot{q}} \, \partial_= w^I (x^=,v^-_q) + \hspace{7.5cm}\\
& + \sum\limits_{k=1}^{4}\left(-i\right)^k \; \frac {(2k-1)!!}{(2k)!!\, (2k)!}
(\theta^-{\gamma}^{I_{k-1}I_{k}}\theta^-)\ldots (\theta^-{\gamma}^{I_{1}I_{2}}\theta^-)\, (\gamma^{I_1I_2}\ldots \gamma^{I_{k-1}I_{k}})_{\dot{q}\dot{p}} (\partial_=)^k \psi_{\dot{p}} +  \nonumber \\
 & + \sum\limits_{k=1}^{3}\left(-i\right)^k \; \frac {(2k)!!}{(2k+1)!!\, (2k+1)!}
(\theta^-{\gamma}^{I_1I_2}\theta^-)\ldots (\theta^-{\gamma}^{I_{k-1}I_{k}}\theta^-) (\tilde{\gamma}{}^{I_1I_2}\ldots \tilde{\gamma}{}^{I_{k-1}I_{k}}\tilde{\gamma}{}^I\theta^-)_{\dot q} (\partial_=)^{k+1} w^I   . \nonumber
\end{eqnarray}
Hence, as it was stated already in
\cite{Galperin:1992pz}, this constrained superfield describes on-shell degrees of freedom of SYM multiplet.

\subsubsection{ On-shell SUSY and supersymmetric invariant of the (linearized) 10D SYM }

\label{On-shellSG}

The  supersymmetry transformations of the leading components of bosonic and fermionic on-shell superfields
\begin{eqnarray}\label{Psi0=psi}
\Psi_{\dot{q}}\vert_0 =\psi_{\dot{q}} \; ,\qquad  W^I\vert_0 =w^I\; .
\end{eqnarray}
can be extracted from  the superfield equations (\ref{D+Psi=gV}) and (\ref{D+W=gdPsi}).
 They have the form already announced in (\ref{susy=8+8a}),
\begin{eqnarray}\label{susy=8+8}
\delta_{\epsilon}\psi_{\dot q } = \epsilon^{-{q}}  \gamma^I_{q\dot{q}}  \; \partial_{=} w^I \; , \qquad \delta_{\epsilon} w^I = 2i \epsilon^{-{q}} \gamma^I_{q\dot{q}}  \psi_{\dot q }\; .  \qquad
\end{eqnarray}
It is also not difficult to find the following supersymmetric invariant
\begin{eqnarray}\label{susy=8+8inv}
I_{_{10D}}= \int dx^= \left( \partial_{=}w^I\, \partial_{=}w^I + 2i \partial_{=} \psi_{\dot q } \; \psi_{\dot q }\right) \; ,  \qquad
\end{eqnarray}
 and to observe that it can be written in terms of superfields as follows
\begin{eqnarray}\label{susy=8+8=Issp}
I_{_{10D}}=\frac 1 {16} \int dx^=
 D_q^+(\Psi_{\dot q } \overleftrightarrow{D_q^+} \Psi_{\dot q })\vert_0\; .  \qquad
\end{eqnarray}

Eq. (\ref{susy=8+8=Issp}) can be considered as an integral over a $(1,8)$ dimensional subsuperspace $\{ (x^=, \theta^-_q)\}= {\frak L}^{(1|8)}$ of 10D on-shell superspace $\Sigma^{(9|8)}$ but taken with an exotic superspace measure similar to one used in \cite{Tonin:1991ii,Tonin:1991ia} to construct a worldsheet superfield formulation of the heterotic string (a formal discussion on integral forms in superspace and applications to 3d superfield theories can be found in   \cite{Zupnik:1989bw}).  Notice also that our supersymmetric invariants depend on the spinor frame variables
$v_{\alpha q}^{\; -}$ which is possible as these are inert under supersymmetry.

For our discussion below it will be important that one can also write similar supersymmetric invariant containing
two on-shell superfields,  $\Psi_{\dot{q}}(x^=, v^{\; -}_{\alpha q}, \theta^-_q)$ and  $\tilde{\Psi}_{\dot{q}}(x^=, v^{\; -}_{\alpha q}, \theta^-_q)$. Indeed, one can check that
\begin{eqnarray}\label{xtPP=}
{\cal I}_{10D}= \frac 1 {16} \int dx^=
 D_q^+(\widetilde{\Psi}_{\dot q }\overleftrightarrow{D_q^+} \Psi_{\dot q })\vert_0
   \equiv \frac 1 {16} \int dx^= D_q^+(\widetilde{\Psi}_{\dot q }D_q^+ \Psi_{\dot q }+ D_q^+ \widetilde{\Psi}_{\dot q } \; \Psi_{\dot q })\vert_0 = \qquad \\ \label{xtpp=}
 =  \int dx^= \left( \partial_{=}\widetilde{w}{}^I\, \partial_{=}w^I + i \partial_{=} \widetilde{\psi}{}_{\dot q } \; \psi_{\dot q } - i \widetilde{\psi}{}_{\dot q }  \partial_{=} \psi_{\dot q }\right) \; .  \qquad
\end{eqnarray}
is invariant under supersymmetry defined by (\ref{susy=8+8}) and
$$ \delta_{\epsilon}\widetilde{\psi}_{\dot q } = \epsilon^{-{q}}  \gamma^I_{q\dot{q}}  \; \partial_{=} \widetilde{w}^I\; ,  \qquad \delta_{\epsilon}\widetilde{w}^I = 2i \epsilon^{-{q}} \gamma^I_{q\dot{q}}  \widetilde{\psi}_{\dot q }\; . $$

Even more useful for studying superamplitudes will be the expression of the above invariant in terms of Fourier images of fields and superfields
\begin{eqnarray}\label{FurSF}
\widetilde{\Psi}_{\dot q }(\rho^{\#},v^-,\theta^-):= \frac 1 {\sqrt{2\pi }}\int  dx^=
e^{-ix^= \rho^{\#}}\widetilde{\Psi}_{\dot q }(x^=,v^-,\theta^-)= \qquad \nonumber \\ =\widetilde{\psi}_{\dot q }(\rho^{\#},v^-)+i \rho^\#
\theta^-_q\gamma^I_{q\dot q }\widetilde{w}{}^I(\rho^{\#},v^-)+ {\cal O}( \theta^-_q\theta^-_p)
 \; .  \qquad
\end{eqnarray}
The integrand of this expression reads
\begin{eqnarray}\label{DtPDP=}
{\frak I}_{10D}(\rho^\#)= \frac 1 {16}
 D_q^+(\widetilde{\Psi}_{\dot q }(\rho^{\#}) \overleftrightarrow{D_q^+} \Psi_{\dot q }(-\rho^{\#}))\vert_0
   \equiv \frac 1 {16}  D_q^+(\widetilde{\Psi}_{\dot q }D_q^+ \Psi_{\dot q }+ D_q^+ \widetilde{\Psi}_{\dot q } \; \Psi_{\dot q })\vert_0 = \qquad \\ \label{xtpp==}
 =  \widetilde{w}{}^I(\rho^{\#}) (\rho^{\#})^2 w^I(-\rho^{\#}) - 2\rho^{\#} \widetilde{\psi}{}_{\dot q }(\rho^{\#}) \; \psi_{\dot q } (-\rho^{\#}) \;  \qquad
\end{eqnarray}
where we have written only the  argument essential for the discussion below, so that $\widetilde{\Psi}_{\dot q }(\rho^{\#})\equiv \widetilde{\Psi}_{\dot q }(\rho^{\#}, v^-_q,\theta^-_q)$ {\it etc.}, and the covariant derivative has the form
\begin{eqnarray}\label{D=d+rho}
D^{+}_{{q}} =\partial^{+}_{{q}_l} - 2 \rho^{\#}\theta^-_{{q}}  \qquad     \; ,
\qquad \partial^{+}_{{q}}:= \frac \partial{ \partial\theta^-_{{q}}}\; .
\end{eqnarray}
To avoid confusion when reproducing (\ref{DtPDP=}),  we notice that  in the expression $D_q^+\widetilde{\Psi}_{\dot q }(\rho^{\#})$ the derivative
$D^{+}_{{q}}$ does have the form (\ref{D=d+rho}), which can be denoted by $D^{+}_{{q}}(\rho^{\#})$,
while in  $D_q^+{\Psi}_{\dot q }(-\rho^{\#})$ we should use $D^{+}_{{q}}(-\rho^{\#}) =\partial^{+}_{{q}_l} + 2 \rho^{\#}\theta^-_{{q}} $. This reflects the fact that the action of the fermionic covariant derivative is actually defined in $x^=$--coordinate representation, and, its  momentum, $\rho^\#$--representation (\ref{D=d+rho}) is restored from that.

\subsubsection{ On-shell superfields,  superfield equations  and supersymmetric invariant of  linearized 11D SUGRA }

Similarly, observing (\ref{Cliff=16+16=}) one can guess that the linearized 11D supergravity is described by a bosonic  antisymmetric tensor superfield $\Phi^{IJK}=
\Phi^{[IJK]}(x^=, \theta^-_{{q}}, v_{\alpha {q}}^{\; -})$ which obeys the superfield equation \cite{Galperin:1992pz}
\begin{eqnarray}\label{D+Phi=gPsi}
D=11\; : \qquad D^+_{{q}}\Phi^{IJK}  = 3i\gamma^{[IJ}_{qp}  \Psi^{K]}_{p}\, , \qquad \gamma^I_{qp}\Psi^I_{p}=0\; , \qquad
q,p= 1,...,16\; , \quad  I =1,..,8,9\, . \qquad
\end{eqnarray}
The consistency of Eq. (\ref{D+Phi=gPsi}) requires
\begin{eqnarray}\label{D+Psi=11D}
D=11\; : \qquad D^+_{{q}}\Psi^{I}_{p}  = \frac 1 {18} \left(\gamma^{IJKL}_{qp}+ 6 \delta ^{I[J}\gamma^{KL]}_{qp}  \right) \partial_{=}\Phi^{JKL}   + 2  \partial_{=}H_{IJ}\gamma^{J}_{qp}  \; , \qquad
\\ \label{D+h=11D} D^+_{{q}}H_{IJ} = i \gamma^{(I}_{qp} \Psi^{J)}_{p} \; , \qquad H_{IJ}= H_{JI}\; , \qquad H_{II}=0 \; . \qquad
\end{eqnarray}

These equations can be used to extract the supersymmetry transformations of the leading components of superfields
\begin{eqnarray}\label{psi=Psi0}
D=11:\quad \phi_{IJK}:= \phi_{IJK} (x^=,v^-_q)=\Phi_{IJK}\vert_{\theta_q^-=0}\; ,\qquad   \psi^{I}_{p}= \Psi^{I}_{p}\vert_{\theta_q^-=0}\; ,\qquad h_{IJ}= H_{IJ}\vert_{\theta_q^-=0}\; .\qquad
\end{eqnarray}
These are
\begin{eqnarray}\label{susy-phi=11D}
  && \delta_\epsilon \phi_{IJK} = 3i \epsilon^- \gamma_{[IJ}\psi_{K]}\; , \qquad \delta_\epsilon h_{IJ} = i \epsilon^- \gamma_{(I}\psi_{J)}\; , \qquad \nonumber
\\
&& \qquad  \delta_\epsilon \psi^{I}_{q}  = \frac 1 {18} (\epsilon^-\gamma^{IJKL}+ 6 \delta ^{I[J}\epsilon^-\gamma^{KL]})_q \partial_{=}\phi^{JKL}   + 2  \partial_{=}h_{IJ}(\epsilon^-\gamma^{J})_{q}   \; . \qquad
\end{eqnarray}
One can easily find that the following linearized action is invariant (modulo integral of total derivative) under the rigid supersymmetry (\ref{susy-phi=11D}):
\begin{eqnarray}\label{Isusy=11D}
 && I_{_{11D}} = \int dx^= \left(2\partial_{=} h_{IJ} \partial_{=} h_{IJ} + \frac 1 6  \partial_{=}\phi_{IJK} \partial_{=}\phi_{IJK} + i \partial_{=}\psi_{Iq} \, \psi_{Iq}  \right) \; . \qquad
\end{eqnarray}

One can also check that the integrand of (\ref{Isusy=11D}) can be identified with the leading component of the  composed  on-shell superfield $D_q^+(\Psi_{Ip}D_q^+\Psi_{Ip})$,
\begin{eqnarray}\label{DfDf=11D}
\frac {1} {32} D_q^+(\Psi_{Ip}D_q^+\Psi_{Ip})\vert_0 = 2\partial_{=} h_{IJ} \partial_{=} h_{IJ} + \frac 1 6  \partial_{=}\phi_{IJK} \partial_{=}\phi_{IJK} + i \partial_{=}\psi_{Iq} \,\psi_{Iq} \; , \\ \nonumber   \qquad q,p=1,...,16 \; ,  \qquad I,J,K=1,...,9\; ,
\end{eqnarray}
so that our invariant can be written in the form
\begin{eqnarray}\label{Isusy=ssp11D}
 I_{_{11D}} = \frac {1} {32} \int dx^=  D_q^+(\Psi_{Ip}D_q^+\Psi_{Ip})\vert_0  \; . \qquad
\end{eqnarray}
This can be considered as an integral over a superline  ${\frak L}^{(1|16)}=  (x^=, \theta^-_q)\}$ in the
11D on-shell superspace $\Sigma^{(10|16)}$ but taken with an exotic superspace measure (see comments on the 10D case, Eq. (\ref{susy=8+8=Issp})  and below it). Clearly, the value of the integral over the superline depend on the spinor frame variables $v_{\alpha q}^{\; -}$, which  are inert under supersymmetry (\ref{susyAn-Min}).

Like in the 10D SYM case, we can also write the invariant including two different supergravity superfields
\begin{eqnarray}\label{DftDf=11D}
{\cal I}_{11D}&=& \frac {1} {32}\int dx^= D_q^+(\widetilde{\Psi}_{Ip}\overleftrightarrow{D}_q^+\Psi_{Ip})\vert_0 = \qquad \nonumber \\ &=& \int dx^= (4\partial_{=} \widetilde{h}_{IJ} \partial_{=} h_{IJ} + \frac 1 3  \partial_{=}\widetilde{\phi}_{IJK} \partial_{=}\phi_{IJK} + i \partial_{=}\widetilde{\psi}_{Iq} \,\psi_{Iq} - i \widetilde{\psi}_{Iq}  \partial_{=}\psi_{Iq}) \; . \\ \nonumber  && \qquad q,p=1,...,16 \; ,  \qquad I,J,K=1,...,9\; ,
\end{eqnarray}

Below we will generalize the above on-shell superfield representation for the case of three amplitudes
of 10D SYM and linearized 11D supergravity and present a candidate for generalization of the BCFW recurrent relations \cite{Britto:2005fq} for such  superamplitudes.
The equations for supersymmetric invariants, (\ref{xtPP=}) and (\ref{DftDf=11D}),  will be useful when writing these candidate  BCFW-type relations for 10D and 11D superamplitudes \cite{Bandos:2016tsm}.

\section{Superfield representation for  10D SYM and 11D SUGRA amplitudes}

In this section we first introduce the tree superamplitudes of 10D SYM and 11D SUGRA as multiparticle counterparts of the corresponding on-shell superfields (secs. 6.1 and 6.2). These latter suggest the index structure of the amplitudes, the variables they depend on, as well as the set of differential equations these obey.

As we have already stressed at the beginning of Sec. 4, we are working with the on-shell superamplitudes so that we do not use neither Lagrangian nor Feynman rules in their derivation. The experience gained in D=4 suggests to search for a BCFW-type recurrent relations to calculate the on-shell superamplitudes starting from the basic 3-point superamplitudes. Those include on-shell superamplitudes dependent on deformed helicity spinors and fermionic variables. In sec. 6.3 we generalize the BCFW deformations for the case of 10D SYM and 11D SUGRA superamplitudes. The  candidate BCFW recurrent relations are discussed in secs. 6.4-6.6. In this latter part the results are preliminary.

\subsection{Superfield representation of the 10D SYM amplitudes: 10D superamplitudes}

The structure of the constrained on-shell superfields of 10D SYM and of the equations imposed on them,
 (\ref{D+Psi=gV}) and  (\ref{D+W=gdPsi}), suggests to define a set of bosonic superfield amplitudes or bosonic {\it superamplitudes}
 \begin{eqnarray}\label{Sf-Amp-q}
{\cal A}^{(n)}_{I_1 ... I_l... I_n}(\{\rho^{\#}_{(i)}\}; \{v^-_{{q}(i)}\} ; \{\theta^-_{{q}(i)}\}  )=:
{\cal A}^{(n)}_{... I_l...}( ...; k_l, \theta^-_{l}; ...)     \;   \qquad
\end{eqnarray}
depending, in addition to spinor helicity variables $v^-_{\alpha {q}(i)}$
 and `energies' $\rho^{\#}_{(i)}$, on $n$-sets of fermionic 8s-spinor variables $\theta^-_{{q}(i)}$, and carrying a 8v index for each of the scattered particles. The {\it r.h.s.} of (\ref{Sf-Amp-q}) shows the schematic notation which we will use below when it cannot result in a confusion. The leading component ($\theta^-_{{q}(i)}=0$ `value') of this superfield amplitude is a particular case of the 'component' amplitude in    (\ref{S-multA}), the purely bosonic amplitude
  ${\cal A}^{(n)}_{... I_l...}(\{\rho^{\#}_{(i)}\}; \{v^-_{{q}(i)}\} ) $,
where  multidotes also denote the 8v indices.

The superamplitudes (\ref{Sf-Amp-q}) obey the following  equations
\begin{eqnarray}\label{D+A=GA}
 D^{+(l)}_{{q}_l} {\cal A}^{(n)}_{... I_l...}( ...; k_l, \theta^-_{l}; ...)
= 2i \gamma^{I_l}_{q_l\dot{q}_l} {\cal A}^{(n)}_{... \dot{q}_l...}( ...; k_l, \theta^-_{l}; ...)\; \, ,
\end{eqnarray}
which are suggested by the equations (\ref{D+W=gdPsi}) for the on-shell superfields. In (\ref{D+A=GA})
the fermionic covariant derivatives  have the form ({\it cf.} (\ref{D=d+rho}))
\begin{eqnarray}\label{D+l=d+rho-th}
D^{+(l)}_{{q}_l} =\partial^{+(l)}_{{q}_l} - 2 \rho^{\#}_{(l)}\theta^-_{{q}(l)}  \qquad     \; ,
\qquad \partial^{+(l)}_{{q}_l}:= \frac \partial{ \partial\theta^-_{{q}(l)}}\;
\end{eqnarray}
and obey
\begin{eqnarray}\label{D+D+=rho}
{}\{ D^{+(l)}_{{q}}, D^{+(j)}_{{p}} \} = - 4 \delta_{l\, l'}\delta_{{q}{p}}  \, \rho^{\#}_{(l)}\; .
\end{eqnarray}
The fermionic superamplitudes  in the {\it r.h.s.} of (\ref{D+A=GA}),
$$
{\cal A}_{... \dot{q}_l...} (...; k_l, \theta^-_{l}; ...)= {\cal A}^{(n)}_{I_1...I_{l-1} \dot{q}_lI_{l+1}... I_n}(\{\rho^{\#}_{(i)}, v^-_{{q}(i)}; \theta^-_{\dot{q}(i)}\}  )\, , $$
are defined by contractions of Eq. (\ref{D+A=GA}) with $\gamma^{I_l}_{q_l \dot{p}_l}$ and, by  consistency of (\ref{D+A=GA}), obey
\begin{eqnarray}\label{D+Adq=GA}
D^{+(l)}_{{q}_l} {\cal A}^{(n)}_{... \dot{q}_l ...}( ...; k_l, \theta^-_{l}; ...)
=   i(-)^{\Sigma_l} \gamma^{I_l}_{q_l\dot{q}_l} \rho^{\#}_{l}{\cal A}^{(n)}_{... I_l ...}( ...; k_l, \theta^-_{l}; ...)\; .
\end{eqnarray}
The sign factors $(-)^{\Sigma_l} $ are introduced here to make the equation applicable for a more general case, where some of the indices denoted by multi-dots are fermionic ($c$-spinor ones). In the case under consideration $(-)^{\Sigma_l} =1$; see Eq.  (\ref{Sl=10D}) below  for generic case.
Eqs. (\ref{D+A=GA}) and (\ref{D+Adq=GA}) imply
\begin{eqnarray}\label{Sf-Amp=dec}
{\cal A}^{(n)}_{... \dot{q}_l...}( ...; k_l, \theta^-_{l}; ...)={\cal A}^{(n)}_{... \dot{q}_l...}( ...; k_l; ...) +
(-)^{\Sigma_l} \theta^-_{{q}l} \gamma^{I_l}_{q_l\dot{q}_l} {\cal A}^{(n)}_{... I_l...}( ...; k_l; ...)+ {\cal O}( \theta^-_{l}\gamma^{JK}\theta^-_{l} )    \; , \qquad
\end{eqnarray}
where  ${\cal A}^{(n)}_{... \dot{q}_l...}( ...; k_l; ...)=
{\cal A}^{(n)}_{... \dot{q}_l ...}( ...; k_l, 0; ...)$ and  ${\cal A}^{(n)}_{... I_l...}( ...; k_l; ...)= {\cal A}^{(n)}_{... I_l...}( ...; k_l, 0; ...)$ may  depend on other fermionic  coordinates, $ \theta^-_{j}$ with $j\not=l$. By analogy with decomposition of the on-shell superfield, we can show that the terms of higher order in $l$-th fermionic coordinate, ${\cal O}( \theta^-_{l}\gamma^{JK}\theta^-_{l} )$, are expressed through ${\cal A}^{(n)}_{... q_l...}( ...; k_l; ...)$, ${\cal A}^{(n)}_{... I_l...}( ...; k_l; ...)$ and $\rho^{\#}_{l}$.

Using the algebra of covariant derivatives, one can extend the above line of arguments and to obtain the equations for the  generic superamplitude,
\begin{eqnarray}\label{cA-Q1-Qn=10D}
{\cal A}^{(n)}_{Q_1... Q_l...Q_n}(k_{1}, \theta^-_{1}; ...; k_{l}, \theta^-_{l}; ...; k_{n}, \theta^-_{n})=:{\cal A}^{(n)}_{... Q_l...}( ...; k_l, \theta^-_{l}; ...)\; , \qquad
\end{eqnarray}
with a number of 8v and a number of 8c indices,
\begin{eqnarray}
\label{Ql=10D}
Q_l=(I_l, \dot{q}_l)\; ,
\qquad I_l= 1,...,8\; , \qquad \dot{q}_l=1,...,8\; .
\end{eqnarray}
The superfield equations for the generic superamplitude read
\begin{eqnarray}\label{D+Adq=pmGA}
\fbox{$ D^{+(l)}_{{q}_l} {\cal A}^{(n)}_{Q_1... Q_l...Q_n}( ...; k_l, \theta^-_{l}; ...)
= (-)^{\Sigma_l} \Delta_{q_lQ_lP_l} {\cal A}^{(n)}_{Q_1... P_l...Q_n}( ...; k_l, \theta^-_{l}; ...)\;$} \, ,
\end{eqnarray}
where ({\it cf.} (\ref{gqQQ'=10D}))
\begin{eqnarray}\label{Del-qQP=}
\Delta_{qQP}:=  i\gamma^{I}_{q\dot{q}}\left(  2\delta_Q^{I} \delta_P^{\dot{q}}+\rho^\#  \delta_Q^{\dot{q}} \delta_P^{I}  \right) \, ,
\end{eqnarray}
 and the integer ${\Sigma_l}$ counts (modulo 2) a number of the fermionic $\dot{q}_j$ indices among
$Q_{j}$'s with  $j=1,...,(l-1)$ \footnote{\label{Sigma-l} The choice of definition of ${\Sigma_l}$ is not unique. It corresponds to a prescription of the ordering of $\theta^-_{q_ii}$ in the decomposition of the superfield amplitudes. Such type ambiguity is similar to the one appearing in defining the old-fashioned form of the operator approach to the second quantization of non-relativistic fermionic field which can be found e.g. in \cite{Schweber:1961zz}.},
 \begin{eqnarray}\label{Sl=10D}
D=10\, : \qquad {\Sigma_l} =  \sum\limits_{j=1}^{l-1} \frac {(1-(-)^{\varepsilon (Q_j)})}{2}  \, , \qquad
\varepsilon (I_j)=0, \qquad \varepsilon (\dot{q}_j)=1. \qquad
\end{eqnarray}

Eqs. (\ref{D+Adq=pmGA}) contain (\ref{D+Adq=GA}) and
\begin{eqnarray}\label{D+AI=GA}
 D^{+(l)}_{{q}_l} {\cal A}^{(n)}_{... I_l...}( ...; k_l, \theta^-_{l}; ...)
= 2i (-)^{\Sigma_l}   \gamma^{I_l}_{q_l\dot{q}_l} {\cal A}^{(n)}_{... \dot{q}_l...}( ...; k_l, \theta^-_{l}; ...)\; \, .
\end{eqnarray}
(\ref{D+A=GA}) is a  particular case  of this equation in which  all the indices denoted by multidots are bosonic (8v); obviously, it is also correct for the case when the set $Q_1\ldots Q_{(l-1)}$ contains an even number of $\dot{q}_j$'s.

As in the case of (\ref{Sf-Amp=dec}), one can decompose the generic superamplitudes $ {\cal A}^{(n)}_{Q_1... Q_l...Q_n}( ...; k_l, \theta^-_{l}; ...)$ and define {\it $l$-type-component} amplitudes
   $ {\cal A}^{(n)}_{Q_1... Q_l...Q_n}( ...; k_l; ...)= {\cal A}^{(n)}_{Q_1... Q_l...Q_n} \vert_{\theta^-_{l}=0}$ which depend on other fermionic variables $\theta^-_{q j}$ with $j\not=l$. Their  supersymmetry transformations are essentially described by Eqs. (\ref{susy=8A+8A}):  $l$-th projection of supersymmetry mixes them as in  (\ref{susy=8A+8A})  while all the others are essentially given  by supertranslations of fermionic coordinates $\theta^-_{q j}$ with $j\not= l$. Then the combination  (\ref{xtpp=}) of such  $l$-type-component amplitudes is invariant under l-th projection of  supersymmetry  and one can ask  what is the superfield representation of that.
   This appears to coincide with leading $l$-component ($\theta_{l}=0$ 'value') of
\begin{eqnarray}\label{Df-ADfA-}
&& \; D^{+(l)}_{{q}_l} \left({\cal A}^{(n')}_{... \dot{q}_l}(...;k_l, \theta^-_{l})\; \overleftrightarrow{D}^{+(l)}_{{q}_l} {\cal A}^{(n)}_{ \dot{q}_l...}(-k_l, \theta^-_{l}; ...)  \right) \,  , \qquad
\end{eqnarray}
where $ {\cal A}^{(l)} \overleftrightarrow{D}^{+(l)}_{{q}_l} {\cal A}^{(n)}= {\cal A}^{(l)} {D}^{+(l)}_{{q}_l} {\cal A}^{(n)} -
(-)^{\varepsilon ({\cal A}^{(n')})} {D}^{+(l)}_{{q}_l}{\cal A}^{(n')} \;  {\cal A}^{(n)} $ and $\varepsilon ({\cal A}^{(l)})$ denotes the Grassmann parity of the superamplitude ${\cal A}^{(l)}$.

This in its turn  can be considered as a superspace integral   with an exotic superspace measure ({\it cf.} (\ref{susy=8+8=Issp}) and see comment below this equation)
\begin{eqnarray}\label{dxDfADfA}
&&  \int \, dx_l^{=} \, (-)^{\Sigma_l} D^{+(l)}_{{q}_l} \left({\cal A}^{(l)}_{... \dot{q}_l}(...;x_l^{=}, v^-_{{p}l}, \theta^-_{pl})\; \overleftrightarrow{D^{+(l)}_{{q}_l}} {\cal A}^{(n)}_{... \dot{q}_l...}(x_l^{=}, v^-_{pl}, \theta^-_{pl};...)  \right)\vert_{\theta^-_{ql}=0} \,   \qquad
\end{eqnarray}
of the fourier image of the superamplitudes with respect to $\rho^{\#}_l$,
\begin{eqnarray}\label{Ax--=int}
&& {\cal A}^{(n)}_{... \dot{q}_l...}(...;x_l^{=}, v^-_{pl}, \theta^-_{pl};...)
= \int d \rho^{\#}_{l} \;  e^{ ix_l^= \rho^{\#}_l} \; {\cal A}^{(n)}_{... \dot{q}_l...}(...;\rho^{\#}_l, v^-_{pl}, \theta^-_{pl};...)\; . \qquad
\end{eqnarray}

Notice that in (\ref{dxDfADfA}) $D^{+(l)}_{{q}_l} $ is the derivative in the coordinate representation,
\begin{eqnarray}\label{Df:=}
&&  D^{+(l)}_{{q}_l}= \frac {\partial} {\partial  \theta^-_{{q}l}} +2i \theta^-_{{q}l}\frac {\partial} {\partial x^{=}_{l}}\equiv \partial^+_{{q}l} +2i \theta^-_{{q}l}\partial_{=}^{l} \; . \qquad
\end{eqnarray}

\subsection{Tree superamplitudes in D=11}

The on-shell superfield description of linearized 11D  supergravity, as described by  Eqs. (\ref{D+Phi=gPsi}), (\ref{D+Psi=11D}) and (\ref{D+h=11D}), suggests to consider the superfield amplitudes carrying a set of three antisymmetrized
indices of SO(9) group,
\begin{eqnarray}\label{Sf-Amp-IJK}
{\cal A}^{(n)}_{[I_1J_1K_1] ... [I_lJ_lK_l]... [I_nJ_nK_n]}(\{\rho^{\#}_{i}\}; \{v^-_{{q}i}\} ; \{\theta^-_{{q}i}\}  )=:
{\cal A}^{(n)}_{... [I_lJ_lK_l]...}( ...; k_l, \theta^-_{l}; ...)     \; .  \qquad
\end{eqnarray}
Such a bosonic superamplitude obeys $n$ superfield equations \begin{eqnarray}\label{Df-AIJK}
&&  D^{+(l)}_{{q}_l} {\cal A}^{(n)}_{... [I_lJ_lK_l]... }(...;k_l, \theta^-_{l};...) = (-)^{\Sigma_l}
 \, 3i \gamma_{[J_lK_l| q_lp_l} \,  {\cal A}^{(n)}_{... \, |I_l] p_l\, ... }(...;k_l, \theta^-_{l};...) \,  , \quad l=1,..., n\; , \qquad
\end{eqnarray}
where $D^{+(l)}_{{q}_l}$ has the form of (\ref{D+l=d+rho-th}) with $q=1,...,16$, and  the superamplitude in the {\it r.h.s.},
 \begin{eqnarray}\label{fcAIp:=}{\cal A}^{(n)}_{... \, I_l p_l\, ... }(...;k_l, \theta^-_{l};...):=  {\cal A}^{(n)}_{[I_1J_1K_1] ... [I_{l-1}J_{l-1}K_{l-1}] \, I_l  p_l\, [I_{l+1}J_{l+1}K_{l+1}] ... }(...;k_l, \theta^-_{l};...)\; , \qquad
 \end{eqnarray} is fermionic, gamma traceless in $ K_l p_l$ indices,
 \begin{eqnarray}\label{gIqpAIp=0}
&& \gamma^{I_l}_{q_lp_l}  {\cal A}^{(n)}_{... \, I_l p_l\, ... }(...;k_l, \theta^-_{l};...) =0 \,  , \qquad
\end{eqnarray}
and is defined by Eq. (\ref{Df-AIJK}) itself.
The factor $(-)^{\Sigma_l} $ can be set to unity in the case under consideration, but we put it in the equation to make it applicable for the case of more general superamplitudes which we discuss below.

Using the suggestion from the on-shell superfield description of the linearized 11D supergravity in Eqs. (\ref{D+Phi=gPsi})--(\ref{D+h=11D}), we can also introduce a bosonic superfield amplitudes
${\cal A}_{.... [I_jJ_iK_i]... ((I_lJ_l))...}$,
in which, for a part or for all of the scattered particles, the multi-index $[I_lJ_lK_l]$, is replaced by a  pair of symmetrized traceless indices, $((I_lJ_l))$. For all $l$ with these new indices we impose on the superamplitude the equation
\begin{eqnarray}\label{Df-AIJ}
&&  D^{+(l)}_{{q}_l} {\cal A}^{(n)}_{... ((I_lJ_l))... }(...;k_l, \theta^-_{l};...) = (-)^{\Sigma_l}
 \, i \gamma_{(I_l| q_lp_l} \,  {\cal A}^{(n)}_{... \, |J_l) p_l\, ... }(...;k_l, \theta^-_{l};...) \,  , \qquad
\end{eqnarray}
with the same fermionic superamplitude obeying (\ref{gIqpAIp=0}). To be consistent with Eqs. (\ref{Df-AIJK}) and (\ref{Df-AIJ}), the fermionic superamplitudes should obey
\begin{eqnarray}\label{Df-AIq}
 &  (-)^{\Sigma_l} D^{+(l)}_{{q}_l}&  {\cal A}^{(n)}_{... \, I_l p_l\, ... }(...;k_l, \theta^-_{l};...)
 =    2i \rho^\#_{(l)} \gamma_{J_l\, qp}{\cal A}^{(n)}_{... ((I_lJ_l))... }(...;k_l, \theta^-_{l};...) + \qquad \nonumber \\ && {} \qquad  +\frac {i} {18}\rho^\#_{(l)}  \left(\gamma^{I_lJ_lK_lL_l}_{qp}+ 6 \delta ^{I_l[J_l}\gamma^{K_lL_l]}_{qp}  \right)
 {\cal A}^{(n)}_{... [J_lK_lL_l]... }(...;k_l, \theta^-_{l};...)\; . \qquad
\end{eqnarray}

Literally, till now we have discussed superamplitude  all the hidden indices of which,  denoted by ellipses,  are 'bosonic': either antisymmetric triple
 $[I_jJ_jK_j]$,  or  symmetric  and traceless pair $((I_jJ_j))$, in which  cases $\Sigma_l=0$. A more generic amplitude
\begin{eqnarray}\label{cAQ1-Qn=11D}
&& {\cal A}_{Q_1\ldots Q_n} ({k_1},\theta^-_{1};k_2 , \theta^-_{2};\ldots ;{k_n},\theta^-_{n})=:
{\cal A}_{\ldots Q_l \ldots } (\ldots ;k_l , \theta^-_{l};\ldots )
\; ,
\end{eqnarray}
can contain for each particle either the bosonic ($[I_jJ_jK_j]$ or $((I_jJ_j))\;$)  or a fermionic  multiindex (gamma-traceless pair of one SO(9) vector and one SO(9) spinor index, $I_jq_j$),
\begin{eqnarray}\label{Ql=11D}
D=11\, : \qquad Q_l= \left\{ [I_lJ_lK_l]\, , \, ((I_lJ_l))\, ; \, I_lq_l \right\} \; , \qquad I_l,J_l,K_l=1,...,9\, , \qquad q_l=1,...,16\; . \qquad
\end{eqnarray}
Such superfield amplitudes obey Eqs.  (\ref{Df-AIJK}), (\ref{Df-AIJ}) and (\ref{Df-AIq}) with generically nontrivial sign factor $(-)^{\Sigma_l}$:
$\Sigma_l$ can be chosen to be a number of $q_j$ indices in the set of indices corresponding to scattered particles with number $j$ in the interval between $1$ and $l$,
$1\leq j< l$ (see comment in footnote \ref{Sigma-l}),
\begin{eqnarray}\label{Sl=11D}
D=11\, : \qquad {\Sigma_l} =\sum\limits_{j=1}^{l-1} \frac {(1-(-)^{\varepsilon (^{Q_j})})}{2}  \, , \qquad
\varepsilon ( ^{[I_jJ_jK_j]})=0=\varepsilon ( \, ^{((I_jJ_j))}\, ), \qquad \varepsilon (^{I_jq_j})=1. \qquad
\end{eqnarray}

Sometimes it is instructive  to think about  Fourier images of the superamplitudes with respect to $\rho^{\#}_l$,
{\it i.e.} dependent on a coordinate $x_l^=$ instead of momentum (energy) $ \rho^{\#}_l$, {\it e.g.}
\begin{eqnarray}\label{Ax--=int11}
&& {\cal A}^{(n)}_{... I_l{q}_l...}(...;x_l^{=}, v^-_{pl}, \theta^-_{pl};...)
= \int d \rho^{\#}_{l} \; e^{ ix_l^= \rho^{\#}_l} \; {\cal A}^{(n)}_{... I_l {q}_l...}(...;\rho^{\#}_l, v^-_{pl}, \theta^-_{pl};...)\;  . \qquad
\end{eqnarray}
This superamplitude and its counterparts with a bosonic (multi)index corresponding to $l$-th scattered particle
obey the equations  (\ref{Df-AIJK}), (\ref{Df-AIJ}) and (\ref{Df-AIq}) in which
 $l$-th fermionic covariant derivative $D^{+(l)}_{{q}_l} $  has the form of Eq.
 (\ref{Df:=}).

As in sec. \ref{10D=WId}, we can formally discuss the action of supersymmetry transformations on the variables corresponding to $l$-th particle only (like (\ref{susy=8A+8A}) in the case of 10D SYM); the complete supersymmetry transformation will correspond to simultaneous action of all these $l$-supersymmeteries the parameters of which are expressed through the same constant fermionic spinor parameter as in  (\ref{el-q=}).
Let us  consider the following supersymmetric invariant, which can be treated as an integral over the on-shell superspace corresponding to $l$-th of scattered particles with an exotic  superspace  measure
({\it cf.} (\ref{Isusy=ssp11D}) and see comments and references below (\ref{susy=8+8=Issp})))
\begin{eqnarray}\label{dxDfADfA=11}
&& \frac 1 {32}  \int \, dx_l^{=} \, (-)^{\Sigma_l}\, D^{+(l)}_{{q}_l} \left({\cal A}^{(l)}_{... I_l{q}_l}(...;x_l^{=}, v^-_{{p}l}, \theta^-_{pl})\; \overleftrightarrow{D}{}^{+(l)}_{{q}_l} {\cal A}^{(n+2-l)}_{I_l{q}_l...}(x_l^{=}, v^-_{pl}, \theta^-_{pl};...)  \right)\vert_{\theta^-_{ql}=0} \,  . \qquad
\end{eqnarray}
Using Eqs.  (\ref{Df-AIJK}), (\ref{Df-AIJ}) and (\ref{Df-AIq}) we find the equivalent form of this integral:
\begin{eqnarray}\label{dxDfADfA=11=}
&& -i \int \, dx_l^{=} \,  ({\cal A}^{(l)}_{... I_l{q}_l}(...;x_l^{=}, v^-_{{p}l})\; \overleftrightarrow{\partial}_{=}  {\cal A}^{(n+2-l)}_{I_l{q}_l...}(x_l^{=}, v^-_{{p}l};...))\;+  \qquad \nonumber  \\
&&  + \int \, dx_l^{=} \, 4 \partial_={\cal A}^{(l)}_{... ((I_lJ_l))}(...;x_l^{=}, v^-_{{p}l})\; \partial_= {\cal A}^{(n+2-l)}_{((I_lJ_l)) ...}(x_l^{=}, v^-_{{p}l};...)\;+  \qquad
\nonumber  \\
&& + \int \, dx_l^{=} \, \frac 1 3  \partial_={\cal A}^{(l)}_{... [I_lJ_lK_l]}(...; x_l^{=}, v^-_{{p}l})\; \partial_= {\cal A}^{(n+2-l)}_{[I_lJ_lK_l] ...}(x_l^{=}, v^-_{{p}l};...)\;  , \qquad
\end{eqnarray}
 which includes the '$l$-leading components'   ($\theta^-_{p(l)}=0$ value) of the superamplitudes.
 For the Fourier images of the amplitude the integrand of the  above invariant reads
\begin{eqnarray}\label{rhoDfADfA=11}
&& -2  {\cal A}^{(l)}_{... I_l{q}_l}(...;\rho^{\#}_{l} , v^-_{{p}l})\; \rho^{\#}_{l} {\cal A}^{(n+2-l)}_{I_l{q}_l...}(-\rho^{\#}_{l}, v^-_{{p}l};...)\;+  \qquad \nonumber  \\
&&  + 4 {\cal A}^{(l)}_{... ((I_lJ_l))}(...;\rho^{\#}_{l}, v^-_{{p}l})\; (\rho^{\#}_{l})^2  {\cal A}^{(n+2-l)}_{((I_lJ_l)) ...}(-\rho^{\#}_{l}, v^-_{{p}l};...)\;+  \qquad
\nonumber  \\
&& + \, \frac 1 3  {\cal A}^{(l)}_{... [I_lJ_lK_l]}(...; \rho^{\#}_{l}, v^-_{{p}l})\;(\rho^{\#}_{l})^2\; {\cal A}^{(n+2-l)}_{[I_lJ_lK_l] ...}(-\rho^{\#}_{l}, v^-_{{p}l};...)\;  . \qquad
\end{eqnarray}
 It is not difficult to check that this  is invariant under the following 'component' form of the $l$-th supersymmetry which acts on the $l$-leading components of the superfield amplitudes:
 \begin{eqnarray}\label{susy-AIJK}
&& \delta_{\epsilon_{(l)}} {\cal A}^{(n')}_{... [I_lJ_lK_l]... }(...;k_l;...) = (-)^{\Sigma_l}
 \, 3i (\epsilon\, \gamma_{[J_lK_l|})_{p_l} \,  {\cal A}^{(n')}_{... \, |I_l] p_l\, ... }(...;k_l;...) \,  , \qquad \\ \label{susy-AIJ}
&&  \delta_{\epsilon_{(l)}}{\cal A}^{(n')}_{... ((I_lJ_l))... }(...;k_l;...) = (-)^{\Sigma_l}
 \, i(\epsilon\, \gamma_{(I_l|})_{ p_l} \,  {\cal A}^{(n')}_{... \, |J_l) p_l\, ... }(...;k_l;...) \,  , \qquad \\ \label{susy-AIq}
 && \delta_{\epsilon_{(l)}} {\cal A}^{(n')}_{... \, I_l p_l\, ... }(...;k_l;...)
 =    2i  (-)^{\Sigma_l}\rho^\#_{l} (\epsilon\, \gamma_{J_l})_{p_l}{\cal A}^{(n')}_{... ((I_lJ_l))... }(...;k_l;...) + \qquad \nonumber \\ && {} \qquad  +\frac {i} {18} (-)^{\Sigma_l}\rho^\#_{(l)}   \left(\epsilon\, \gamma^{I_lJ_lK_lL_l}+ 6 \delta ^{I_l[J_l}\epsilon\gamma^{K_lL_l]} \right){}_{p_l} \,
 {\cal A}^{(n')}_{... [J_lK_lL_l]... }(...;k_l;...)\; . \qquad
\end{eqnarray}

The counterpart of (\ref{dxDfADfA=11=})  for the $\rho^\#_l$ dependent superamplitudes reads
\begin{eqnarray}\label{Df-ADfA=11D}
&& \fbox{$ \;(-)^{\Sigma_l}\,  D^{+(l)}_{{q}_l} \left({\cal A}^{(l)}_{... J_l{p}_l}(\ldots; k_l, \theta^-_{l})\; \overleftrightarrow{D}{}^{+(l)}_{{q}_l} {\cal A}^{(n+2-l)}_{J_lp_l...}(-k_l, \theta^-_{l}; ...)  \right) \;$}\vert_{\theta^-_{l}=0}\,  , \qquad
\end{eqnarray}
where $k_l$ is expressed by (\ref{k=pv-v-11}) and
$D^{+(l)}_{{q}_l}$ has the form of (\ref{D+l=d+rho-th}). This invariant of $l$-th (projection of the) supersymmetry is constructed from  two superfield amplitudes including the same variables of one  'internal'  particle as the last ($l$-th) arguments of the first and first arguments of the last superamplitude.  The data (variables and indices) of other $l-1$ particles in the first amplitude are generically different from the data of other $(n+1-l)$ particles in the second amplitude.  The indices describing 'helicity' of $l$-th particle are fermionic and contracted, $l$-th fermionic variables coincide and $l$-th momentum differs by a sign in two amplitudes.
Speaking in terms of energies and spinor moving frame variables, true arguments of the amplitudes, both amplitudes depend on the same $v_{\alpha q(l)}^{\; -}$ but on $+\rho^{\#}_l$ and $-\rho^{\#}_l$ respectively.
Notice that this implies that in the expression (\ref{D+l=d+rho-th}) for $D^{+(l)}_{{q}_l}$ acting on the amplitude with momentum $-k_l$ the sign in front of the second term  should be also inverted.


The above invariants, as well as their straightforward generalizations including the momentum-dependent multipliers
  will be useful to  describe a candidate for  generalization of the BCFW recurrent relations \cite{Britto:2005fq} for 11D tree on-shell  superamplitudes. But before turning to this, we should understand the 11D generalization of the BCFW--deformation of the helicity spinors  and of the fermionic variables.

\subsection{ Generalization of the BCFW deformations for 10D and 11D superamplitudes }

\subsubsection{Bosonic part of D=10 BCFW deformation and its 11D generalization.  }

The  D=10 counterparts of the {BCFW deformation} (\ref{BCFWln=4D}), (\ref{BCFWl1=4D}) of the helicity spinors of the massless particles  were presented  in \cite{CaronHuot:2010rj}. Our spinor moving frame form of the spinor helicity formalism will allow us to simplify them and also to obtain the generalization of the BCFW deformation to the 11D case.

As D=10 spinor helicity variables from \cite{CaronHuot:2010rj} are identified with spinor frame variables by (\ref{spol=l=}), the  {\it spinor frame  counterparts of the BCFW deformation} from  \cite{CaronHuot:2010rj} has the form
\begin{eqnarray}\label{BCWF=vnM}
\widehat{v_{\alpha {q}n}^{\; -}}= v_{\alpha {q}n}^{\; -} + z \;
\sqrt{\frac {\rho^{\#}_{1}}{\rho^{\#}_{n}}} v_{\alpha {p}1}^{\; -} \; {\bb M}_{ {p}{q}}
\; , \qquad \\ \label{BCWF=v1M} \widehat{v_{\alpha {q}1}^{\; -}}= v_{\alpha {q}1}^{\; -} - z \;
 \sqrt{\frac {\rho^{\#}_{n}}{\rho^{\#}_{1}}}\; {\bb M}_{{q}{p}}\;  v_{\alpha {p}n}^{\; -}
\; , \qquad \end{eqnarray}
with $\alpha=1,..,16$, ${}\quad q,p=1,...,8$ and an arbitrary $z\in {\bb C}$.
The same equation {\it describes also the 11D version of the  BCFW deformation}  provided
$\alpha$ and $q,p$ are considered as $SO(1,10)$ and $SO(9)$ spinor indices respectively,
$\alpha=1,..., 32$ and $q,p=1,...,16$.

The shift of the spinor frame variables, (\ref{BCWF=vnM}) and (\ref{BCWF=v1M})
 results in shifting the momentum of the first and of the $n$--th particle on a complex light-like vector $q^a$
 orthogonal to both $ k_{(1)}^{a}$ and $ k_{(n)}^{a}$,
\begin{eqnarray}\label{BCWF=hp}
\widehat{k_{1}^{a}}= k_{1}^{a} -z q^a \; , \qquad \widehat{k_{n}^{a}}= k_{n}^{a} +z q^a \; , \qquad
\end{eqnarray}
\begin{eqnarray}\label{qq=0=}
q_aq^a=0\; , \qquad q_ak_{1}^{a} =0\; , \qquad q_ak_{n}^{a}= 0\; , \qquad
\end{eqnarray}
provided we choose
\begin{eqnarray}\label{Mpq=}
{\bb M}_{{q}{p}} = - \frac {1} { {\sqrt{\rho^{\#}_{1}\rho^{\#}_{n} }} (u_{1}^=u^{=}_{n}) }\; {(v_{{q}1}^{\; -} \, \tilde{q}\!\!\!/{} v_{{p}n}^{\; -})}  \; . \qquad
\end{eqnarray}
Here and below
\begin{eqnarray}
\label{tq=qts}
\tilde{q}\!\!\!/{}^{ \alpha\beta}:= q^a
\tilde{\Gamma}_a^{ \alpha\beta} \; , \qquad {q}\!\!\!/{}_{ \alpha\beta}:= q_a
{\Gamma}^a_{ \alpha\beta} \; . \qquad
\end{eqnarray}
It is easy to check that the light-likeness of $q^a$, (\ref{qq=0=}), implies the nilpotency of the
matrix  ${\bb M} $ (\ref{Mpq=}): $\;  {\bb M}^T {\bb M}=0={\bb M} {\bb M}^T $ or more explicitly
\begin{eqnarray}\label{MMT=0}
{\bb M}_{rp} {\bb M}_{r{q}} =0\;  ,\qquad {\bb M}_{{q}r } {\bb M}_{{p}r} =0\;  .\qquad
\end{eqnarray}
We can also write the expression for light-like complex vector in terms of deformation matrix,
\begin{eqnarray}\label{q:=M}
D=10: \qquad q^a= \frac 1 {4}\,
{\sqrt{\rho^{\#}_1\rho^{\#}_n}} \;  v_{{q}1}^{\; -}\tilde{\sigma}^a{\bb M}_{{q}{p}}v_{{p}n}^{\; -}
\; , \qquad  \\ \label{q:=M11}
D=11: \qquad q^a= \frac 1 {8}\,
{\sqrt{\rho^{\#}_1\rho^{\#}_n}} \;  v_{{q}1}^{\; -}\tilde{\Gamma}^a{\bb M}_{{q}{p}}v_{{p}n}^{\; -}
\; . \qquad
\end{eqnarray}

The nilpotency condition (\ref{MMT=0}) guarantees that the shifted spinor  frame variables obeys the characteristic constraints, Eqs. (\ref{k=pv-v-}), (\ref{k=pv-v-11}) with shifted light-like momenta $k_{(1)}$ and $k_{(n)}$, (\ref{BCWF=hp}), or equivalently, (\ref{u==v-v-}) with shifted light-like $u^{=a}_{(1)}$ and
$u^{=a}_{(n)}$,
\begin{eqnarray}\label{BCWF=hu--}
\widehat{u_{1}^{=a}}= u_{1}^{=a} -\frac {z q^a} {\rho^\#_{1}} \; , \qquad \widehat{u_{n}^{=a}}= u_{n}^{=a} +\frac {z q^a} {\rho^\#_{n}} \; . \qquad
\end{eqnarray}

Notice that (\ref{BCWF=vnM}) and (\ref{BCWF=v1M}) imply
\begin{eqnarray}\label{hp1+hpn=}
\widehat{k^a_1}+\widehat{k^a_n}= k^a_1+k^a_n\;
\qquad
\end{eqnarray}
so that
\begin{eqnarray}\label{hPij=}
\widehat{P}_{ij}(z)=\hat{k_i}+...+\hat{k_j}\;
\qquad
\end{eqnarray}
is independent of $z$ if both $1$ and $n$ or neither of them lay between $i$ and $j$. With our specific choice of two special points to be (1) and (n), this  means just that $\widehat{P}_{ij}(z)$ is independent on $z$ if    $1<i <j<n$, but keeps $z$ dependence if e.g.  $1<j<i<n$ .

\bigskip

\subsubsection{Fermionic  part of the D=10 and D=11 BCFW-type deformations}

To wite a candidate for the 10D and 11D generalizations of the BCFW-type recurrent relations \cite{Britto:2005fq,ArkaniHamed:2008gz}  for our constrained superamplitudes, we need to understand how the BCFW deformations, acting on the 10D and 11D momenta as (\ref{BCWF=hp}), and on the spinor helicity (spinor moving frame) variables as in  (\ref{BCWF=vnM}) and (\ref{BCWF=v1M}), act of the fermionic variables $\theta^-_{{p}(i)}$. The study of Clifford superfield formalism in \cite{CaronHuot:2010rj} makes clear  that such an action should be nontrivial and actually suggests its possible form:
\begin{eqnarray}\label{BCFW=thn}
\widehat{ \theta^-_{{p}(n)}}= \theta^-_{{p}(n)}+ z \,\theta^-_{{q}(1)} \, {\bb M}_{{q}{p}} \, \sqrt{\frac {\rho^{\#}_{(1)}}{\rho^{\#}_{(n)}}} \; , \qquad \\ \label{BCFW=th1}
\widehat{ \theta^-_{{q}(1)}}= \theta^-_{{q}(1)}- z \, \sqrt{\frac {\rho^{\#}_{(n)}}{\rho^{\#}_{(1)}}}  \, {\bb M}_{{q}{p}} \, \theta^-_{{p}(n)}\;  \qquad \end{eqnarray}
(remember that $q,p=1,...,8$ for D=10 while $q,p=1,...,16$ for D=11).

This implies the following transformations of the covariant derivatives (\ref{D+l=d+rho-th}):
\begin{eqnarray}\label{BCFW=D+n}
\widehat{ D^+_{{p}(n)}}= D^+_{{p}(n)}+ z \, D^+_{{q}(1)} \, {\bb M}_{{q}{p}} \, \sqrt{\frac {\rho^{\#}_{(n)}}{\rho^{\#}_{(1)}}} \; , \qquad \\ \label{BCFW=D+1}
\widehat{ D^+_{{q}(1)}}= D^+_{{q}(1)}- z \, \sqrt{\frac {\rho^{\#}_{(1)}}{\rho^{\#}_{(n)}}}  \, {\bb M}_{{q}{p}} \, D^+_{{p}(n)}\; . \qquad \end{eqnarray}
Due to the nilpotency of the matrix ${\bb M}$, (\ref{MMT=0}), the shifted derivatives also obey the algebra (\ref{D+D+=rho}),
\begin{eqnarray}\label{hD+hD+=rho}
{}\{ \widehat{D}{}^{+(l)}_{{q}_l}, \widehat{D}{}^{+(j)}_{{p}_{j}} \} = - 2 \delta_{l\, j}\delta_{{q}_l{p}_{j}}  \, \rho^{\#}_{(l)}\; .
\end{eqnarray}

It is convenient to introduce a schematic notation
\begin{eqnarray}\label{DnMth1=}
 & D_{(1)}{\bb M}\theta_{(n)}:=  \,  \sqrt{\frac {\rho^{\#}_{(n)}}{\rho^{\#}_{(1)}}}\,  D^+_{{q}(1)} \, {\bb M}_{{q}{p}}\theta^-_{{q}(n)}  \; , \qquad
 \theta_{(1)}{\bb M}D_{(n)}:= \, \sqrt{\frac {\rho^{\#}_{(1)}}{\rho^{\#}_{(n)}}}\, \theta^-_{{q}(1)} {\bb M}_{{q}{p}} \, D^+_{{p}(n)} \; .
\end{eqnarray}
Then one can equivalently write the definition of the fermionic deformation, (\ref{BCFW=thn}), (\ref{BCFW=th1}), and
its consequences,  (\ref{BCFW=D+n}) and (\ref{BCFW=D+1}),  in a more universal form
\begin{eqnarray}\label{BCFW=thl}
\widehat{ \theta^-_{{p}(l)}} &=& e^{-zD_{(1)}{\bb M}\theta_{(n)} -z \theta_{(1)}{\bb M}D_{(n)} }\; \theta^-_{{p}(l)}\; , \qquad  \\
\label{BCFW=D+l}
\widehat{ D^+_{{p}(l)}} &=& e^{-zD_{(1)}{\bb M}\theta_{(n)} -z \theta_{(1)}{\bb M}D_{(n)} }\; D^+_{{p}(l)} e^{zD_{(1)}{\bb M}\theta_{(n)} +z \theta_{(1)}{\bb M}D_{(n)} }\; . \qquad \end{eqnarray}

The BCFW deformation
of our  10D and 11D  superamplitudes
 $${\cal A}_{....}(k_1,\theta_1 \ldots, k_n,\theta_1 )=  {\cal A}^{(n)}_{....}(\rho^{\#}_{(1)}, v^-_{{q}(1)},\theta^-_{{q}(1)}; \ldots \rho^{\#}_{(n)}, v^-_{{q}(n)},\theta^-_{{q}(n)} )\equiv {\cal A}^{(n)}_{... }(\{\rho^{\#}_{(i)}\}; \{v^-_{{q}(i)}\} ; \{\theta^-_{{q}(i)}\}  ) $$
 can be described by the equation
\begin{eqnarray}\label{cA-Sf=BCFW}
{\cal A}_{z ....}(\ldots, \widehat{k_{(i)}}, \widehat{\theta_{(i)}}  ... )  = e^{zD_{(1)}{\bb M}\theta_{(n)} +z \theta_{(1)}{\bb M}D_{(n)} }\;
 {\cal A}_{....}(\widehat{k_1}(z),k_2,\ldots, k_{(n-1)},\widehat{k_n}(z))
\; \; .
\qquad
\end{eqnarray}

\subsection{Candidate BCFW recurrent relations for 10D SYM superamplitudes}

The BCFW shifted D=10 superamplitude (\ref{cA-Sf=BCFW}) obeys the superfield equations (\ref{D+Adq=pmGA}) with BCFW shifted fermionic covariant derivatives,
\begin{eqnarray}\label{hD+hA=GA}
\widehat{D^{+(l)}_{{q}_l}}{{\cal A}}_z^{(n)}{}_{_{... {\dot{q}_l} ...}}( ...; \widehat{k_l}, \widehat{\theta^-_{l}}; ...)
= (-)^{\Sigma_l}\gamma^{I_l}_{q_l\dot{q}_l} {{\cal A}}_z^{(n)}{}_{_{... I_l...}}( ...; \widehat{k_l}, \widehat{\theta^-_{l}}; ...)\; ,
\end{eqnarray}
or, more schematically,
\begin{eqnarray}\label{hD+hA=GAs}
\widehat{D^{+(l)}_{{q}_l}}\widehat{{\cal A}}_z^{(n)}{}_{_{... {\dot{q}_l} ...}}
= (-)^{\Sigma_l}\gamma^{I_l}_{q_l\dot{q}_l} \widehat{{\cal A}}_z^{(n)}{}_{_{... I_l...}}\; ,
\end{eqnarray}
where
$\widehat{{\cal A}}_{\;z}^{(n)}{}_{_{... I_l...}}:= {{\cal A}}_{\;z}^{(n)}{}_{_{... I_l...}}( ...; \widehat{k_l}, \widehat{\theta^-_{l}}; ...)$ is related to
${{\cal A}}^{(n)}{}_{_{... I_l...}}( ...;  \widehat{k_l}(z), \theta^-_{l}; ...)$ by Eq. (\ref{cA-Sf=BCFW}).

 A natural candidate for
 generalization of the BCFW recursion relations \cite{Britto:2005fq} for the case of 10D tree superfield amplitudes is
\begin{eqnarray}\label{cA-Sf=rBCFW}
&{\cal A}_{Q_1\ldots Q_n}& ({k_1},\theta^-_{(1)};k_2 , \theta^-_{(2)};\ldots ;{k_n},\theta^-_{(n)})= \qquad\nonumber \\ & = &  \sum\limits_{l=2}^n  \frac{(-)^{\Sigma_{(l+1)}}}{32 \widehat{\rho}{}^{\#}(z_l)} D^+_{{q}(z_l)}\left(
{\cal A}_{z_l \; Q_1\ldots Q_l\,  \dot{q}}(\widehat{k_1}, \widehat{\theta^-_{(1)}};k_2, \theta^-_{(2)}; \ldots ;{k_l}, \theta^-_{(l)};\widehat{{P_l}}(z_l),\Theta^-) \right. \times
\qquad \\ \nonumber
&&  \left. \;
\times \frac {1}{({P_l})^2} \overleftrightarrow{D}{}
^+_{{q}(z_l)} {\cal A}_{z_l\;  \dot{q}\,  Q_{l+1}\ldots Q_n}(-\widehat{{P_l}}(z_l),\Theta^-; k_{l+1}, \theta^-_{(l+1)};\ldots ; k_{n-1}, \theta^-_{(n-1)}; \widehat{k_n}, \widehat{\theta^-_{(n)}})\right)\vert_{\Theta^-=0} .
\end{eqnarray}
Let us write the relation (\ref{cA-Sf=rBCFW}) a bit more explicitly in terms of original amplitudes,
\begin{eqnarray}\label{cA-Sf=rrBCFW}
&&{\cal A}_{Q_1\ldots Q_n} ({k_1},\theta^-_{1};k_2 , \theta^-_{2};\ldots ;{k_n},\theta^-_{n}) =   \sum\limits_{l=2}^n  \frac{(-)^{\Sigma_{(l+1)}}}{32 \widehat{\rho}{}^{\#}(z_l)} \times \qquad\nonumber
\\ && {}\qquad  \times D^+_{{q}(z_l)}\left( e^{z_lD_{(1)}{\bb M}\theta_{(n)} +z_l \theta_{(1)}{\bb M}D_{(n)} }\;
{\cal A}_{ Q_1\ldots Q_l \, \dot{q}}(\widehat{k_1}(z_l),\theta^-_{1};k_2,\theta_2; \ldots ;{k_l},\theta_l;\widehat{P_{l}}(z_l),\Theta^-) \right. \times
\nonumber \\
&&  \left. \qquad {}\qquad
\times \frac {1}{(P_{l})^2} \overleftrightarrow{D}{}^+_{{q}(z_l)} {\cal A}_{  \dot{q}\, Q_{l+1}\ldots Q_n}(-\widehat{P_{l}}(z_l),\Theta^-; k_{l+1},\theta_{l+1};\ldots ;\widehat{k_n}(z_l),\theta_l)\right)\vert_{\Theta^-=0}
\; .
\end{eqnarray}
Here
\begin{eqnarray}\label{kSl=}
{P_{l}^a}=- \sum\limits_{m=1}^l  {k_m^a}\; , \qquad
\end{eqnarray}
the deformation parameter is chosen to be
\begin{eqnarray} \label{zl:=}
 z_l:= \frac {P_{l}^a P_{l \,a}} {2P_{l}^b q_b}
\end{eqnarray}
and
\begin{eqnarray}\label{hkSl==}
\widehat{P_{l}^a}(z)=- \sum\limits_{m=1}^l  \widehat{k_m^a}(z)= {P_{l}^a} - 2z \sqrt{\rho^{\#}_1\rho^{\#}_n}  \; v_{q1}^{-}\tilde{\sigma^a}{\bb M}_{{q}{p}}v_{{p}n}^{-}
=: {P_{l}^a} - z  q^a \;
\end{eqnarray}
with $ v_{{q}1}^{-}\tilde{\sigma^a}v_{{p}n}^{-}:= v_{\alpha {q}}{}^{\! -}_{(1)}\tilde{\sigma^a}{}^{\alpha\beta}v_{\beta{p}}{}^{\! -}_{(n)}$ and $q_a$ defined by
Eq. (\ref{q:=M}). Eq. (\ref{hkSl==}) implies that
$(\widehat{P_{l}}(z))^2= (P_{l})^2 -2z P_{l}\cdot q$, so that at $z=z_l$ (\ref{zl:=}) $\widehat{P_{l}^a}$ becomes light--like
\begin{eqnarray}\label{hkSl2zl=0}
(\widehat{P_{l}}(z_l))^2=0\; , \qquad z_l:= \frac {(P_{l})^2} {2P_{l}\cdot q}\; . \;
\end{eqnarray}
As a result, both amplitudes in the {\it r.h.s.} of (\ref{cA-Sf=rBCFW}) are on the mass shell. To express them in terms of spinor helicity formalism  we introduce spinor frame variables $\widehat{v_{\alpha {q}}^{\! -}}(z_l)$ related to the light-like
$\widehat{P_{l}}{}^a(z_l)$ by the counterpart of  Eq. (\ref{k=pv-v-}),
\begin{eqnarray}\label{hkSlzl=11}
\widehat{P_{l}{}^a}(z_l)\sigma_{a\alpha\beta} = 2 \widehat{\rho}{}^{\#}(z_l)\,  \widehat{v_{\alpha {q}}^{\; -} } (z_l) \widehat{v_{\beta {q}}^{\; -}} (z_l) \; , \qquad \widehat{\rho}{}^{\#}(z_l) \widehat{v^-_{{q}}} (z_l) \tilde{\sigma}{}^{a}\widehat{v^-_{{p}}}(z_l)= \widehat{P_{l}}{}^a(z_l) \delta_{{q}{p}}\; . \qquad
\end{eqnarray}
Notice that $\widehat{\rho}{}^{\#}(z_l)$ defined in this equations, enters the denominators of the terms in the r.h.s.
of (\ref{cA-Sf=rBCFW}), in the combination $\frac{(-)^{\Sigma_{(l+1)}}}{32 \widehat{\rho}{}^{\#}(z_l)}$, where coefficient $1/32$ is  needed to reproduce the (presumably) correct
purely bosonic limit (the counterpart of the relation described for all D in the introduction of \cite{ArkaniHamed:2008gz}).

Finally, $D^+_{{q}(z_l)}$ in (\ref{cA-Sf=rBCFW}) is the covariant derivative with respect to
$\theta^-_{\dot{q}}$ (in the last argument of first multiplier and first argument of the last multiplier in the {\it r.h.s.} of this equation) constructed with the use of $\widehat{\rho}{}^{\#}(z_l)$,
\begin{eqnarray}\label{D+=Sl}
D^+_{{q}(z_l)}= \frac {\partial }  {\partial \theta^-_{{q}}} - 2\widehat{\rho}{}^{\#}(z_l)  \theta^-_{\dot{q}}
\; . \qquad
\end{eqnarray}

\subsection{On validity of higher dimensional BCFW relations}

The derivation of the BCFW recurrent relations \cite{Britto:2005fq} and their generalizations \cite{ArkaniHamed:2008yf,Cheung:2008dn,Cheung:2009dc,ArkaniHamed:2008gz} uses essentially the fact that the deformed amplitude as a function of $z$ does not have a pole at $z=\infty$.

We did not study whether this is the case for our constrained superamplitudes and amplitudes. However,
generic D-dimensional results of \cite{ArkaniHamed:2008yf,Cheung:2008dn,Cheung:2009dc,ArkaniHamed:2008gz}
suggest to hope that  the above recurrent relations for tree 10D SYM superamplitudes (and the below relations for 11D SUGRA) are  valid at least
\begin{itemize}
\item
in the case if the amplitude in the l.h.s. has at least one vector particle (at least one graviton), which is taken to be the first one, with deformed momentum,
\item and after we contract the first vector index $I_1$ (first multi-index $(I_1J_1))$) with the deformation vector $q_{(1)}^I$ (\ref{q=uiqi1}) (with direct product of deformation vectors $q_{(1)}^I q_{(1)}^J$).
\end{itemize}
This latter implies that the deformation 10-vector $q^a= u^{aI}_1 q_{(1)}^I$ (direct product of two copies of  its 11D counterpart) is considered to be a polarization vector (tensor) of the scattered vector particle (graviton). This can be always achieved as $q^a$ has the properties characteristic for  a polarization vector. However, in contradistinction to 4D case, in higher $D$ there exists an essential freedom in choosing such a deformation vector.

Below, in sec. 8, we will actually observe  a problem related to an essential dependence of the amplitude calculated with BCFW prescription  on the deformation vector, and conclude that it might reflect either inconsistency/need for modification of our candidate recurrent relations or
incompleteness of our prescription for BCFW deformation. We suggest that this latter might be improved on the line of interplay of our present approach and the analytic superamplitude formalism of \cite{Bandos:2017zap}.

Keeping all these in mind, we nevertheless have written the  most generic form of the candidate BCFW relations for 10D SYM (and, below, for 11D SUGRA), with the aim to study their properties and to gain suggestions for further development of the formalism. In particular, below   we use them to obtain  candidate 4 particle 10D SYM amplitudes with  4 and with 2 fermionic legs which allow us to observe the above mentioned problem of deformation vector dependence.

\subsection{Candidate BCFW recurrent relation for the  11D supergravity amplitudes}

The candidate BCFW-type recurrent relation for tree superfield amplitudes of 11D supergravity reads
\begin{eqnarray}\label{cA-Sf=rBCFW11D}
&& {\cal A}_{Q_1\ldots Q_n} ({k_1},\theta^-_{(1)};k_2 , \theta^-_{(2)};\ldots ;{k_n},\theta^-_{(n)})= \hspace{7cm}\nonumber \\ & &\quad =  \sum\limits_{l=2}^n \frac{(-)^{\Sigma_{(l+1)}}}{64 (\widehat{\rho}{}^{\#}(z_l))^2}\left(D^+_{{q}(z_l)}\left(
{\cal A}_{z_l \; Q_1\ldots Q_l \; Jp}(\widehat{k_1}, \widehat{\theta^-_{(1)}};k_2, \theta^-_{(2)}; \ldots ;{k_l}, \theta^-_{(l)};\widehat{P_{l}}(z_l),\theta^-) \right. \right. \times
\qquad \\ \nonumber
&&  \left. \left. \qquad
\times \frac {1}{(P_{l})^2} \overleftrightarrow{D}{}^+_{{q}(z_l)} {\cal A}_{z_l\; Jp\;  Q_{l+1}\ldots Q_n}(-\widehat{P_{l}}(z_l),\theta^-; k_{l+1}, \theta^-_{(l+1)};\ldots ; k_{n-1}, \theta^-_{(n-1)}; \widehat{k_n}, \widehat{\theta^-_{(n)}})\right)\right)_{\theta^-_q=0}
\; .
\end{eqnarray}
 This looks very much the same as  $D=10$ candidate super-BCFW relation (\ref{cA-Sf=rBCFW}), up to that the superindices take now different values (\ref{Ql=11D}),
\begin{eqnarray}\label{Ql==11D}
D=11\, : \qquad Q_l= \left\{ [I_lJ_lK_l]\, , \, ((I_lJ_l))\, ; \, I_lq_l \right\} \; , \qquad I_l,J_l,K_l=1,...,9\, , \qquad q_l=1,...,16\; ,\qquad
\end{eqnarray}
so that, in particular, the fermionic multindex of the deformed amplitudes which we sum on is the set of an SO(9) vector and SO(9) spinor indices, $Iq$ (instead of 8c index $\dot{q}=1,..,8$ in the 10D case).

For the convenience of a reader interested in 11D case only,  let us explain the notation, although this is almost identical to the one in 10D case, Eqs. (\ref{kSl=}) --(\ref{D+=Sl}). The reader who feels sufficient the reference to notation desceibed above for 10D case might pass directly to the next Sec. \ref{Towards}.

In (\ref{cA-Sf=rBCFW11D})
\begin{eqnarray}\label{kSl==}
{P_{l}^a}=- \sum\limits_{m=1}^l  {k_m^a}\;  \qquad
\end{eqnarray}
is the (minus) sum of momenta of 'first' $l$ of scattered (super)particles
in the first superamplitude,
\begin{eqnarray}
 \label{hkSl=}
 \widehat{P_{l}^a}(z)=- \sum\limits_{m=1}^l  \widehat{k_m^a}(z)
= {P_{l}^a} - z  q^a\;  \qquad
\end{eqnarray}
is the (minus) sum of 'first' $l$ deformed momenta,
$q^a$
is the deformation vector (\ref{q:=M11}) obeying   (\ref{qq=0=}) as far as the matrix ${\bb M}_{{q}{p}}$
is nilpotent, (\ref{MMT=0}) (see also (\ref{Mpq=})).
Finally
\begin{eqnarray}
\label{zl:==}
z_l:= {P_{l}^a P_{l \,a}}/ ({2P_{l}^b q_b})\; , \qquad
\end{eqnarray}
is the value of the deformation parameter $z$ at which $\widehat{P_{l}}{}^a(z)$ becomes light-like,
\begin{eqnarray}\label{Plzl2=0}
(\widehat{P_{l}}(z_l))^2=0\; , \qquad z_l:= {(P_{l})^2}/ ({2P_{l}\cdot q})\; . \;
\end{eqnarray}
Thus both amplitudes in the {\it r.h.s.} of (\ref{cA-Sf=rBCFW11D}) are on the mass shell.

Furthermore, the bosonic arguments of the on-shell amplitudes are energies $\rho^\#_{(i)}$ and
spinor harmonics $v_{\alpha q(i)}^{\; -}$
 related to light-like momenta $k_{(i)}^a$  by (\ref{k=pv-v-11}); just for shortness in (\ref{cA-Sf=rBCFW11D}), we hide this writing instead the  dependence on the momenta. In particular, the last bosonic argument of the first amplitudes and the first argument of the second amplitude in the {\it r.h.s.} of (\ref{cA-Sf=rBCFW11D}) are actually  pairs of energy $\pm\widehat{\rho^{\#}}(z_l)$  and spinor frame variables $v_{\alpha {q}}^{\; -}(z_l)$ related to
$\widehat{P_{l}}{}^a(z_l)$ of (\ref{hkSl=}) by Eqs. (\ref{k=pv-v-11}),
\begin{eqnarray}\label{hkSlzl=}
\widehat{P_{l}{}^a}(z_l)\Gamma_{a\alpha\beta} = 2 \widehat{\rho^{\#}}(z_l)\,  v_{\alpha {q}}{}^{\!\! -}(z_l) v_{\beta {q}}{}^{\!\! -}(z_l) \; , \qquad \nonumber \\
\widehat{P_{l}}{}^a(z_l)\delta_{qp}  = \widehat{\rho^{\#}}(z_l)\,  v^{-}_{{q}}(z_l)\tilde{\Gamma}^{a} v^{-}_{p}(z_l) \; . \qquad
\end{eqnarray}
$\widehat{\rho}{}^{\#}(z_l)$ defined in this equations enters the denominators of the terms in the r.h.s.
of (\ref{cA-Sf=rBCFW11D}) with coefficient $64$. This is needed to reproduce a (presumably) correct
purely bosonic limit of the candidate  BCFW relations  in the assumption of simple relation between amplitudes and superamplitudes.


Finally, $D^+_{{q}(z_l)}$ in (\ref{cA-Sf=rBCFW11D}) is the covariant derivative with respect to $\theta^-_{{q}}$ (in the last argument of first multiplier and first argument of the last multiplier in the {\it r.h.s.} of (\ref{cA-Sf=rBCFW11D})) constructed with the use of $\widehat{\rho^{\#}}(z_l)$ of (\ref{hkSlzl=}),
\begin{eqnarray}\label{D+=Slrho}
D^+_{{q}(z_l)}= \frac {\partial }  {\partial \theta^-_{{q}}} - 2\widehat{\rho^{\#}}(z_l)  \theta^-_{{q}}
\; . \qquad
\end{eqnarray}

Notice that the structure of the terms in r.h.s. of  (\ref{cA-Sf=rBCFW11D}),  $D^+_{{q}(z_l)}\left(
{\cal A}_{z_l \;\ldots  Jp}\overleftrightarrow{D}{}^+_{{q}(z_l)} {\cal A}_{z_l\; Jp\;  \ldots }\right)$,
 can be treated as an integration over the fermionic  variable $\theta^-_q$ in (\ref{D+=Slrho}) with an exotic measure similar to one used in \cite{Tonin:1991ii,Tonin:1991ia} to construct a worldsheet superfield formulation of the heterotic string (see  \cite{Zupnik:1989bw} for formal discussion of similar exotic measures).

\section{Towards calculation of  11D and 10D amplitudes}
\label{Towards}
In this section we elaborate a bit more  the spinor frame form of the spinor helicity formalism and present some details which will be useful for amplitude and superamplitude calculations. A simplest application will be described in the next Sec. \ref{4-fermions} for the case of D=10 SYM (as we have already noticed, these will indicate some problems of our candidate BCFW recurrent relations). In contrast, the equations of this section are written explicitly for D=11 case, in which $q, p=1,.., 16$, $I=1,...,9$.
The discrepancy with D=10 case, in which $I=1,...,8$, ${}\quad q,p=1,...,8$ and $\dot{q},\dot{p}=1,...,8$ replace $q$ and $p$ in certain expressions,  is described explicitly when it is not evident.

\subsection{Candidate BCFW-type relation for 4-point superamplitudes in D=11}

To gain a feeling of the structure of the candidate recurrent relations (\ref{cA-Sf=rBCFW11D}), let us first write it's version for 4-particle (4-supergraviton) tree superamplitude in D=11, selecting 1-st and 4-th particle variables to be deformed,
\begin{eqnarray}\label{cA4=11D}
&& {\cal A}_{Q_1Q_2Q_3Q_4} ({k_1},\theta^-_{(1)};k_2 , \theta^-_{(2)};k_3 , \theta^-_{(3)};{k_4},\theta^-_{(4)})= \hspace{7cm}\nonumber \\ & &\quad =  \frac{(-)^{Q_1+Q_2}}{64 (\widehat{\rho}{}^{\#}(z_{12}))^2} \left(D^+_{{q}(z_{12})}\left(
{\cal A}_{z_{12} \; Q_1Q_2 \; Jp}(\widehat{k_1}, \widehat{\theta^-_{(1)}};k_2, \theta^-_{(2)};\widehat{P_{12}}(z_{12}),\theta^-) \right. \right. \times
\qquad  \nonumber \\ \nonumber
&&  \left. \left. \qquad
\times \frac {1}{(P_{12})^2} \overleftrightarrow{D}{}^+_{{q}(z_{12})} {\cal A}_{z_{12}\; Jp\;  Q_{3} Q_4}(-\widehat{P_{12}}(z_{12}),\theta^-; k_{3}, \theta^-_{(3)}; \widehat{k_4}, \widehat{\theta^-_{(4)}})\right)\right)_{\theta^-_q=0}
+   \nonumber \\ & &\quad +  \frac{(-)^{Q_1+Q_3+ Q_2Q_3}}{64 (\widehat{\rho}{}^{\#}(z_{13}))^2} \left(D^+_{{q}(z_{13})}\left(
{\cal A}_{z_{12} \; Q_1Q_3 \; Jp}(\widehat{k_1}, \widehat{\theta^-_{(1)}};k_3, \theta^-_{(3)};\widehat{P_{13}}(z_{13}),\theta^-) \right. \right. \times
\qquad \nonumber  \\
&&  \left. \left. \qquad
\times \frac {1}{(P_{13})^2} \overleftrightarrow{D}{}^+_{{q}(z_{13})} {\cal A}_{z_{13}\; Jp\;  Q_{2} Q_4}(-\widehat{P_{13}}(z_{13}),\theta^-; k_{2}, \theta^-_{(2)}; \widehat{k_4}, \widehat{\theta^-_{(4)}})\right)\right)_{\theta^-_q=0}
\; .
\end{eqnarray}
Here the multiindices $Q_i$ take values (\ref{Ql=11D}), and the remaining notations are explained in the previous section.

The recurrent relation for the 4-point superamplitude for $D=10$ SYM has the same form as (\ref{cA4=11D}) but with $1/64$ multiplies replaced by
$1/32$, $q=1,..., 8$ and the multiindices taking values as in (\ref{Ql=10D}).

A more explicit form of the relation for 11D superamplitude can be obtained by calculating the action of $D^+_{{q}(z_{12})}$ derivatives and using the superfield equations   (\ref{Df-AIJK}),  (\ref{Df-AIJ}) and  (\ref{Df-AIq}).
For instance for the superamplitude with 4 graviton multi-indices we obtain
 \begin{eqnarray}\label{cA4g=11Ds}
&& \hspace{-1cm} {\cal A}_{(I_1J_1)\ldots (I_4J_4)} ({k_1},\theta^-_{(1)};k_2 , \theta^-_{(2)};k_3 , \theta^-_{(3)};{k_4},\theta^-_{(4)})= \hspace{7cm}\nonumber \\ &  =& \;
  -{\cal A}_{z_{12} \; (I_1J_1)(I_2J_2) \; Jp}(\widehat{k_1}, \widehat{\theta^-_{(1)}};k_2, \theta^-_{(2)};
 \widehat{\rho^{\#}}(z_{12}),  v^{-}_{q}(z_{12}), 0)    \nonumber \\ \nonumber
&&  \qquad
\times \frac {1}{(P_{12})^2 {\widehat{\rho}{}^{\#}(z_{12})}}  {\cal A}_{z_{12}\; Jp\;  (I_3J_3)(I_4J_4)}(-\widehat{\rho^{\#}}(z_{12}),  v^{-}_{q}(z_{12}), 0; k_{3}, \theta^-_{(3)}; \widehat{k_4}, \widehat{\theta^-_{(4)}})
  \\ && + 2
  {\cal A}_{z_{12} \; (I_1J_1)(I_2J_2) \; (IJ)}(\widehat{k_1}, \widehat{\theta^-_{(1)}};k_2, \theta^-_{(2)};
 \widehat{\rho^{\#}}(z_{12}),  v^{-}_{q}(z_{12}), 0)    \nonumber \\ \nonumber
&&  \qquad
\times \frac {1}{(P_{12})^2}  {\cal A}_{z_{12}\; (IJ)\;  (I_3J_3)(I_4J_4)}(- \widehat{\rho^{\#}}(z_{12}),  v^{-}_{q}(z_{12}), 0; k_{3}, \theta^-_{(3)}; \widehat{k_4}, \widehat{\theta^-_{(4)}})\nonumber
 \\  && + \frac {1}{6}
  {\cal A}_{z_{12} \; (I_1J_1)(I_2J_2) \; [IJK]}(\widehat{k_1}, \widehat{\theta^-_{(1)}};k_2, \theta^-_{(2)};
 \widehat{\rho^{\#}}(z_{12}),  v^{-}_{q}(z_{12}), 0)    \nonumber \\ \nonumber
&&  \qquad
\times \frac {1}{(P_{12})^2}  {\cal A}_{z_{12}\; [IJK]\;  (I_3J_3)(I_4J_4)}(- \widehat{\rho^{\#}}(z_{12}),  v^{-}_{q}(z_{12}), 0; k_{3}, \theta^-_{(3)}; \widehat{k_4}, \widehat{\theta^-_{(4)}}) \\ && +
(2\longleftrightarrow 3)\; .
\end{eqnarray}
All the superamplitudes in the r.h.s. are
superfields as functions of 'external variables' (i.e. depend on two of four
$\theta^-_{q(1)}$, $\ldots$, $\theta^-_{q(4)}$) but leading component of a complete superamplitude with respect to the 'internal' fermionic variables, i.e. are taken at zero value of the fermionic variable corresponding to the
'internal' line, $\Theta^-_q=0$.

Let us recall that \begin{itemize} \item the subindex $z_{12}$  of the superamplitudes indicate the value of the deformation parameter (\ref{zl:==}) used in $\widehat{k_1}=\widehat{k_1}(z_{12})$,  $\widehat{\theta^-_{(1)}}= \widehat{\theta^-_{(1)}}(z_{12})$ and $\widehat{k_4}=\widehat{k_4}(z_{12})$,  $\widehat{\theta^-_{(4)}}= \widehat{\theta^-_{(4)}}(z_{12})$,
\item the arguments  $k_{a(i)}$ actually indicate  the dependence on corresponding 'energy' and spinor moving frame variables, $\rho^{\#}_i$ and $v_{\alpha q(i)}^{\; -}$;
\item in contrast, in the case of 'internal line' the dependence on $\widehat{\rho^{\#}}(z_{12})$ and $  v^{-}_{q}(z_{12})$ (of
(\ref{hkSlzl=}) with $P_{l}=P_{12}$) is indicated explicitly.
\end{itemize}

\subsection{On 4-particle  and 3-particle amplitudes in D=11}

The superamplitude in the left hand side of (\ref{cA4g=11Ds}) provides a superfield generalization of the well--known 4-graviton amplitude calculated in 11D directly in \cite{Deser:2000xz} and having the structure similar to one which had been known from zero slop limit of type II superstring \cite{Green:1981xx,Green:1981ya,Schwarz:1982jn,Gross:1986iv}. In our spinor moving frame version of the spinor helicity formalism it reads
\begin{eqnarray}\label{cAg4=11D}
&& {\cal A}_{((I_1J_1))... ((I_4J_4))}(k_1, k_2, k_3, k_4)= - \frac {4\kappa^2} {stu} K^{((I_1|_1\ldots ((I_4|_4}  \;  K^{|J_1))_1\ldots |J_4))_4} = \nonumber
\\ && - \frac {4\kappa^2} {stu} t_8^{a_1\ldots a_8 } t_8^{b_1\ldots b_8 }
k_{a_1(1)}u_{a_2(1)}^{\;\;((I_1|_1} \ldots k_{a_7(4)}u_{a_8(4)}^{\;\; ((I_4|_4}\; k_{b_1(1)}u_{b_2(1)}^{\;\;|J_1))_1} \ldots k_{b_7(4)}u_{b_8(4)}^{\;\;|J_4))_4}
\; , \end{eqnarray}
where
\begin{eqnarray}
 \label{kai=}
&& k_{a(i)}= \rho^{\#}_{(i)} u_{a(i)}^{=} \; , \qquad
\end{eqnarray}
$s, t, u$ are standard Mandelstam variables (see below for their expression in moving frame formalism),
\begin{eqnarray} \label{KIJKL=}
 K^{I_1\ldots I_4} &=&  t_8^{a_1\ldots a_8 }k_{a_1(1)}u_{a_2(1)}^{I_1} \ldots k_{a_7(4)}u_{a_8(4)}^{I_4} \\ \nonumber &=& \rho^{\#}_{(1)}\ldots \rho^{\#}_{(4)}\;  t_8^{a_1\ldots a_8 }u_{a_1(1)}^{\; =}u_{a_2(1)}^{I_1} \ldots u_{a_7(4)}^{\; =}u_{a_8(4)}^{I_4}
\;
\end{eqnarray}
and tensor $t_8$  is defined in \cite{Green:1981xx,Green:1981ya}. A compact expression for this tensor is
\begin{eqnarray}\label{t8F4=}
&t_8^{a_1\ldots a_8 }&F^{(1)}_{a_1a_2} \ldots F^{(4)}_{a_7a_8}=
\frac{1}{3}(tr (F^{(1)}F^{(2)}F^{(3)}F^{(4)})+ (2\leftrightarrow 3) +   (1\leftrightarrow 4)) - \qquad  \\ \nonumber
&& - \frac{1}{12}(tr (F^{(1)}F^{(2)}) tr (F^{(3)}F^{(4)}) +tr (F^{(1)}F^{(3)}) tr (F^{(2)}F^{(4)}) +
tr (F^{(1)}F^{(2)}) tr (F^{(3)}F^{(4)}) )\; .
\end{eqnarray}
where $F^{(1)}_{a_1a_2}, \ldots , F^{(4)}_{a_7a_8}$ are arbitrary antisymmetric tensor fields.

It is also instructive to compare (\ref{cAg4=11D}) with the $D=10$ SYM 4-gluon tree amplitude \cite{Green:1981xx,Green:1981ya,Schwarz:1982jn}. Omitting the color factor ($\rm{Tr} (T_1T_2T_3T_4)$ in
\cite{Gross:1986iv}) it reads
\begin{eqnarray}\label{cAg4=10D}
&& {\cal A}^{10D}_{\check{I}\check{J}\check{K}\check{L}}
= - \frac {2g^2} {st} K^{\check{I}_1\check{I}_2\check{I}_3\check{I}_4}=
-  \frac {2g^2} {st}\, \rho^{\#}_{(1)}\ldots \rho^{\#}_{(4)}\;  t_8^{a_1\ldots a_8 }
u_{a_1(1)}^{\; =}u_{a_2(1)}^{\check{I}} \ldots u_{a_7(4)}^{\; =}u_{a_8(4)}^{\check{L}}\; , \qquad \\ \nonumber && \qquad  \check{I},\check{J},\check{K}, \check{L}=1,...,8\; .
\end{eqnarray}
Here and below in this section we have denoted the $SO(8)$ vector indices by hatted symbols, $\check{I}=1,...,8$, to distinguish it from SO(9) indices,
$I,J=1,...,9$ often situated close.

To make the expression (\ref{cAg4=10D})  more illustrative, one has to use (\ref{t8F4=}) with the on-shell field strengths (\ref{Fab=kuwI}),
\begin{eqnarray}\label{FabI=}
F_{ab}^{\check{I}(i)} = k_{[a|(i)}u_{|b](i)}^{\check{I}}= \rho^\#_{(i)}u_{[a|(i)}^{\; =}u_{|b](i)}^{\check{I}}\; , \qquad
\check{I}=1,...,8\; ,
\end{eqnarray}
and writes  (\ref{cAg4=10D}) as ${\cal A}^{10D}_{\check{I}_1\check{I}_2\check{I}_3\check{I}_4}
=
-  \frac {2g^2} {st}\,   t_8^{a_1\ldots a_8 }
F_{a_1a_2\, 1}^{\check{I}_1} \ldots F_{a_7a_8\, 4}^{\check{I}_4}$. In string theory (the 8-dimensional version of) this comes from the path integral over zero modes of fermionic variables which can be described by  SO(8) spinors  $\eta_{\dot{q}}$, ${\dot{q}}=1,...,8$,
\begin{eqnarray}\label{t8F4==}
&t_8^{\check{I}_1\ldots \check{I}_8 }& F_{\check{I}_1\check{I}_2}{}^{(1)} \ldots F_{\check{I}_7\check{I}_8}{}^{(4)}
=\propto  \int d^8\eta_{\dot{q}} \; \exp \left\{ \sum_{i=1}^4 \eta\tilde{\gamma}{}^{\check{I}_i\check{J}_i}\eta  \; F_{\check{I}_1\check{I}_2}{}^{(i)} \right\}\; ,
\end{eqnarray}
\cite{Gross:1986iv}
where $\tilde{\gamma}{}^{\check{I}\check{J}}_{\dot{q}\dot{p}}= {\gamma}{}^{[\check{I}}_{q\dot{q}} {\gamma}{}^{\check{J}]}_{{q}\dot{p}}$ and ${\gamma}{}^{\check{I}}_{{q}\dot{p}}$ are SO(8) Klebsh-Gordan coefficients,
$\check{I}, \check{J}=1,..,8$.

It is natural to assume that  (\ref{cAg4=11D}) is given by leading ($\theta^-_{q(1)}=0$, ..., $\theta^-_{q(4)}=0$)
component of (\ref{cA4g=11Ds}). Then (\ref{cAg4=11D}) should be reproducible form the
following recurrent relation
 \begin{eqnarray}\label{cA4g=11D0}
&& \hspace{-1cm} {\cal A}_{(I_1J_1)\ldots (I_4J_4)} ({k_1},k_2 ,k_3 , {k_n})= \hspace{7cm}\nonumber \\ &  =&  2
  {\cal A}_{z_{12} \; (I_1J_1)(I_2J_2) \; (IJ)}(\widehat{k_1}, k_2;
 \widehat{\rho^{\#}}(z_{12}),  v^{-}_{q}(z_{12}))     \nonumber \\ \nonumber
&&  \qquad
\times \frac {1}{(P_{12})^2}  {\cal A}_{z_{12}\; (IJ)\;  (I_3J_3)(I_4J_4)}(- \widehat{\rho^{\#}}(z_{12}),  v^{-}_{q}(z_{12}); k_{3}, \widehat{k_4})\nonumber
 \\  && + \frac {1}{6}
  {\cal A}_{z_{12} \; (I_1J_1)(I_2J_2) \; [IJK]}(\widehat{k_1}, k_2;
 \widehat{\rho^{\#}}(z_{12}),  v^{-}_{q}(z_{12}))    \nonumber \\ \nonumber
&&  \qquad
\times \frac {1}{(P_{12})^2}  {\cal A}_{z_{12}\; [IJK]\;  (I_3J_3)(I_4J_4)}(- \widehat{\rho^{\#}}(z_{12}),  v^{-}_{q}(z_{12}); k_{3}, \widehat{k_4})\\ && +
(2\longleftrightarrow 3)\; .
\end{eqnarray}
In writing this we have used the evident fact that the amplitude with odd number of fermions vanishes, in particular
${\cal A}_{z_{12} \; (I_1J_1)(I_2J_2) \; Jp}(\widehat{k_1}, k_2,
 P_{12}) \equiv 0  $ (which is not correct in the case of superamplitudes).

 Furthermore it is reasonable to assume that
$ {\cal A}_{z_{12} \; (I_1J_1)(I_2J_2) \; [IJK]}(\widehat{k_1}, k_2; P_{12})$ also vanishes
as otherwise it would produce after dimensional reduction, a nonvanishing amplitude of creation
of a single scalar particle in the collision of two gravitons; even with complex momenta the existence of such an amplitude is counterintuitive. Using this hypothesis to omit the second term and its $(2\leftrightarrow 3)$ counterpart, we arrive at the pure bosonic BCFW relation for four graviton amplitudes
\begin{eqnarray}\label{cA4g=11D0g}
&& \hspace{-2cm}{\cal A}_{(I_1J_1)\ldots (I_4J_4)} ({k_1},k_2 ,k_3 , {k_n})=\hspace{7cm}\nonumber \\ &  =& \;\;   2
  {\cal A}_{z_{12} \; (I_1J_1)(I_2J_2) \; (IJ)}(\widehat{k_1}, k_2;
 \widehat{\rho^{\#}}(z_{12}),  v^{-}_{q}(z_{12}))     \nonumber \\ \nonumber
&&  \hspace{2cm}
\times \frac {1}{(P_{12})^2}  {\cal A}_{z_{12}\; (IJ)\;  (I_3J_3)(I_4J_4)}(- \widehat{\rho^{\#}}(z_{12}),  v^{-}_{q}(z_{12}); k_{3}, \widehat{k_4})\nonumber
\\ && +  2
  {\cal A}_{z_{12} \; (I_1J_1)(I_3J_3) \; (IJ)}(\widehat{k_1}, k_3;
 \widehat{\rho^{\#}}(z_{13}),  v^{-}_{q}(z_{13}))     \nonumber \\
&&  \hspace{2cm}
\times \frac {1}{(P_{13})^2}  {\cal A}_{z_{13}\; (IJ)\;  (I_2J_2)(I_4J_4)}(- \widehat{\rho^{\#}}(z_{13}),  v^{-}_{q}(z_{13}); k_{2}, \widehat{k_4})\; .
\end{eqnarray}
In different formalism  such type relations in a gravity field theory at arbitrary $D$ were discussed in \cite{ArkaniHamed:2008gz}

Three graviton amplitude is known to be  \cite{Sannan:1986tz} (in \cite{Schwarz:1982jn} one can find also the $\alpha^\prime$ correction to this tensor)
\begin{eqnarray}\label{cAg3=D}
&& {\cal A}_{((I_1J_1)) ((I_2J_2))((I_3J_3))}(k_1, k_2, k_3)= \kappa
  \, t^{((I_1|_1((I_2|_2((I_3|_3}  \;  t^{|J_1))_1|J_3))_3 |J_3))_3} \nonumber
\\ &&  \qquad =   \kappa \, t^{a_1a_2a_3 }(k_1,k_2,k_3)  t^{b_1b_2b_3 }(k_1,k_2,k_3)
u_{a_1(1)}^{((I_1} u_{b_1(1)}^{J_1))} u_{a_2(2)}^{((I_2}u_{b_2(2)}^{J_2))}  u_{a_3(3)}^{((I_3}u_{b_3(3)}^{J_3))}
\; , \end{eqnarray}
where
\begin{eqnarray}\label{t3=D}
t^{I_1I_2I_3}&=&  t^{abc} (k_1,k_2,k_3) u_{a(1)}^{I_1}u_{b(2)}^{I_2}u_{c(3)}^{I_3}
\\ & =& (k_{(2)}u_{(1)}^{I_1})\, (u_{(2)}^{I_2}u_{(3)}^{I_3})+ (k_{(3)}u_{(2)}^{I_2})\, (u_{(1)}^{I_1}u_{(3)}^{I_3})+ (k_{(1)}u_{(3)}^{I_3})\, (u_{(1)}^{I_1}u_{(2)}^{I_2})
\; , \end{eqnarray}
and
\begin{eqnarray}\label{tabc=D}
&&  t^{abc} (k_1,k_2,k_3) = k_{(2)}^a\eta^{bc} + k_{(3)}^b\eta^{ac}+ k_{(1)}^c\eta^{ab}
\; . \end{eqnarray}
Notice the cyclic symmetry property of the t-tensor: $t^{abc} (k_1,k_2,k_3)=t^{bca} (k_2,k_3,k_1)$.

The structure of the 'Chern-Simons' term of the 11D supergravity action and the form of the on-shell field strength of the 3-form gauge field (called 'formon' in \cite{Deser:2000xz}), $F_{abcd}^{\; IJK}= k_{[a}u_b^Iu_c^Ju_{d]}^K$ (\ref{FabcdIJK=}),  suggests the following structure of the 'three formon' amplitude
\begin{eqnarray}\label{cAFFA=}
&&  {\cal A}_{[I_1J_1K_1]\; [I_2J_2K_2]\; [I_3J_3K_3]}(k_1;k_2;k_3)= \qquad \\
&&  \propto
\epsilon^{abc_1c_2\ldots c_8 c_9} \;  k_{1a} k_{2b}\; u_{c_1(1)}^{I_1}u_{c_2(1)}^{J_1}u_{c_3(1)}^{K_1}\, u_{c_4(2)}^{I_2}u_{c_5(2)}^{J_2}u_{c_6(2)}^{K_2} \, u_{c_7(3)}^{I_3}u_{c_8(3)}^{J_3}u_{c_9(3)}^{K_3}\; .\qquad
\end{eqnarray}
To check that this amplitude possess the cyclic symmetry property characteristic for the bosonic particles ${\cal A}(1,2,3)={\cal A}(3,1,2)={\cal A}(2,3,1)$, we have to take into account the momentum conservation $k_{a1}+k_{a2}+k_{a3}=0$.
It is also, symmetric under exchange of complete sets of two particle data, e.g. it implies
${\cal A}_{[I_1J_1K_1]\; [I_2J_2K_2]\; [I_3,J_3,K_3]}(k_1;k_2;k_3)=
{\cal A}_{[I_2J_2K_2]\; [I_1J_1K_1]\;  [I_3J_3K_3]}(k_2;k_1;k_3)$, etc.

Some  amplitudes of the 10D SYM  will be discussed in the next section 8.

The above 11D discussion makes transparent that the bosonic amplitudes are expressed through contractions of the vector harmonics describing the frames associated with different particles, typically
$u^=_{ai}u^{aJ}_j$ and $u^I_{ai}u^{aJ}_j$; similarly, the  fermionic amplitudes involve contractions of spinor harmonics from different spinor frames.

Thus, to proceed with amplitude calculations in our formalism, we should understand better the relation between spinor frames associated to different particles. Below we address this problem, present an explicit parametrization of $j$-th frame in term of $i$-th frame and a number of parameters, and also find a gauge fixing conditions for the auxiliary gauge symmetries acting on the spinor frames, which makes the relation of different frames especially simple.

\subsection{Relation between spinor frames associated to different external particles}

For a possible reader convenience, let us begin this section by recalling  that, in our spinor frame form of the spinor helicity formalism, the light-like external 11D momenta $k_{i}^a$ are expressed thought  bilinear of spinor helicity variables, $\lambda_{\alpha q i}= \sqrt{\rho^{\#}_{i}}  v_{\alpha qi}^{\; - }$,
 \begin{eqnarray}\label{kia=Up11}
k_{i}^a =  \rho^{\#}_{i} u_{i}^{a=} \, ,  \qquad  u_{i}^{a=} = \frac 1 {16} v^-_{qi} \Gamma^av^-_{qi}\; ,
\end{eqnarray}
where  $ v_{\alpha q(i)}^{\; - }$ obey the constraints
\begin{eqnarray}\label{u--=v-v-11-2}
u^=_{ai}
\Gamma^a_{\alpha\beta}= 2 v_{\alpha qi}^{\; - } v_{\beta qi}^{\; - } \; , \qquad
  v^-_{{q}i} \tilde{\Gamma}_{a}v^-_{{p}i}= u^=_{a} \delta_{{q}{p}}  \qquad
\end{eqnarray}
and are considered as the spinor frame variables (or Lorentz harmonics) (\ref{harmVi=D}), related to vector frame (vector harmonics) (\ref{harmUi=}), (\ref{uu=0})--(\ref{uui=0}), by   (\ref{v+v+=u++})--(\ref{uIs=v-v+}) (with  $q=1,...,8$, $\dot{q}=1,...,8$, $I=1,..,8$  when D=10 and $\dot{q}=q=1,...,16$, $I=1,..,9$ when D=11).

To proceed with clarification of the structure of (\ref{cA4=11D}) and similar amplitudes, we need to describe the relations between spinor  frame variables associated to different particles.

For the spinor helicity variables this problem might seem to be difficult, but knowing its relation to Lorentz harmonics and using the group theoretical meaning of these one can solve it.
Indeed, as Lorentz harmonics are elements of $Spin(1,D-1)$ group valued matrix (\ref{harmVi=D}), they are related with another set of
  Lorentz harmonics by an $Spin(1,D-1)$ transformations.
Allowing for both positive and negative values of the 'energy' variables $\rho_i^{\#}$ and $\rho_j^{\#}$,
we can write the relation between the harmonics describing  spinor frames of $j$-th and $i$-th 11D particles
in the form
\begin{eqnarray}\label{v-j=+Kv-i}
v_{\alpha qj}^{\; -} &=& e^{-\beta_{ji}} {\cal O}_{ji{q}{p}}  \left(
v_{\alpha  p i}^{\; -}+ {1\over 2} K^{=I}_{ji}  \gamma^I_{pp'} v_{\alpha  p' i}^{\; +}
\right) \; , \qquad  \\
\label{v+j=+Kv-}
v_{\alpha  qj}^{\; +}&=& e^{\; \beta_{ji}} {\cal O}_{ji{q}{p}}
v_{\alpha {p}i}^{\;+} +  {1\over 2}  K^{\# I}_{ji}  v_{\alpha p (j)}^{\; -} \gamma^I_{p{q}}
  \; . \qquad
\end{eqnarray}
In these equations ${\cal O}_{(ji){q}{p}}$ is an element of $Spin(9)$ ($\subset Spin(1,10)$),  i.e.
$16\times 16$ matrix which obeys\footnote{Notice that the second term in (\ref{v+j=+Kv-}) contains
$v_{\alpha p (j)}^{\; -} $.
A more explicit expression in terms of $v_{\alpha pi}^{\; \pm} $ reads  $v_{\alpha  qj}^{\; +}=  e^{\; \beta_{ji}} {\cal O}_{ji{q}{p}} \left(
v_{\alpha {p'}i}^{\;+} \left(\delta_{{p'}{p}} + \frac 1 4 K^{=I}_{ji} \tilde{K}{}^{\# J}_{ji} ({\gamma}{}^I{\gamma}^J)_{{p'}{p}} \right)+ {1\over 2} \tilde{K}{}^{\# I}_{ji}  v_{\alpha p' (i)}^{\; -} \gamma^I_{p'{p}}\right)$  with $\tilde{K}{}^{\# I}_{(ji)}= K^{\# J}_{ji}{\cal O}_{ji}^{JI} e^{-2\beta_{ji}}$ (see (\ref{tK++:=})).}
\begin{eqnarray}\label{OIJ=OGO}
 {\cal O}_{jiqr} {\cal O}_{ji{p}{s}} \gamma^{I}_{rs}= \gamma^{J}_{qp} {\cal O}^{JI}_{ji}
 \; ,  \qquad
\end{eqnarray}
where $9\times 9$ matrix ${\cal O}^{JI}_{(ji)}$ belongs to $SO(9)$ group. The scale factor $ e^{-\beta_{ji}} $ is considered as an element of $SO(1,1)\subset SO(1,10)$.
It is instructive to calculate  the contractions
\begin{eqnarray}\label{v-iv-j=}
v_{pi}^{-\alpha }
v_{\alpha qj}^{\; -} &=&- \frac 1 2 e^{-\beta_{ij}}  K^{= I}_{ij} ({\cal O}_{ij}\gamma^I)_{{p}{q}} =\frac 1 2 e^{-\beta_{ji}}  K^{= I}_{ji} ({\cal O}_{ij}\gamma^I)_{{q}{p}} \; , \qquad  \\ \label{v+iv-j=}
v_{pi}^{+\alpha }
v_{\alpha qj}^{\; -} &=& e^{+\beta_{ij}} {\cal O}_{ij}{}_{{p}{p'}}   \left( \delta_{{p'}q} + \frac 1 4   \tilde{K}{}^{\# I}_{ij}K^{= I}_{ij} (\gamma^I\gamma^J)_{{p'}{q}}
\right)= e^{-\beta_{ji}} {\cal O}_{ji}{}_{{p}{q}}
 \; ,  \qquad \\ \label{tK++:=} && {}\qquad \tilde{K}{}^{\# I}_{ji}= K^{\# J}_{ji}{\cal O}_{ji}^{JI} e^{-2\beta_{ji}}\; .  \qquad
\end{eqnarray}

Eqs. (\ref{v+j=+Kv-}) and (\ref{v-iv-j=}) are  written for 11D case, while their 10D counterparts carry dotted spinor indices,
\begin{eqnarray}
\label{v+z=+Kv-10}
D=10:   && \nonumber \\ && v_{\alpha  \dot{q} j}^{\; +}=  e^{\; \beta_{ji}} \tilde{{\cal O}}_{ji\dot{q}\dot{p}} \left(
v_{\alpha \dot{p'}i}^{\;+} \left(\delta_{\dot{p'}\dot{p}} + \frac 1 4 K_{ji}^{=I} K_{ji}^{\# J} (\tilde{\gamma}{}^I{\gamma}^J)_{\dot{p'}\dot{p}} \right)+ {1\over 2} K^{\# I}_{ji}  v_{\alpha p' i}^{\; -} \gamma^I_{p'\dot{p}}
\right)\; , \qquad \\ \label{v-iv-j=10D}
 && v_{\dot{p}i}^{-\alpha }
v_{\alpha qj}^{\; -} =- \frac 1 2 e^{-\beta_{ij}}  K^{= I}_{ij} (\tilde{{\cal O}}_{ij}\tilde{\gamma}^I)_{\dot{p}{q}} =\frac 1 2 e^{-\beta_{ji}}  K^{= I}_{ji} ({\cal O}_{ij}\gamma^I)_{{q}\dot{p}} \; , \qquad  \\
 \label{OIJ=OGO10}
 && \qquad {\cal O}_{jiqp} \tilde{{\cal O}}_{ji\dot{q}\dot{p}} \gamma^{I}_{q\dot{q}}= \gamma^{J}_{q\dot{q}}
 {\cal O}^{JI}_{ji}  \; .  \qquad
\end{eqnarray}

For the vector harmonics  we have the following relations
\begin{eqnarray}\label{u--=KuI=11}
u^=_{aj} &=& e^{-2\beta_{ji}}\left(u^=_{ai} + \frac 1 4 (\vec{K}^{=}_{ji} )^2 u^\#_{ai}  + K^{=I}_{ji}u^I_{ai} \right)
 \; , \qquad \\
 \label{u++=Kui=11}
u^\#_{aj} &=& e^{+2\beta_{ji}}\left(u^\#_{ai} \left(1+ \frac 1 2 K^{=I}_{ji} \tilde{K}{}^{\# I}_{ji}+ \frac 1 {16} (\vec{K}{}^{=}_{ji})^2 (\vec{K}{}^{\#}_{ji})^2\right) + {1\over 4} u^=_{ai} (\tilde{K}{}^{\#  I}_{ji})^2  + \right. \qquad \nonumber \\ && \qquad  \left. +  u^I_{a(i)} \left( \tilde{K}{}^{\#  I}_{ji}+ \frac 1 4 K^{=I}_{ji} (\vec{K}{}^{\#}_{ji})^2 \right)   \right)
, \; \qquad \\
 \label{uI=K--ui}
u^I_{aj} &=& \left(u^K_{ai}
 \left(\delta^{KJ} + \frac 1 2 K^{=K}_{ji} \tilde{K}{}^{\#  J}_{ji} \right)
+ \right. \qquad \nonumber \\ && \qquad + \left. \frac 1 2 u^\#_{ai} \left( K^{= J }_{ji}+ \frac 1 4 \tilde{K}{}^{\#  J}_{ji} (\vec{K}{}^{=}_{ji})^2 \right)  +  \frac 1 2 u^=_{ai}  \tilde{K}{}^{\#  J}_{ji}
  \right){\cal O}^{JI}_{ji} \; . \qquad
\end{eqnarray}
where $(\vec{K}^{=}_{ji} )^2= {K}^{=I}_{ji} {K}^{=I}_{ji}$ and ${\cal O}^{JI}_{ji}\in SO(9)$ ($\in SO(8)$ for D=10) is defined in (\ref{OIJ=OGO}) ((\ref{OIJ=OGO10}) for D=10). Finally $K^{\# J}_{(ji)}$ is the parameter of $K_{(D-2)}$ symmetry and $K^{= I}_{(ji)}$ parametrize the coset $\frac {SO(1,D-1) }{[SO(1,1)\otimes SO(D-2)]\subset \!\!\!\!\times K_{(D-2)}}$.

The above equations simplify essentially if we fix the gauge with respect to
$K_{(D-2)i}$ symmetries acting on different spinor frames (\ref{harmVi=D})  by setting
\begin{eqnarray}\label{K++Iij=0}
K^{\# I}_{ji}=0 \; . \qquad
\end{eqnarray}
To be convinced that this is possible, one should notice that, say, $K^{\# I}_{ji}= K^{\# I}_{j1}-K^{\# I}_{i1}$ so that the set of independent relations in (\ref{K++Iij=0}) can be chosen to be $K^{\# I}_{j1}=0$. In the gauge (\ref{K++Iij=0})  the relations between sets of vector harmonics  (\ref{u--=KuI=11})--(\ref{uI=K--ui})
simplify to
\begin{eqnarray}\label{u--=KuI=11D}
u^=_{aj} &=& e^{-2\beta_{ji}}\left(u^=_{ai} + \frac 1 4 (\vec{K}^{=}_{ji} )^2 u^\#_{ai}  + K^{=I}_{ji}u^I_{ai} \right)
 \; , \qquad \\
 \label{u++=Kui=11D}
u^\#_{aj} &=& e^{+2\beta_{ji}}u^\#_{ai}
, \; \qquad \\
 \label{uI=K--ui=11D}
u^I_{aj} &=& \left(u^J_{ai} + \frac 1 2 \vec{K}^{=J}_{ji} u^\#_{ai}\right)
 {\cal O}^{JI}_{ji} \; , \qquad
\end{eqnarray}
and  (\ref{v+j=+Kv-}) acquires the form
\begin{eqnarray}
\label{v+j=+Kv-=11D}
v_{\alpha  qj}^{\; +}&=& e^{\; \beta_{ji}} {\cal O}_{ji{q}{p}}
v_{\alpha {p}i}^{\;+}
  \; . \qquad
\end{eqnarray}
For completeness, let us write explicitly also the gauge fixed expression for
the complete set of 10D Lorentz harmonics
\begin{eqnarray}\label{v-j=+Kv-i=10D}
v_{\alpha qj}^{\; -} &=& e^{-\beta_{ji}} {\cal O}_{(ji){q}{p}}  \left(
v_{\alpha  p i}^{\; -}+ {1\over 2} K^{=I}_{ji}  \gamma^I_{p\dot{p}} v_{\alpha  \dot{p} i}^{\; +}
\right) \; , \qquad  \\
\label{v+j=+Kv-=10D}
v_{\alpha  \dot{q}j}^{\; +}&=& e^{\; \beta_{ji}} {\cal O}_{ji\dot{q}\dot{p}}
v_{\alpha \dot{p}i}^{\;+}
  \; , \qquad
\end{eqnarray}
and for the elements of the inverse spinor frame matrix (\ref{harmV-1=D}),
\begin{eqnarray}\label{v-j-1=+Kv-i=10D}
v_{\dot{q}j}^{-\alpha} &=& e^{-\beta_{ji}} {\cal O}_{ji\dot{q}\dot{p}}  \left(
v_{  \dot{p} i}^{ -\alpha}- {1\over 2} K^{=I}_{ji}  v_{\alpha  {p} i}^{+\alpha}\gamma^I_{p\dot{p}}
\right) \; , \qquad  \\
\label{v+j=Ov+=10D}
v_{{q}j}^{+\alpha}&=& e^{\; \beta_{ji}} {\cal O}_{ji{q}{p}}
v_{ {p}i}^{+\alpha}
  \; .\qquad
\end{eqnarray}
In the next sec. \ref{ref-sp-fr=sec} we consider stronger gauge fixing conditions for the auxiliary gauge symmetries acting on the spinor frame variables, in which these are expressed through the $(D-2)$ parameters $K^{=I}_{ji} $ only.

\subsection{Reference spinor frame and complete gauge fixing of the auxiliary  gauge symmetries}
\label{ref-sp-fr=sec}

It is convenient to introduce an auxiliary spinor frame $(v_{\alpha q}^{\; -}, v_{\alpha q}^{\; +})$ and associated vector frame
$(u^=_a, u^\#_a, u^I_a)$. Any of the spinor harmonics  $(v_{\alpha qj}^{\; -}, v_{\alpha qj}^{\; +})$ and associated vector harmonics
$(u^=_{aj},$ $u^\#_{aj}, u^I_{aj})$ are related with reference spinor frame and reference frame by (\ref{v-j=+Kv-i}),  (\ref{v+j=+Kv-}), (\ref{u--=KuI=11}), (\ref{u++=Kui=11}), (\ref{uI=K--ui}) with omitted index $i$. Then they are parametrized by the set of
$K^{= I}_{j}$, $e^{-\beta_j}$, ${\cal O}_{j}^{IJ}=({\cal O}_{j})^{-1\; JI}$, $K^{\# I}_{j}$, in which the last three subsets of parameters correspond to the gauge symmetry transformations. These are used as identification
relations which allow to consider the sets of harmonic variables  $(v_{\alpha q(i)}^{\; -}, v_{\alpha q(i)}^{\; +})$  as homogeneous coordinates of the celestial sphere.  We call them {\it auxiliary gauge symmetries} and  can conventionally fix them by setting
\begin{eqnarray}\label{K++=0gauge}
 K^{\# I}_{i}=0
 \; , \qquad{\cal O}_{i}^{IJ}=\delta^{IJ}, \qquad e^{-\beta_i}=1 . \qquad
\end{eqnarray}
Then any 11D spinor frame can be expressed through the auxiliary spinor frame by
\begin{eqnarray}\label{v-=v+Kv=G}
v_{\alpha q(i)}^{\; -} &=&
v_{\alpha  q }^{\; -}+ {1\over 2} K^{=I}_{i}  \gamma^I_{qp} v_{\alpha  p}^{\; +}
\; , \qquad
v_{\alpha {q}(i)}^{\;+}= v_{\alpha  q}^{\; +}
 \; . \qquad
\end{eqnarray}
The corresponding relations for inverse 11D harmonics read
\begin{eqnarray}
\label{v--1=G} v^{+\alpha}_{q(i)}&=&
v^{+\alpha}_{q}
 \; ,  \qquad
v^{-\alpha}_{q(i)}=
v^{-\alpha}_{q}-  {1\over 2} K^{= I}_{(i)}  \gamma^I_{q{p}}v^{+\alpha}_{p}
\; , \qquad
\end{eqnarray}
so that Eqs. (\ref{v-iv-j=}) and (\ref{v+iv-j=}) drastically simplify
\begin{eqnarray}\label{v+iv-j=G}
v_{p(i)}^{+\alpha }
v_{\alpha q(j)}^{\; -} &=& \delta_{pq}
 \; ,  \qquad \\  \label{v-iv-j=G} v_{p(i)}^{-\alpha }
v_{\alpha q(j)}^{\; -} &=&
\frac 1 2 \,   K^{= I}_{ji}   \gamma^I_{{p}{q}} \; , \qquad  K^{= I}_{ji} := K^{= I}_{j}- K^{= I}_{i}
 \; .  \qquad
\end{eqnarray}

The frame vectors are decomposed in the basis provided by  the auxiliary frame as
\begin{eqnarray}\label{u--=KuI=11G}
u^=_{a(i)}= u^=_{a}   + K^{=I}_{(i)}u^I_{a} + \frac 1 4 (\vec{K}^{=}_{(i)} )^2 u^\#_{a}
 \; , \qquad \\
 \label{uI=K--uiG}
u^I_{a(i)} = u^I_{a} + \frac 1 2  K^{= I }_{(i)}\, u^\#_{a}\; , \qquad \\ \label{u++=KuiG}
u^\#_{a(i)} =u^\#_{a} \; . \qquad
\end{eqnarray}
These expressions for vector harmonics apply for both 10D and 11D cases, while the
(\ref{v-=v+Kv=G}) and (\ref{v--1=G}) are written for D=11.

Although the D=10 relations can be easily restored from (\ref{v-=v+Kv=G}) and (\ref{v--1=G}), for possible reader convenience we write them explicitly:
\begin{eqnarray}\label{v-=v+Kv=G=10}
v_{\alpha q(i)}^{\; -} &=&
v_{\alpha  q }^{\; -}+ {1\over 2} K^{=I}_{i}  \gamma^I_{q\dot{p}} v_{\alpha  \dot{p}}^{\; +}
\; , \qquad
v_{\alpha \dot{q}(i)}^{\;+}= v_{\alpha  \dot{q}}^{\; +}
 \; , \qquad\\
\label{v--1=G=10} v^{+\alpha}_{q(i)}&=&
v^{+\alpha}_{q}
 \; ,  \qquad
v^{-\alpha}_{\dot{q}(i)}=
v^{-\alpha}_{\dot{q}}-  {1\over 2} K^{= I}_{(i)}  v^{+\alpha}_{q}\gamma^I_{{q}\dot{q}}
\; . \qquad
\end{eqnarray}

Let us apply the above gauge fixing expressions to study the momentum conservation conditions.

\subsection{Momentum conservation and Mandelstam variables}

In our formalism the momentum conservation in the scattering of $n$ massless particles implies
\begin{eqnarray}
\label{Sriui--=0}
\sum\limits_{i=1}^n \rho^{\#}_{(i)} u^{=}_{a(i)}  =0 \; . \qquad
\end{eqnarray}
Contracting this equation with the basic vectors of the reference frame we can split it into three sets which look  especially simple in  the gauge (\ref{K++=0gauge}),
\begin{eqnarray}
\label{Sri=0}
&& \sum\limits_{i=1}^n \rho^{\#}_{(i)}  =0 \; , \qquad \\ \label{SriKi=0}
&& \sum\limits_{i=1}^n \rho^{\#}_{(i)} K^{=I}_{(i)}  =0 \; , \qquad \\ && \label{SriKi2=0} \sum\limits_{i=1}^n \rho^{\#}_{(i)} (K^{=I}_{(i)})^2  =0 \; . \qquad
\end{eqnarray}

In the case of 4-point amplitude we find that
the expressions for the Mandelstam variables are sufficiently simple already with the generic parametrization of the frame variables (\ref{u--=KuI=11})--(\ref{uI=K--ui}). Keeping all the gauge symmetry unfixed, we obtain from (\ref{u--=KuI=11}) and (\ref{kia=Up11})
\begin{eqnarray}\label{s=}
s= (k_1+k_2)^2= 2\rho^{\#}_1\rho^{\#}_2 u^{=a}_{2}u^{=}_{a1} =  \rho^{\#}_1\rho^{\#}_2 e^{-2\beta_{21}} (\vec{K}{}^=_{21})^2 \; , \qquad \\ \label{t=}
t= (k_1+k_3)^2= 2\rho^{\#}_1\rho^{\#}_3 u^{=a}_{3}u^{=}_{a1} =  \rho^{\#}_1\rho^{\#}_3 e^{-2\beta_{31}} (\vec{K}{}^=_{31})^2 \; , \qquad \\ \label{u=}
u= (k_1+k_4)^2= 2\rho^{\#}_1\rho^{\#}_4 u^{=a}_{4}u^{=}_{a1} =  \rho^{\#}_1\rho^{\#}_4 e^{-2\beta_{41}} (\vec{K}{}^=_{41})^2 \; . \qquad
\end{eqnarray}

Notice that the   denominators in the first and in the second terms of the BCFW-type relations for the four point
amplitudes and superamplitudes,  (\ref{cA4g=11D0g}) and (\ref{cA4g=11Ds}),
are equal to Mandelstam variable $s$ and $t$, respectively
\begin{eqnarray}\label{P122=}
&& P^2_{12}=s
= 2\rho^{\#}_1\rho^{\#}_2 u^{=a}_{2}u^{=}_{a1}= 2 e^{-\beta_{21}-\beta_{43}} \sqrt{\rho^{\#}_1\rho^{\#}_2\rho^{\#}_3\rho^{\#}_4} \sqrt{(\vec{K}{}^=_{21})^2(\vec{K}{}^=_{43})^2}
\; , \qquad \\  \label{P132=}
&& P^2_{13}= t
=2\rho^{\#}_1\rho^{\#}_2 u^{=a}_{3}u^{=}_{a(1)}
\; .  \qquad
\end{eqnarray}
In the second equality of (\ref{P122=}) we have used the conservation of the momentum which implies
$s=P^2_{12}=P^2_{43}=\sqrt{P^2_{12}P^2_{43}}$.

The set of  the arguments of the (super)amplitudes in the {\it r.h.s}'s of
(\ref{cA4g=11D0g}) and (\ref{cA4g=11Ds}) include (the variables related to) the deformed versions
of (\ref{P122=}) and (\ref{P132=}), $P^a_{12}(z_{12})$ and $P^a_{13}(z_{13})$, which are dependent on the complex null-vector $q_a$ obeying (\ref{qq=0=}).  Let us discuss  the representation of this and of the special values $z_{12}$, $z_{13}$ of the deformation parameter in our spinor frame approach.

\subsection{Studying the BCFW-like deformation with the gauge fixed spinor frames }

In the case of 4-particle amplitude, the deformation 11--vector/10--vector  $q_a$, a complex null-vector orthogonal to the 1-st and the 4-th light-like momenta, (\ref{qq=0=}), can be decomposed on the  frame related to any of four light-like  momenta (\ref{kia=Up11}). The decompositions  on 1-st and 4-th frames cannot contain terms with $u^{\#}_{a(1)}$ and  $u^{\#}_{a(4)}$, respectively. Generically we can also assume the absence of the terms proportional to $u^{=}_{a(1)}$ and  $u^{=}_{a(4)}$ (this is to say, to $k_{a(1)}$ and $k_{a(4)}$) so that
\begin{eqnarray}\label{q=uiqi1}
q_a=  - u^{I}_{a(1)} q^{I}_{(1)} = - u^{I}_{a(4)} q^{I}_{(4)} \; , \qquad
(\vec{q}_{(1)})^2=q_{(1)}^Iq_{(1)}^I=0\; , \qquad (\vec{q}_{(4)})^2=q_{(4)}^Iq_{(4)}^I=0\; . \qquad
\end{eqnarray}
In the gauge (\ref{K++Iij=0}), in which the relation between vectors from different frames are described
by Eqs. (\ref{u--=KuI=11D})--(\ref{uI=K--ui=11D}), the components $ q^{I}_{(4)}$ and $ q^{I}_{(1)}$ are related by $SO(D-2)$ rotation
\begin{eqnarray}\label{q4=Oq1}
q^{I}_{(4)}=
{\cal O}^{JI}_{41} q^{I}_{(1)}\;  \qquad
\end{eqnarray}
 and obey, besides the null-conditions  (\ref{q=uiqi1}),
\begin{eqnarray}\label{K41q1=0} K^{=I}_{41} q^{I}_{(1)}=0 \; , \qquad
 K^{=I}_{14} q^{I}_{(4)}= 0 . \qquad
\end{eqnarray}

\bigskip

Now, one can easily check that the special values $z_{12}$ and $z_{13}$ of the deformation parameter $z$, for which   $\widehat{P^a_{12}}(z)$ and $\widehat{P^a_{13}}(z)$ become light-like, read  (see (\ref{zl:==}))
\begin{eqnarray}\label{z12=}
z_{12}= -\frac { \rho^{\#}_{(1)}(u^{=}_{(1)}u^{=}_{(2)})}{q_{(1)}^I \, (u^{I}_{(1)}u^{=}_{(2)}) }= \frac { \rho^{\#}_{(1)}(\vec{K}{}^{=}_{(21)})^2}{2  K^{=I}_{(21)}q_{(1)}^I} \qquad \Rightarrow \qquad
(\widehat{P_{12}}(z_{12}))^2=0\; , \qquad \\ \label{z13=}
z_{13}= -\frac { \rho^{\#}_{(1)}(u^{=}_{(1)}u^{=}_{(3)})}{q_{(1)}^I \, (u^{I}_{(1)}u^{=}_{(3)}) }= \frac { \rho^{\#}_{(1)}(\vec{K}{}^{=}_{(31)})^2}{2  K^{=I}_{(31)}q_{(1)}^I}\qquad \Rightarrow \qquad
(\widehat{P_{13}}(z_{13}))^2=0\; . \qquad
\end{eqnarray}

Using (\ref{q=uiqi1}),  the nilpotent matrix (\ref{Mpq=}) can be written as
\begin{eqnarray}\label{Mpq=qI}
{\bb M}_{{q}{p}}& = &- \frac {1} { {\sqrt{\rho^{\#}_{(1)}\rho^{\#}_{(n)} }} (u_{(1)}^=u^{=}_{(n)}) }\;
(v_{{p'}(1)}^{-\alpha}
v_{\alpha {q}(n)}^{\; -}) \gamma^I_{p'p} q_{(1)}^I  = \qquad \\ \nonumber
& = & -\frac {1} { {\sqrt{\rho^{\#}_{(1)}\rho^{\#}_{(n)} }} (u_{(1)}^=u^{=}_{(n)}) }\;
q_{(n)}^I\gamma^I_{qq'}  (v_{{q'}(n)}^{-\alpha}
v_{\alpha {p}(1)}^{\; -})  \; . \qquad
\end{eqnarray}
Furthermore, Eq.  (\ref{q=uiqi1}) considered together with (\ref{u--=KuI=11}) and  (\ref{BCWF=hu--}), implies
\begin{eqnarray}\label{BCWF=hu--2}
\widehat{u_{(1)}^{=a}}= u_{(1)}^{=a} + u_{(1)}^{Ia}\frac {z q^I_{(1)}} {\rho^\#_{(1)}} \; , \qquad \widehat{u_{(n)}^{=a}}= u_{(n)}^{=a} - u_{(n)}^{Ia} \frac {z q^I_{(n)}} {\rho^\#_{(n)}} \; , \qquad
\end{eqnarray}
which can be recognized as  $S^9$-transformations (see (\ref{v-qi=G-H})) with nilpotent 9-vector parameters
\begin{eqnarray}\label{hK--I=}
\widehat{{\cal K}}{}^{=I}_{(1)}(z)= \frac {z q^I_{(1)}} {\rho^\#_{(1)}} \; , \qquad
\widehat{{\cal K}}{}^{=I}_{(n)}(z)= - \frac {z q^I_{(n)}} {\rho^\#_{(n)}} \;   \qquad
\end{eqnarray}
(here we use the notation $\widehat{{\cal K}}{}^{=I}_{(1)}(z)$ reserving
 $\widehat{{K}}{}^{=I}_{(1)}(z)$ for the parameter of transformation relating the deformed 1-st frame with the auxiliary reference frame).

First of all, this observation allows to conclude that $u_{(1)}^{\# a}$ and $u_{(n)}^{\# a}$ vectors may be taken to be undeformed  and, using (\ref{uI=K--ui}),  to find the deformation of the remaining frame vectors
\begin{eqnarray}
\label{BCWF=huI}
\widehat{u_{(n)}^{Ia}}= u_{(n)}^{Ia} - \frac {z q^I_{(n)}} {2\rho^\#_{(n)}} u_{(n)}^{\# a} \; , \qquad
\widehat{u_{(1)}^{Ia}}= u_{(1)}^{Ia} + \frac {z q^I_{(1)}} {2\rho^\#_{(1)}} u_{(1)}^{\# a}
  \; . \qquad
\end{eqnarray}

Secondly, this implies that the deformation of the spinor frame variables, (\ref{BCWF=vnM}) and (\ref{BCWF=v1M}), can be written in the form
\begin{eqnarray}\label{BCWF=vnK}
\widehat{v_{\alpha {q}(n)}^{\; -}}= v_{\alpha {q}(n)}^{\; -} + z \;
\sqrt{\frac {\rho^{\#}_{(1)}}{\rho^{\#}_{(n)}}} v_{\alpha {p}(1)}^{\; -} \; {\bb M}_{ {p}{q}} =v_{\alpha  q (n)}^{\; -} - \frac {z}{2\rho^\#_{(n)}} q^I_{(n)}  \gamma^I_{qp} v_{\alpha  p (n)}^{\; +}
 \; , \qquad \\
  \label{BCWF=v1K} \widehat{v_{\alpha {q}(1)}^{\; -}}= v_{\alpha {q}(1)}^{\; -} - z \;
 \sqrt{\frac {\rho^{\#}_{(n)}}{\rho^{\#}_{(1)}}}\; {\bb M}_{{q}{p}}\;  v_{\alpha {p}(n)}^{\; -}=
v_{\alpha  q (1)}^{\; -} + \frac {z}{2\rho^\#_{(1)}} q^I_{(1)}  \gamma^I_{qp} v_{\alpha  p (1)}^{\; +}\; . \qquad
\end{eqnarray}

At $z=z_{12}$ and $z=z_{13}$ of (\ref{z12=}) and (\ref{z13=}), the momenta  $\widehat{P^a_{12}}(z)$ and $\widehat{P^a_{13}}(z)$ become light-like and can be expressed in terms of 'energies' and the associated spinor  frame variables by
\begin{eqnarray}\label{hP12=vv}
\widehat{P_{12}^a}(z_{12})\Gamma_{a\alpha\beta} = 2 \widehat{\rho}^{\#}(z_{12})\,  v_{\alpha {q} (12)}^{\; -} v_{\beta {q} (12)}^{\; -} \; , \qquad
 \widehat{\rho}^{\#}(z_{12}) v^-_{{q}\,{12}} \tilde{\Gamma}{}^{a}v^-_{{p}\, {12}}= \widehat{P_{12}^a}(z_{12}) \delta_{{q}{p}} \;  \qquad
\end{eqnarray}
and its ${}_{12}\mapsto {}_{13}$ counterparts. Factorizing
$ \widehat{P_{12}}\!\!\!\!\!\!\!\!/{}\quad (z_{12}) := \widehat{P_{12}^a}(z_{12})\Gamma_{a}=  \widehat{k_{1}}\!\!\!\!\!/\;
(z_{12})+  {k_{2}}\!\!\!\!\!/=  {k_{1}}\!\!\!\!\!/
+  {k_{2}}\!\!\!\!\!/-  z_{12} {q}\!\!\!/$, one finds
\begin{eqnarray}\label{v12-=+Kv-1}
v_{\alpha q(12)}^{\; -} &=& e^{-\beta_{(12)1}}   \left(
v_{\alpha  q(1)}^{\; -}+ {1\over 2} K^{=I}_{(12)1}  \gamma^I_{qp} v_{\alpha  p (1)}^{\; +}
\right) \; , \end{eqnarray}
with
\begin{eqnarray}\label{b12-1=}
e^{-\beta_{(12)1}} &=& \sqrt{\left| \frac {\rho^{\#}_1 + e^{-2\beta_{21}}\rho^{\#}_2}{\widehat{\rho}^{\#}(z_{12})}\right|} \; , \qquad \\
\label{KI12-1=}
 K^{=I}_{(12)1}=   \frac { e^{-2\beta_{21}}\rho^{\#}_2 }{\rho^{\#}_1 + e^{-2\beta_{21}}\rho^{\#}_2}\widehat{K}{}^{=I}_{21}(z_{12})\; &,&
  \qquad\widehat{K}{}^{=I}_{21}(z_{12})= {K}{}^{=I}_{2}-\widehat{K}{}^{=I}_{1}(z_{12})= {K}{}^{=I}_{21} -z_{12}q^I\; .
  \qquad
\end{eqnarray}
Notice that to describe  the deformed frame it is convenient to relax a bit the
 gauge (\ref{K++=0gauge}) allowing for $e^{-2\beta_{(12)1}}\not= 1$.
Eqs. (\ref{v12-=+Kv-1}) implies
\begin{eqnarray}\label{u--12=Ku1}
u^=_{a(12)}(z_{12}) &=& e^{-2\beta_{(12)1}}\left(u^=_{a(1)} + \frac 1 4 (\vec{K}^{=}_{(12)1} )^2 u^\#_{a(1)}  + K^{=I}_{(12)1}u^I_{a(1)} \right)= \frac {\widehat{P}_{12}(z_{12})} { \widehat{\rho^\#}(z_{12})}
 \; , \qquad \\
 \label{u++12=Ku1}
u^\#_{a(12)}(z_{12}) &=& e^{+2\beta_{(12)1}}u^\#_{a(1)}
, \; \qquad \\
 \label{uI=K--ui2}
u^I_{a(12)}(z_{12}) &=& u^I_{a(1)}+ \frac 1 2 u^\#_{a(1)} K^{= I }_{(12)1} \; . \qquad
\end{eqnarray}

\medskip

\subsection{3- and 4- particle kinematics. Momentum conservation }

The momentum conservation condition for our four point amplitudes reads
\begin{eqnarray}\label{P1-4=0}
 \rho^{\#}_1u^{=a}_{(1)}+ \rho^{\#}_2u^{=a}_{(2)}+ \rho^{\#}_3u^{=a}_{(3)} + \rho^{\#}_4u^{=a}_{(4)}=0\; ,
 \qquad
\end{eqnarray}
Generically, this can be split into
\begin{eqnarray}\label{Pu+=0}
 && \rho^{\#}_1+ \tilde{\rho}^{\# 1}_{2} + \tilde{\rho}^{\# 1}_{3}+ \tilde{\rho}^{\# 1}_{4}=0\; ,\qquad \\
 \label{Pu-=0}
 && \tilde{\rho}^{\# 1}_{2} (\vec{K}{}^{=}_{21})^2+ \tilde{\rho}^{\# 1}_{3}(\vec{K}{}^{=}_{31})^2+ \tilde{\rho}^{\# 1}_{4}(\vec{K}{}^{=}_{41})^2 =0 \; ,\qquad \\
 \label{Pu+=I}
 && \tilde{\rho}^{\# 1}_{2} K^{=I}_{21}+ \tilde{\rho}^{\# 1}_{3}K^{=I}_{31}+ \tilde{\rho}^{\# 1}_{4}K^{=I}_{41} =0\; ,\qquad
 \qquad
\end{eqnarray}
where we have used the notation
\begin{eqnarray}\label{rij=}
\tilde{\rho}^{\# i}_{j}:=  e^{-2\beta_{ji}}\rho^{\#}_{j}
\; . \qquad
\end{eqnarray}

The three particle kinematics is more restrictive and a nontrivial solution of the momentum conservation conditions,
(\ref{Sriui--=0}) with $n=3$,
exists for deformed, complexified  momenta only.
Let us write this for the case of BCFW-deformed first momentum and using the
notation $P^a_{12}(z_{12})=\rho^{\#}(z_{12})u^{=a}(z_{12})$ for the third momentum:
\begin{eqnarray}\label{P1-3=0}
 \rho^{\#}_1\widehat{u^{=a}_{1}}+ \rho^{\#}_2u^{=a}_{2}+ {\rho^{\#}}(z_{12}){u^{=a}}(z_{12})=0\; .
 \qquad
\end{eqnarray}
In the gauge (\ref{K++=0gauge}) all the three frames are related with the reference frame by
(\ref{u--=KuI=11G}), (\ref{uI=K--uiG}), (\ref{u++=KuiG}) with complex vector parameters
\begin{eqnarray}\label{hK1=}
\widehat{K}{}^{=I}_{(1)} := \widehat{K}{}^{=I}_{(1)}(z_{12})= {K}{}^{=I}_{(1)} + \frac {z_{12}q_{(1)}^I}{\rho^{\#}_1} = {K}{}^{=I}_{(1)} + q_{(1)}^I \frac {(\overrightarrow{K}{}^{=}_{(2)}-\overrightarrow{K}{}^{=}_{(1)})^2  }{\, 2\overrightarrow{q}_{(1)} (\overrightarrow{K}{}^{=}_{(2)}-\overrightarrow{K}{}^{=}_{(1)})}
 \;  \qquad
\end{eqnarray}
and ${K}{}^{=I}_{(12)}(z_{12})$ which we are going to find now. This is to say
\begin{eqnarray}\label{hu--1=G}
&& \widehat{u}{}^=_{a1} = u^=_{a} + \widehat{K}{}^{=I}_{1}u^I_{a} + \frac 1 4 (\widehat{\vec{K}}{}^{=}_{1} )^2 u^\#_{a}
 \; , \qquad \\ \label{u--2=G}
&& u^=_{a2} = u^=_{a}   + K^{=I}_{2}u^I_{a} + \frac 1 4 (\vec{K}^{=}_{2} )^2 u^\#_{a}
 \; , \qquad \\ \label{u--12=G}
&& u^=_{a(12)}(z_{12}) = u^=_{a} + K^{=I}_{(12)}(z_{12}) u^I_{a} + \frac 1 4 (\vec{K}^{=}_{(12)}(z_{12})  )^2 u^\#_{a}
 \; . \qquad
\end{eqnarray}
Substituting these expressions into Eq. (\ref{P1-3=0}), one splits this into the set of three equations ({\it cf.} (\ref{Sri=0}), (\ref{SriKi=0}) and  (\ref{SriKi2=0}))
\begin{eqnarray}\label{P3r=0}
 && \rho^{\#}(z_{12}) +\rho^{\#}_1 +{\rho}_2^{\# }=0 \; ,\qquad  \\
 \label{P3rK=0}
 && \rho^{\#}(z_{12})   K^{=I}_{(12)}(z_{12})+\rho^{\#}_1  \widehat{K}{}^{=I}_{1}  + \rho^{\#}_2{K}{}^{=I}_{2} =0\; , \qquad \\
 \label{P3rK2=0}
 &&  \rho^{\#}(z_{12})   (\overrightarrow{K}{}^{=I}_{(12)}(z_{12}))^2+\rho^{\#}_1  (\widehat{\overrightarrow{K}}{}^{=I}_{1} )^2 + \rho^{\#}_2 (\overrightarrow{K}{}^{=I}_{2})^2=0\; , \qquad
\end{eqnarray}
which imply
\begin{eqnarray}\label{rz12=}
 && \rho^{\#}(z_{12}) =-\rho^{\#}_1 - {\rho}_2^{\# }
  \; ,\qquad  \\
 \label{K12z12=}
 &&   K^{=I}_{(12)}(z_{12})= \frac {\rho^{\#}_1}{\rho^{\#}_1 + {\rho}_2^{\# }} \;  \widehat{K}{}^{=I}_{1}  + \frac {\rho^{\#}_2
 }{\rho^{\#}_1 + {\rho}_2^{\# }} \;{K}{}^{=I}_{2}
 \; , \qquad
\end{eqnarray}
 and
 \begin{eqnarray}
 \label{hK1-K2=null}
 &&
(\widehat{\overrightarrow{K}}{}^{=}_{12} )^2=0\; , \qquad \widehat{K}{}^{=I}_{21}= ({K}{}^{=I}_{2} -\widehat{K}{}^{=I}_{1})
 \; . \qquad
\end{eqnarray}
This latter however is satisfied identically due to $\vec{q}_{(1)}^2\equiv {q}_{(1)}^I{q}_{(1)}^I =0$ (see (\ref{hK1=})).

For the application it is convenient to write (\ref{K12z12=}) also in the form
\begin{eqnarray}
\label{K12z12-hK1=}
 &&   K^{=I}_{(12)1}(z_{12}):= K^{=I}_{(12)}(z_{12})- \widehat{K}{}^{=I}_{1} = \frac {\rho^{\#}_2}{\rho^{\#}_1 + {\rho}_2^{\# }} \; \widehat{K}{}^{=I}_{21}
 \; , \qquad \\ \label{K12z12-hK2=}
 &&   K^{=I}_{(12)2}(z_{12}):= K^{=I}_{(12)}(z_{12})- {K}{}^{=I}_{2} = - \frac {\rho^{\#}_1}{\rho^{\#}_1 + {\rho}_2^{\# }} \; \widehat{K}{}^{=I}_{21}
 \; . \qquad
\end{eqnarray}
Then  (\ref{hK1-K2=null}) implies nilpotency of $K^{=I}_{(12)1}(z_{12})$ and $K^{=I}_{(12)2}(z_{12})$:
\begin{eqnarray}
\label{K12z12-2=0}
 &&   (K^{=I}_{(12)1}(z_{12}))^2=0 \; , \qquad (K^{=I}_{(12)2}(z_{12}))^2=0 \; . \qquad
\end{eqnarray}

\subsection{$t_3$ tensor }

Now we are ready to calculate the  expression for $t_3$ tensor (\ref{t3=D})  with arguments
$\widehat{k_{a(1)}}$, $k_{a(2)}$ and $-\widehat{k_{a(1)}}-k_{a(2)}$, a square of which determines, through (\ref{cAg3=D}), the
first partial amplitude, $$ {\cal A}_{z_{12} \; (I_1J_1)(I_2J_2) \; (IJ)}={\cal A}_{z_{12} \; (I_1J_1)(I_2J_2) \; (IJ)} (\widehat{k_1}, k_2;
 \widehat{\rho^{\#}}(z_{12}),  v^{-}_{q}(z_{12}))$$ in the r.h.s. of (\ref{cA4g=11D0g}). The gauge invariant expression is not simple
\begin{eqnarray}\label{t3=uuq+}
&& t^{I_1I_2I_3}(\widehat{k_{(1)}}, k_{(2)},-\widehat{k_{(1)}}-k_{(2)})   = \qquad \nonumber \\
&& \qquad   - \rho^\#_{(1)} (u^=_{(1)}u_{(2)}^{I_2}) \delta^{I_1I_3}  - z_{12} q^J_{(1)} (u^J_{(1)}u_{(2)}^{I_2}) \delta^{I_1I_3}- z_{12}  (u^{I_1}_{(1)}u_{(2)}^{I_2}) q_{(1)}^{I_3} \qquad \nonumber \\ && \qquad  +  \rho^\#_{(2)} (u^{I_1}_{(1)}u_{(2)}^{=}) (u_{(2)}^{I_2}u_{(1)}^{I_3})
 + \rho^\#_{(1)} K^{=I_3}_{(12)1} (u^{I_1}_{(1)}u_{(2)}^{I_2})  + \frac {\rho^\#_{(2)}}{2} K^{=I_1}_{(12)1} (u^{I_3}_{(1)}u_{(2)}^{I_2}) (u^{\#}_{(1)}u_{(2)}^{=})  \qquad \nonumber \\ && \qquad
+ \frac {z_{12}}{2} q_{(1)}^{I_1}(u^{\#}_{(1)}u_{(2)}^{I_2})  K^{=I_3}_{(12)1} -
\frac {(z_{12})^2}{\rho^\#_{(1)}} q_{(1)}^{I_1}
(u^{\#}_{(1)}u_{(2)}^{I_2}) q_{(1)}^{I_3}\; . \qquad  \end{eqnarray}
However, in the gauge (\ref{K++=0gauge}), and after the use of the momentum conservation conditions (\ref{rz12=}) and (\ref{K12z12=}), this simplifies
essentially:
\begin{eqnarray}\label{t3=uuq-G}
 && t^{I_1I_2I_3}(\widehat{k_{(1)}}(z_{12}), k_{(2)},-\widehat{k_{(1)}}(z_{12})-k_{(2)})   =
\nonumber \\ && \qquad =  \delta^{I_2I_3}  \rho^\#_{(2)} \widehat{K}^{=I_1}_{21}
  + \rho^\#_{(1)} \widehat{K}^{=I_2}_{21}\delta^{I_1I_3} +   \delta^{I_1I_2}  \frac { \rho^\#_{(1)}\rho^\#_{(2)}}  { \rho^\#_{(1)}+\rho^\#_{(2)}}  \widehat{K}^{=I_3}_{21}\; . \qquad  \end{eqnarray}
Here the  complex SO(9) vector
\begin{eqnarray}\label{hK21=}
 \widehat{K}^{=I}_{21}:=  \widehat{K}^{=I}_{21}(z_{12})&=&  {K}^{=I}_{2} - \widehat{K}^{=I}_{1}=
 \nonumber \\
 &=& {K}^{=I}_{2} - {K}^{=I}_{1} - q_{(1)}^I \frac {(\overrightarrow{K}{}^{=}_{(2)}-\overrightarrow{K}{}^{=}_{(1)})^2  }{\, 2\overrightarrow{q}_{(1)} (\overrightarrow{K}{}^{=}_{(2)}-\overrightarrow{K}{}^{=}_{(1)})}
 \; \qquad  \end{eqnarray}
\bigskip
is null (\ref{hK1-K2=null}) because  its square is proportional to $\vec{q}_{(1)}^2 =0$.

\subsection{Supermomentum }

In construction of superamplitudes one might want to use a sum of  $\theta^-_{q(i)}$ with different $i$-th.
The construction of this is hampered by the fact that $\theta^-_{q(i)}$'s with different values of $i$ are transformed
by different $ [SO(1,1)_i \otimes SO(D-2)_i]$ symmetry groups.

The $\prod\limits_{i=1}^n [SO(1,1)_i \otimes SO(D-2)_i]$ invariant sum of all  $\theta^-_{q(i)}$ does exist but carries a $Spin(1,D-1)$ index $\alpha$ with twice more values (32 in D=11) than $Spin(D-2)$ index $q$ (16 in D=11),
\begin{eqnarray}\label{qf:=}
\mathfrak{q}_{\alpha} &:=& \sum\limits_{i=1}^n \rho^{\#}_{(i)}
v_{\alpha q(i)}^{\; -} \theta^-_{p(i)} . \qquad
\end{eqnarray}
Following the custom of D=4 amplitude literature we call this 'supermomentum'; it is superpartner of the total momentum $\sum_{i=1}^n p_{a(i)}$ and hence is  supersymmetric invariant as far as the total momentum is conserved, $\sum_{i=1}^n p_{a(i)} =0$:
\begin{eqnarray}\label{susy-qf:=}
 \delta_\epsilon \mathfrak{q}_{\alpha} = \epsilon^\beta \sum\limits_{i=1}^n \rho^{\#}_{(i)}
v_{\alpha q(i)}^{\; -}v_{\beta q(i)}^{\; -} = \frac 1 2  \Gamma^{a}_{\alpha\beta}\epsilon^\beta \sum\limits_{i=1}^n p_{a(i)} =0
 \; .   \qquad
\end{eqnarray}

The supersymmetry invariance is equivalent to stating that, when $\sum\limits_{i=1}^n p_{a(i)} =0$, the supermomentum can be  equivalently written in the form
\begin{eqnarray}\label{qf:=-}
\mathfrak{q}_{\alpha} &:=& \sum\limits_{i=1}^n \rho^{\#}_{(i)}
v_{\alpha q(i)}^{\; -} \left(\theta^-_{p(i)}- \Xi^\beta v_{\beta q(i)}^{\; -}\right)  \qquad
\end{eqnarray}
with an arbitrary fermionic spinor $\Xi^\beta$. This spinor can identified with global supersymmetry parameter and the above equation makes transparent that this can be used to cancel the contribution of one
of the fermionic variable, say  $\theta^-_{p(1)}$. Actually,  as we will see in no time,  the supermomentum depends on $(n-2)$ linear combinations of $n$ fermionic variables.

In the gauge (\ref{K++=0gauge}) we can obtain a simple decomposition of  supermomentum  on the  auxiliary reference spinor frame,
\begin{eqnarray}\label{q-al=G}
\mathfrak{q}_{\alpha} = v_{\alpha q}^{\; -}  \sum\limits_{i=1}^n \rho^{\#}_{(i)}  \theta^-_{q(i)}  + \frac 1 2
v_{\alpha p}^{\; +}\gamma^I_{pq}
  \sum\limits_{i=1}^n \rho^{\#}_{(i)} K^{=I}_{(i)}\theta^-_{q(i)}\; .   \qquad
\end{eqnarray}
We can split this in a Lorentz covariant manner on
\begin{eqnarray}\label{q+q=G}
\mathfrak{q}^+_q &=& v^{+\alpha}_q \mathfrak{q}_{\alpha}=  \sum\limits_{i=1}^n \rho^{\#}_{(i)}  \theta^-_{q(i)}
= \sum\limits_{i=2}^n \rho^{\#}_{(i)}  (\theta^-_{q(i)} - \theta^-_{q(1)} )\; , \qquad
\\ \label{q-q=G}
\mathfrak{q}^-_q &=& v^{-\alpha}_q \mathfrak{q}_{\alpha}=
  \frac 1 2
 \gamma^I_{qp}
  \sum\limits_{i=1}^n \rho^{\#}_{(i)} K^{=I}_{(i)}\theta^-_{p(i)}=
  \frac 1 2
 \gamma^I_{qp}
  \sum\limits_{i=2}^n \rho^{\#}_{(i)} K^{=I}_{(i)}(\theta^-_{p(i)}- \theta^-_{p(1)} )\; ,   \qquad
\end{eqnarray}
where the last parts of equalities are obtained by using the momentum conservation conditions ({\it cf.} (\ref{qf:=-})).

We can also use one of the spinor frame associated to a scattering particle instead of the reference spinor frame thus obtaining the supermomentum projections
\begin{eqnarray}\label{q+qj=G}
\mathfrak{q}^+_{qj} &=& v^{+\alpha}_{qj} \mathfrak{q}_{\alpha}=  \sum\limits_{i=1, \, i\not= j}^n \rho^{\#}_{(i)}  \theta^-_{q(i)}
= \sum\limits_{i\not= j, 1 }^n \rho^{\#}_{(i)}  (\theta^-_{q(i)} - \theta^-_{q(1)} )\; , \qquad
\\ \label{q-qj=G}
\mathfrak{q}^-_{qj} &=& v^{-\alpha}_{qj} \mathfrak{q}_{\alpha}=
  \frac 1 2
 \gamma^I_{pq}
  \sum\limits_{i=1; i\not= j}^n \rho^{\#}_{(i)} K^{=I}_{(ij)}\theta^-_{q(i)}=
  \frac 1 2
 \gamma^I_{qp}
  \sum\limits_{i\not= j, 1}^n \rho^{\#}_{(i)} K^{=I}_{(ij)}(\theta^-_{p(i)}- \theta^-_{p(1)} )\; .   \qquad
\end{eqnarray}
These relations make transparent that actually the supermomentum depends on only $(n-2)$ (linear combinations) of $n$ fermionic variables.
Furthermore, the projection $\mathfrak{q}^-_{qj}$  (\ref{q-qj=G}) looks more interesting as it contains more information about momenta of scattered particle and also because it is defined with the use of spinor helicity variables only, while  $\mathfrak{q}^+_{qj}$ is defined with the use of $j$-th complementary spinor $v^{+\alpha}_{qj}$ and thus is not covariant under $K_{9(i)}$ symmetry ((\ref{K8i}) with $\dot{q}=q=1,...,16$).

\subsubsection{Invariants from the projections of supermomentum }

Using (\ref{q-q=G}) or (\ref{q-qj=G}) we can introduce the covariant delta functions for integration over 16 fermionic variables
\begin{eqnarray}\label{delta16-=}
\delta^{16}(\mathfrak{q}^-)= (\mathfrak{q}^-)^{\wedge 16} := \frac 1 {16!} \epsilon^{q_1\ldots q_{16}} \mathfrak{q}^-_{q_1} \ldots \mathfrak{q}^-_{q_{16}}\; ,
\end{eqnarray}
which has the weight -16 under auxiliary $SO(1,1)_0$ gauge symmetry.

Notice also the existence of weight -8 SO(9) invariant
\begin{eqnarray}\label{q8=}
(\mathfrak{q}^-)^{\wedge 8}:= \frac 1 {c_8} \epsilon^{\, I_1I_2\ldots \, I_{9}} \, (\mathfrak{q}^- \gamma^{I_1I_2 I_3}\mathfrak{q}^-)\, (\mathfrak{q}^- \gamma^{I_4 I_5}\mathfrak{q}^-)\, \ldots (\mathfrak{q}^- \gamma^{I_8 I_9}\mathfrak{q}^-)\,\; .
\end{eqnarray}
In it we prefer to chose  the coefficient
\begin{eqnarray}\label{c8=}
{c_8} = \sqrt{\frac 1 {16!} \epsilon^{\, I_1I_2\ldots \, I_{9}} \,  \epsilon^{\, J_1J_2\ldots \, J_{9}} \, \epsilon^{q_1\ldots q_{16}}  \gamma^{I_1I_2 I_3}_{q_1q_2}\gamma^{I_4 I_5}_{q_3q_4}\, \ldots \gamma^{I_8 I_9}_{q_7q_8} \gamma^{J_1J_2 J_3}_{q_9q_{10}}\gamma^{J_4 J_5}_{q_{11}q_{12}}\, \ldots
\gamma^{J_8 J_9}_{q_{15}q_{16}} }\, 
\end{eqnarray}
which makes (\ref{q8=})  the exact square root of the fermionic delta function
\begin{eqnarray}\label{q8q8=q16}
(\mathfrak{q}^-)^{\wedge 8} \, (\mathfrak{q}^-)^{\wedge 8} = (\mathfrak{q}^-)^{\wedge 16}\equiv \delta^{16}(\mathfrak{q}^-) \; .
\end{eqnarray}
One can also introduce a weight -4 supersymmetric invariant
\begin{eqnarray}\label{q4=qgg2}
(\mathfrak{q}^-)^{\wedge 4} = \frac 1 {c_4}   (\mathfrak{q}^-\gamma^{IJ}\mathfrak{q}^-) \; (\mathfrak{q}^-\gamma^{IJ}\mathfrak{q}^-)
 \; .
\end{eqnarray}

Notice that in D=10 the counterpart of (\ref{q-q=G})
\begin{eqnarray}\label{q-q=G=10D}
\mathfrak{q}^-_{\dot{q}} &=& v^{-\alpha}_{\dot{q}} \mathfrak{q}_{\alpha}=
  \frac 1 2
 \gamma^I_{q\dot{q}}
  \sum\limits_{i=1}^n \rho^{\#}_{(i)} K^{=I}_{(i)}\theta^-_{q(i)}=
  \frac 1 2
 \gamma^I_{q\dot{q}}
  \sum\limits_{i=2}^n \rho^{\#}_{(i)} K^{=I}_{(i)}(\theta^-_{q(i)}- \theta^-_{q(1)} )\; ,   \qquad
\end{eqnarray}
carries the dotted index $\dot{q}=1,...,8$ and
the counterpart of (\ref{q4=qgg2}) will be playing the role of square root of the SO(8) invariant  fermionic delta function, while the counterpart of (\ref{q8=}),
\begin{eqnarray}\label{q8=10D}
D= 10 :  \quad  (\mathfrak{q}^-)^{\wedge 8}:= \frac 1 {c'_8}  \epsilon^{\, I_1I_2\ldots \, I_{8}} \, (\mathfrak{q}^- \tilde{\gamma}^{I_1I_2 }q^-)\, (\mathfrak{q}^- \tilde{\gamma}^{I_3 I_4}\mathfrak{q}^-)\, \ldots (\mathfrak{q}^- \tilde{\gamma}^{I_7 I_8}\mathfrak{q}^-)= \delta^8(\mathfrak{q}^-) \,\;
\end{eqnarray}
with an appropriate choice of $c'_8$, coincides with this fermionic delta function.

\bigskip

\subsubsection{Invariants from the complete spinorial supermomentum }

One can also construct some Lorentz invariant combinations of the complete supermomentum with Lorentz group spinor indices.
In $D=10$ the product of two copies  of fermionic spinors belongs to {\bf 120}  antisymmetric tensor representation of the Lorentz group. This is to say
$ \mathfrak{q}_{\alpha}\mathfrak{q}_{\beta}= \propto  (\mathfrak{q})^{\wedge 2}_{abc}{\sigma}^{abc}_{\alpha\beta}
$ with
\begin{eqnarray}\label{q2L=10D}
D= 10 :  \quad
 (\mathfrak{q})^{\wedge 2}_{abc} :=  \mathfrak{q}_{\alpha}\tilde{\sigma}_{abc}^{\alpha\beta}\mathfrak{q}_{\beta}\; ,
\end{eqnarray}
so that there exists a unique 4-th order invariant
\begin{eqnarray}\label{q4L=10D}
D= 10 :  \quad  (\mathfrak{q})^{\wedge 4\, (s)}= (\mathfrak{q})^{\wedge 2\; abc} (\mathfrak{q})^{\wedge 2}_{abc}\; .
\end{eqnarray}
This is the trace of the forth  order second rank tensor
\begin{eqnarray}\label{q4Lt=10D}
&&
 (\mathfrak{q})^{\wedge 4\, (t)}{}_a{}^b:=   (\mathfrak{q})^{\wedge 2}_{acd}  (\mathfrak{q})^{\wedge 2\; bcd} \;
\end{eqnarray}
which can be also used to form invariants of $4m$ degree with  $m=2,3,4$,
\begin{eqnarray}\label{q8L=10D}
D= 10 &:&  \quad  (\mathfrak{q})^{\wedge 4m\, (s)}= (\mathfrak{q})^{\wedge 4\, (t)}{}_a{}^{b_1}\ldots  (\mathfrak{q})^{\wedge 4\, (t)}{}_{b_{m-1}}{}^a=:
Tr ((\mathfrak{q})^{\wedge 4\, (t)})^{m}\; , \qquad m=2,3,4 . \qquad
\end{eqnarray}
Clearly, $m=4$ invariant is proportional to delta function of the spinorial supermomentum, while $m=2$ invariant can be considered as square root of this latter,
\begin{eqnarray}\label{q8L2=delta}
D= 10 &:&  \quad   (\mathfrak{q})^{\wedge 8\, (s)}\,  (\mathfrak{q})^{\wedge 8\, (s)}\propto  (\mathfrak{q})^{\wedge 16\, (s)}\propto \delta^{16}( \mathfrak{q}_{\alpha})  \; . \qquad
\end{eqnarray}

In $D=11$ the product  of two copies of a fermionic spinor decomposes on three irreducible representations, {\bf 32}$\times${\bf 32}= {\bf 1} + {\bf 165}+ {\bf 330}. This is to say, their exists a second order invariant
\begin{eqnarray}\label{q2Ls=11D}
D= 11 :  \quad
 (\mathfrak{q})^{\wedge 2 (s)} :=  \mathfrak{q}_{\alpha}C^{\alpha\beta}\mathfrak{q}_{\beta}\; , \qquad
\end{eqnarray}
and  two second order tensors constructed from the spinorial supermomentum bilinears
\begin{eqnarray}\label{q2L=11D}
D= 11 :  \quad
 (\mathfrak{q})^{\wedge 2}_{abc} :=  \mathfrak{q}_{\alpha}\tilde{\Gamma}_{abc}^{\alpha\beta}\mathfrak{q}_{\beta}\; , \qquad  (\mathfrak{q})^{\wedge 2}_{abcd} :=  \mathfrak{q}_{\alpha}\tilde{\Gamma}_{abcd}^{\alpha\beta}\mathfrak{q}_{\beta}\; . \qquad
\end{eqnarray}
This can be used to construct invariants of higher order. In particular there exist two 4-th order invariants,
\begin{eqnarray}\label{q4L=11D}
D= 10 :  \quad  (\mathfrak{q})^{\wedge 4\, (s1)}= (\mathfrak{q})^{\wedge 2\; abc} (\mathfrak{q})^{\wedge 2}_{abc}\; , \qquad (\mathfrak{q})^{\wedge 4\, (s2)}= (\mathfrak{q})^{\wedge 2\; abcd} (\mathfrak{q})^{\wedge 2}_{abcd}\;
\end{eqnarray}
and  the following
 interesting 6-th order invariant
\begin{eqnarray}\label{q6Ls=11D}
D= 11 :  \quad
 (\mathfrak{q})^{\wedge 6(s)} :=  \epsilon^{a_1\ldots a_{11}}  (\mathfrak{q})^{\wedge 2}_{a_1a_2 a_3}(\mathfrak{q})^{\wedge 2}_{a_4a_5 a_6a_7}(\mathfrak{q})^{\wedge 2}_{a_8a_9 a_{10}a_{11}}\; . \qquad
\end{eqnarray}
We also have  two second rank fourth order tensors, $(\mathfrak{q})^{\wedge 4\, (t1)}{}_b{}^a=  (\mathfrak{q})^{\wedge 2}_{bcd} (\mathfrak{q})^{\wedge 2\; acd}$ and $(\mathfrak{q})^{\wedge 4\, (t2)}{}_b{}^a= (\mathfrak{q})^{\wedge 2}_{bcde} (\mathfrak{q})^{\wedge 2\; acde} $ the  traces of which give
(\ref{q4L=11D}),  so that zoo of the 11D supersymmetric and Lorentz invariants constructed from supermomentum is even reacher that the 10D one.
 In particular there exists an interesting  (although not unique) 8-order invariant
\begin{eqnarray}\label{q8Ls=11D}
D= 11 :  \quad
 (\mathfrak{q})^{\wedge 8(s)} :=  \epsilon^{a_1\ldots a_{11}}  (\mathfrak{q})^{\wedge 2}_{a_1}{}^{bc}  (\mathfrak{q})^{\wedge 2}_{bc a_2 a_3} (\mathfrak{q})^{\wedge 2}_{a_4a_5 a_6a_7}(\mathfrak{q})^{\wedge 2}_{a_8a_9 a_{10}a_{11}}\;  \qquad
\end{eqnarray}
and a number of 16-th order invariants, all of which can be treated as square roots of the fermionic delta function
$\delta^{32}( \mathfrak{q}_{\alpha})$.

\section{ Studying the candidate  BCFW-type recurrent relations for 4-particle amplitudes of 10D SYM}
\label{4-fermions}

To check consistency and completeness of
our BCFW deformation and candidate BCFW-type recurrent relations, in this section  we will try to obtain on their basis 4-particle amplitudes of 10D SYM with 4 and 2 fermionic legs. Unfortunately, this calculation indicates a problem: the candidate amplitudes obtained from D=10 BCFW procedure suffer an unwanted dependence on the deformation vectors. We nevertheless find useful to describe these calculations as they might suggest a resolution of the issue which is probably general for higher dimensional BCFW-type constructions.

\subsection{Candidate BCFW for 4-fermionic amplitude in 10D SYM from super-BCFW for 4-point superamplitude}

The candidate BCFW relation for the  four fermionic  superamplitude of 10D SYM is
\begin{eqnarray}\label{cA4f=10Ds}
&& {\cal A}_{\dot{q}_1\dot{q}_2\dot{q}_3 \dot{q}_4} ({k_1},\theta^-_{(1)};k_2 , \theta^-_{(2)};k_3 , \theta^-_{(3)};{k_4},\theta^-_{(4)})= \hspace{7cm}\nonumber \\ & &\quad =  \frac{1}{16 (\widehat{\rho}{}^{\#}(z_{12}))^2} \left(D^+_{{q}(z_{12})}\left(
{\cal A}_{z_{12} \; \dot{q}_1\dot{q}_2\dot{p}}(\widehat{k_1}, \widehat{\theta^-_{(1)}};k_2, \theta^-_{(2)};\widehat{P_{12}}(z_{12}),\Theta^-) \right. \right. \times
\qquad  \nonumber \\ \nonumber
&&  \left. \left. \qquad
\times \frac {1}{(P_{12})^2} \overleftrightarrow{D}{}^+_{{q}(z_{12})} {\cal A}_{z_{12}\; \dot{p}\dot{q}_3 \dot{q}_4}(-\widehat{P_{12}}(z_{12}),\Theta^-; k_{3}, \theta^-_{(3)}; \widehat{k_4}, \widehat{\theta^-_{(4)}})\right)\right)_{\Theta^-_q=0}
 \nonumber \\ & &\quad -  \frac{1}{16 (\widehat{\rho}{}^{\#}(z_{13}))^2} \left(D^+_{{q}(z_{13})}\left(
{\cal A}_{z_{13} \;\dot{q}_1\dot{q}_3\dot{p}}(\widehat{k_1}, \widehat{\theta^-_{(1)}};k_3, \theta^-_{(3)};\widehat{P_{13}}(z_{13}),\Theta^-) \right. \right. \times
\qquad \nonumber  \\
&&  \left. \left. \qquad
\times \frac {1}{(P_{13})^2} \overleftrightarrow{D}{}^+_{{q}(z_{13})} {\cal A}_{z_{13}\; \dot{p} \dot{q}_2\dot{q}_4}(-\widehat{P_{13}}(z_{13}),\Theta^-; k_{2}, \theta^-_{(2)}; \widehat{k_4}, \widehat{\theta^-_{(4)}})\right)\right)_{\Theta^-_q=0}
\; .
\end{eqnarray}
After applying the covariant derivatives, this expression can be written in the form
 \begin{eqnarray}\label{cA4f=10s}
&& \hspace{-1cm} {\cal A}_{\dot{q}_1\dot{q}_2\dot{q}_3 \dot{q}_4} ({k_1},\theta^-_{(1)};k_2 , \theta^-_{(2)};k_3 , \theta^-_{(3)};{k_4},\theta^-_{(4)})= \hspace{7cm}\nonumber \\ &  =& \;
 - 2\; {\cal A}_{z_{12} \; \dot{q}_1\dot{q}_2\dot{p}}(\widehat{k_1}, \widehat{\theta^-_{(1)}};k_2, \theta^-_{(2)};
 \widehat{\rho^{\#}}(z_{12}),  v^{-}_{q}(z_{12}), 0)    \nonumber \\ \nonumber
&&  \qquad
\times \frac {1}{(P_{12})^2 {\widehat{\rho}{}^{\#}(z_{12})}}  {\cal A}_{z_{12}\; \dot{p}\, \dot{q}_3\dot{q}_4}(-\widehat{\rho^{\#}}(z_{12}),  v^{-}_{q}(z_{12}), 0; k_{3}, \theta^-_{(3)}; \widehat{k_4}, \widehat{\theta^-_{(4)}})
  \\ && +
  {\cal A}_{z_{12} \; \dot{q}_1\dot{q}_2\; I}(\widehat{k_1}, \widehat{\theta^-_{(1)}};k_2, \theta^-_{(2)};
 \widehat{\rho^{\#}}(z_{12}),  v^{-}_{q}(z_{12}), 0)    \nonumber \\ \nonumber
&&  \qquad
\times \frac {1}{(P_{12})^2}  {\cal A}_{z_{12}\; I\; \dot{q}_3\dot{q}_4}(- \widehat{\rho^{\#}}(z_{12}),  v^{-}_{q}(z_{12}), 0; k_{3}, \theta^-_{(3)}; \widehat{k_4}, \widehat{\theta^-_{(4)}}) -
(2\longleftrightarrow 3)\; .
\end{eqnarray}

In the assumption that amplitudes are reproduced as leading components of superamplitudes,
the BCFW relation for the 4-fermionic amplitude of 10D SYM reads
 \begin{eqnarray}\label{cA4f=10D}
&& \hspace{-1cm} {\cal A}_{\dot{q}_1\dot{q}_2\dot{q}_3 \dot{q}_4} ({k_1};k_2;k_3 ;{k_4})= \hspace{7cm}\nonumber \\ &  =& \;
   {\cal A}_{z_{12} \; \dot{q}_1\dot{q}_2\; I}(\widehat{k_1};k_2, ;
 \widehat{\rho^{\#}}(z_{12}),  v^{-}_{q}(z_{12}))    \nonumber \\
&&  \qquad
\times \frac {1}{(P_{12})^2 }  {\cal A}_{z_{12}\; I\, \dot{q}_3\dot{q}_4}(-\widehat{\rho^{\#}}(z_{12}),  v^{-}_{q}(z_{12}); k_{3}, ; \widehat{k_4})
 -
(2\longleftrightarrow 3)\; ,
\end{eqnarray}
where we have taken into account that the amplitudes of the processes with odd number of fermions vanish, in particular
$ {\cal A}_{z_{12} \; \dot{q}_1\dot{q}_2\; \dot{p}}(\widehat{k_1};k_2, ;
 \widehat{\rho^{\#}}(z_{12}),  v^{-}_{q}(z_{12}))\equiv 0$.

\bigskip

\subsection{3-point amplitudes with two fermionic particles in 10D SYM}

In the case of 10D SYM  the expression for tree 3-point amplitudes with two fermionic legs is suggested by light-cone string vertices of   \cite{Schwarz:1982jn}. In  our notation it reads
\begin{eqnarray}\label{cAffI=10D}
&&  {\cal A}_{\dot{q}_1\; \dot{q}_2\; I }
(\rho^{\#}_{(1)} , {v}{}^-_{q(1)} ; \rho^{\#}_{(2)} , {v}{}^-_{q(2)} ; \rho^{\#}_{3} , {v}{}^-_{q3})=   \frac 1 4 {\rho^\#_1 \rho^\#_2}v^{-\alpha}_{\dot{q}_1(1)} v^{-\beta}_{\dot{q}_2(2)}\sigma^a_{\alpha\beta}
u_{a(3)}^I \qquad \nonumber \\  &&\qquad \; =\;   \frac 1 2 {\rho^\#_1 \rho^\#_2}\gamma^I_{p\dot{p}}\left(v^{-\alpha}_{\dot{q}_1(1)} {v}{}^{\; -}_{\alpha p(3)}\; v^{-\beta}_{\dot{q}_2(2)} v^{\; +}_{\beta \dot{p}(3)}+v^{-\alpha}_{\dot{q}_1(1)} v^{\; +}_{\alpha \dot{p}(3)}\; v^{-\beta}_{\dot{q}_2(2)}{v}{}^{\; -}_{\beta p(3)} \right)
\; .\qquad
\end{eqnarray}
In distinction to \cite{Schwarz:1982jn}, here we assume that $\rho^\#_i$, $v^{-\alpha}_{\dot{q}_i(i)}$ obey the momentum conservation; so that the spinor frame variables should be complex/deformed. The form of the amplitude, the multipliers and coefficients in (\ref{cAffI=10D}) can be checked by Ward identities.

Indeed, in the gauge (\ref{K++=0gauge}) the expression (\ref{cAffI=10D})  simplifies to
\begin{eqnarray}\label{cAffI=10G}
&& {\cal A}_{\dot{q}_1\; \dot{q}_2\; I }
(\rho^{\#}_{(1)} , {v}{}^-_{q(1)} ; \rho^{\#}_{(2)} , {v}{}^-_{q(2)} ; \rho^{\#}_{(3)} , {v}{}^-_{q3})=
 \frac 1 4 {\rho^\#_1 \rho^\#_2} \gamma^I_{p\dot{p}}\left( K^{=J}_{31}\gamma^J_{p\dot{q}_1} \delta_{\dot{q}_2\dot{p}} + K^{=J}_{32}\gamma^J_{p\dot{q}_2} \delta_{\dot{q}_1\dot{p}}\right)
\;  \qquad
\end{eqnarray}
and it is straightforward to check that this expression obeys the Ward identity (\ref{Ward10D=A3fff}). Furhtermore, it obeys
 (\ref{Ward10D=A3I1}) which also fixes the form of purely bosonic 3-point amplitude to be
\footnote{To be precise, (\ref{cAffI=10G}) obeys the $\propto \gamma^{IJK}_{q\dot{q}_3}$ part of also (\ref{Ward10D=A3I1}), while its  $\propto \gamma^{I}_{q\dot{q}_3}$ part fixes the form of ${\cal A}_{I_1\; I_2\; I_3 }$.}
\begin{eqnarray}\label{cAIII=10G}
 {\cal A}_{I_1\; I_2\; I_3 } &=&  -\left( \delta^{I_1I_2}\rho^{\#}_{1}K_{13}^{=I_3} +\delta^{I_2I_3}\rho^{\#}_{2}K_{21}^{=I_1}+ \delta^{I_3I_1}\rho^{\#}_{3}K_{32}^{=I_2} \right)=-
  t^{I_1I_2I_3 }
\; . \qquad
\end{eqnarray}
The last part of (\ref{cAIII=10G}) is given by the projection of the $t$-tensor \cite{Schwarz:1982jn}, (\ref{t3=D}).

In (\ref{t3=D}) $u_{a(i)}^{I_i}$ are considered to be polarization vectors of $i$-th particle and the momenta
are assumed to be proportional to the light-like  vectors $u_{a(i)}^=$ from the same frame (\ref{harmUi=}),
$k_{ai}=\rho^\#_iu_{ai}^=$. Then in the gauge (\ref{v-=v+Kv=G}), where
 (\ref{u--=KuI=11G}), (\ref{uI=K--uiG}), (\ref{u++=KuiG})  hold, one finds the gauge fixed form of
$t^{I_1I_2I_3}$ indicated in (\ref{cAIII=10G}).

\subsection{Testing the candidate BCFW by calculating  4-gluino amplitude. Problem of dependence on deformation vector}

As a test of our candidate BCFW relation (\ref{cA4f=10D}), let us use it to calculate the gauge fixed expression for the  tree 4--fermion amplitudes of 10D SYM.  Substituting  (\ref{cAffI=10G}),   one finds
\begin{eqnarray}\label{cA4f=10D=}
&& \hspace{-0.5cm} {\cal A}_{\dot{q}_1\dot{q}_2\dot{q}_3 \dot{q}_4} ({k_1};k_2;k_3 ;{k_4})= \frac{\sqrt{\rho^\#_1\rho^\#_2\rho^\#_3\rho^\#_4}}{8}\hspace{4cm}\nonumber \\ &  \times & \;\left(
\frac {\widehat{{K}^{=J}_{21}}(z_{12})\widehat{ {K}^{=K}_{43}}(z_{12})}{ \sqrt{(\vec{K}^{=}_{21})^2(\vec{K}^{=}_{43})^2}}
 \; \left( (\tilde{\gamma}^J\gamma^I)_{\dot{q}_1\dot{q}_2} - \frac{2\rho^\#_1}{\rho^\#_1+\rho^\#_2}\;   \delta^{JI}\delta_{\dot{q}_1\dot{q}_2}\right)\;
 \left((\tilde{\gamma}^K\gamma^I)_{\dot{q}_3\dot{q}_4} - \frac{2\rho^\#_4 }{\rho^\#_3+\rho^\#_4}\;  \delta^{KI}\delta_{\dot{q}_3\dot{q}_4}\right) \right.\;\nonumber \\&& \;
\left. - \frac {\widehat{{K}^{=J}_{31}}(z_{13}) \widehat{{K}^{=K}_{42}}(z_{13})}{\sqrt{(\vec{K}^{=}_{31})^2(\vec{K}^{=}_{42})^2}}
 \; \left( (\tilde{\gamma}^J\gamma^I)_{\dot{q}_1\dot{q}_3} - \frac{ 2\rho^\#_1 }{\rho^\#_1+\rho^\#_3}\; \delta^{JI} \delta_{\dot{q}_1\dot{q}_3}\right)\;
 \left((\tilde{\gamma}^K\gamma^I)_{\dot{q}_2\dot{q}_4} - \frac{2\rho^\#_4 }{\rho^\#_2+\rho^\#_4}\;  \delta^{KI}\delta_{\dot{q}_2\dot{q}_4}\right) \right)  \; .\nonumber \\
 {}
\end{eqnarray}
Here
\begin{eqnarray}\label{hK21=F}
\widehat{K}{}^{=I}_{21}(z_{12}) = {K}{}^{=I}_{21} - \frac {z_{12}q_{(1)}^I}{\rho^{\#}_1} = {K}{}^{=I}_{21} - q_{(1)}^I \frac {(\overrightarrow{K}{}^{=}_{21})^2 }{\, 2\overrightarrow{q}_{(1)} \overrightarrow{K}{}^{=}_{21}}
 \;  , \qquad \widehat{K}{}^{=I}_{31}(z_{13}) =  {K}{}^{=I}_{31} - q_{(1)}^I \frac {(\overrightarrow{K}{}^{=}_{31})^2 }{\, 2\overrightarrow{q}_{(1)} \overrightarrow{K}{}^{=}_{31}}
 \;  , \qquad \nonumber \\ {}\\
 \label{hK43=F}\widehat{K}{}^{=I}_{43}(z_{12}) = {K}{}^{=I}_{43} - \frac {z_{12}q_{(4)}^I}{\rho^{\#}_4} = {K}{}^{=I}_{43} - q_{(4)}^I \frac {(\overrightarrow{K}{}^{=}_{21})^2 }{\, 2\overrightarrow{q}_{(1)} \overrightarrow{K}{}^{=}_{21}}
 \;  , \qquad  \widehat{K}{}^{=I}_{42}(z_{13}) =  {K}{}^{=I}_{42} - q_{(4)}^I \frac {(\overrightarrow{K}{}^{=}_{31})^2 }{\, 2\overrightarrow{q}_{(1)} \overrightarrow{K}{}^{=}_{31}}
 \;  , \qquad \nonumber \\ {}
\end{eqnarray}
and $q_{(1)}^I$, $q_{(4)}^I$ are projectors of the same complex null ten-vector (\ref{qq=0=}) on the first and fourth frames, respectively,
$$ q^a= u^{aI}_1q_{(1)}^I=u^{aI}_4 q_{(4)}^I\; , \qquad  q^aq_a=0\; , \qquad  q^au_{a1}^==0=  q^au_{a4}^=\; . $$
The SO(8) complex  vectors $q_{(1)}^I$ and $q_{(4)}^I$ has vanishing squares (\ref{q=uiqi1}) which implies that the vectors (\ref{hK21=F}) and (\ref{hK43=F})  are also null.

The above calculation  demonstrates either incompleteness  of our prescription for BCFW deformation or a need to improve our candidate BCFW recurrent relations. The problem is that the amplitude calculated with the candidate  BCFW relations happens to be apparently  dependent on the deformation null-vector $q^a$ obeying (\ref{qq=0=}). On one hand, using the experience of the 4D calculation, we might expect a necessity to specify some particular solution of (\ref{qq=0=}) in terms of spinor helicity/spinor frame variables corresponding to 1-st and $n$-th ($4$-th) particle. (We will discuss such a possibility below).

On the other hand, this might indicate that a change of
prescription or modification of the proposed candidate BCFW type recurrent relations is necessary.
Notice that we did not prove the validity of these relations, but only proposed them as a reasonable candidate.  Moreover, the general D-dimensional arguments  of \cite{ArkaniHamed:2008yf,Cheung:2008dn} did not allow to prove the existence of BCFW relations for n point n--fermionic amplitudes: they argued their existence for a particular contractions with the deformation vectors (!) of an amplitude with at least one gluon in the case of YM and at least one graviton in the case of gravity (chosen to have a deformed momentum to this end).
Thus the first thing to check now is what will be the situation with such an amplitude in 10D SYM case.

\subsection{From the candidate BCFW relation to an expression for 10D SYM amplitude with two bosonic and two fermionic legs.
Deformation vector dependence.  }

Let us try to reproduce the 4 point tree amplitude with two bosonic and two fermionic legs from the candidate BCFW relation for 10D SYM. We start from the
general relation (which has not been proved to be valid but are obtained from the reasonable candidate (\ref{cA-Sf=rBCFW}))
  \begin{eqnarray}\label{cA2f2b=10D}
&& \hspace{-1cm} {\cal A}_{I_1\dot{q}_2\dot{q}_3 I_4} ({k_1};k_2;k_3 ;{k_4})= \hspace{7cm}\nonumber \\ &  =& \;
   2 {\cal A}_{z_{12} \; I_1\dot{q}_2\dot{q}}(\widehat{k_1};k_2 ; \widehat{\rho^{\#}}(z_{12}),  v^{-}_{q}(z_{12}))    \nonumber \\
&&  \qquad
\times \frac {1}{(P_{12})^2 \widehat{\rho^\# } (z_{12})}  {\cal A}_{z_{12}\;\dot{q}\dot{q}_3 I_4}(-\widehat{\rho^{\#}}(z_{12}),  v^{-}_{q}(z_{12}); k_{3} ; \widehat{k_4})
 -
(2\longleftrightarrow 3)\; .
\end{eqnarray}
Let us stress that, although it is not reflected in the notation for {\it  l.h.s.}, the {\it  r.h.s.} of this relation apparently depends on the deformation vector $q^a= -q^I_{(1)}u^{aI}_1 =-q^I_{(4)}u^{aI}_4$ (\ref{q=uiqi1}). This is the origin of a problem which is possibly common for higher dimensional
generalizations of BCFW relations: for  $D>4$ the deformation vector is not fixed uniquely\footnote{\label{q=coset} The freedom in choice of a normalized complex null vector defined up to a phase transformation can be associated to the coset
$\frac {SO(D-2)}{SO(D-4)\otimes U(1)}$ of dimension $2(D-4)$. See e.g. \cite{Bandos:2017zap}.}.
In $D=4$ case the deformation vector is constructed from the helicity spinors of two particles with deformed momenta in a unique way \cite{Britto:2005fq} and the problem does not occur.

Substituting the expressions (\ref{cAffI=10D}) (the cyclic property of the amplitudes implies
${\cal A}_{\dot{q}_1\; \dot{q}_2\; I }(1,2,3)={\cal A}_{I\dot{q}_1\; \dot{q}_2\;}(3,1,2)$) and using
 (\ref{P122=})  and $\widehat{v^{-\alpha}_{\dot{q}}}(z_{12}) \widehat{v^{-\beta}_{\dot{q}}}(z_{12})=\frac {1} {2} \widehat{u^=_a}(z_{12})\tilde{\sigma}^{a\alpha\beta}$
(see (\ref{u==v-v-})) for the contraction of the variables corresponding to intermediate state,
we arrive at
\begin{eqnarray}\label{cA2f2b=10D=}
&& \hspace{-1cm} {\cal A}_{I_1\dot{q}_2\dot{q}_3 I_4} ({k_1};k_2;k_3 ;{k_4}\vert q)= \hspace{7cm}\nonumber \\
&  =& \;  \frac 1 {16s}\, \rho^{\#}_{(2)} \rho^{\#}_{(3)}
\widehat{u^{I_1}_{a1}}v^{-\alpha}_{\dot{q}_22} (\sigma^a\tilde{\sigma}^b{\sigma}^c)_{\alpha\beta}v^{-\beta}_{\dot{q}_33}\widehat{u^=_b}(z_{12})\widehat{u^{I_4}_{c4}}
- (2\leftrightarrow 3)
\; . \qquad
\end{eqnarray}
Here we have explicitly written  the dependence of {\it l.h.s.} on the complex deformation null--vector (\ref{q=uiqi1}). This enters  $\widehat{u^=_b}(z_{12})= \frac 1 {\rho^{\#}_1+\rho^{\#}_2} (\widehat{k_{1a}}+k_{2a})$ and also $\widehat{u^{I_1}_{a1}}$  and $\widehat{u^{I_1}_{a1}}$ (\ref{BCWF=huI}) (which
are also taken at $z=z_{12}$).
Furthermore, as one can see from (\ref{z12=}), a contraction of the deformation vector also enters the denominator of $z_{12}$.

At this point we would like to exploit  the result of \cite{ArkaniHamed:2008yf,Cheung:2008dn} stating that the BCFW type recurrent relations are valid for a particular type of amplitude. Particularly, they are valid for calculation of some amplitudes of the processes with vector particles, one of which, say the first, is chosen to have a deformed momentum. More specifically, BCFW relations are valid for contraction of such  amplitude  with the deformation vector on its first vector index.
In our formalism this corresponds to the contraction of the first $SO(D-2)$ vector index of the amplitude with  $q_{(1)}^{I_1}$ component of the deformation vector. Thus a potentially valid contraction of the relation  (\ref{cA2f2b=10D=}) reads
\begin{eqnarray}\label{qcA2f2b=10D}
&& \hspace{-1cm} q_{(1)}^{I_1}{\cal A}_{I_1\dot{q}_2\dot{q}_3 I_4} ({k_1};k_2;k_3 ;{k_4}\vert q)= \hspace{7cm}\nonumber \\
&  =& \;  \frac 1 {16s}\, \rho^{\#}_{(2)} \rho^{\#}_{(3)}
q_a v^{-\alpha}_{\dot{q}_22} (\sigma^a\tilde{\sigma}^b{\sigma}^c)_{\alpha\beta}v^{-\beta}_{\dot{q}_33}\widehat{u^=_b}(z_{12})\widehat{u^{I_4}_{c4}}
- (2\leftrightarrow 3)
\;  \qquad \nonumber \\ &  =& \; - \frac 1 {16s}\, \frac {\rho^{\#}_{(2)} \rho^{\#}_{(3)}} {\rho^{\#}_{1} +\rho^{\#}_{2}}
\; q_a v^{-\alpha}_{\dot{q}_22} (\sigma^a\tilde{\sigma}^b{\sigma}^c)_{\alpha\beta}v^{-\beta}_{\dot{q}_33}(k_{b1}+k_{b2})\left(u^{I_4}_{c4} -
\frac {2z_{12}}{2\rho^{\#}_4}q_{(4)}^{I_4}u^{\#}_{c4}
\right)
- (2\leftrightarrow 3)
\; .\qquad
\end{eqnarray}
The passage to the second equality in (\ref{qcA2f2b=10D}) uses the momentum conservation in the first amplitude in the r.h.s. of (\ref{cA2f2b=10D}), and the fact that the contribution of the deformation vector in $\widehat{k_{b1}}$ vanishes identically ($q_a  (\sigma^a\tilde{\sigma}^b{\sigma}^c)q_b\equiv 0$).
In \cite{ArkaniHamed:2008yf,Cheung:2008dn} the interpretation of the relations similar to (\ref{qcA2f2b=10D}) is based on identification  of the deformation vector with polarization vector of the first particle.

Now, the 'covariantization' by the method of \cite{Cheung:2009dc}, namely restoration of the ($SO(D-2)$) covariant amplitude by just extracting a coefficient contracted with $q_{(1)}^{I_1}$, na\"{\i}vly would give us
\begin{eqnarray}\label{cA2f2b=cA+q}
&& \hspace{-1cm} {\cal A}_{I_1\dot{q}_2\dot{q}_3 I_4} ({k_1};k_2;k_3 ;{k_4}\vert q)=
{\cal A}_{I_1\dot{q}_2\dot{q}_3 I_4} ({k_1};k_2;k_3 ;{k_4})- \frac {2z_{12}}{2\rho^{\#}_4}q_{(4)}^{J}D^{\# J}_4 {\cal A}_{I_1\dot{q}_2\dot{q}_3 I_4} ({k_1};k_2;k_3 ;{k_4})
\; ,\qquad
\end{eqnarray}
where $D^{\# J}_i = u^{\#}_{ai}\frac \partial {\partial u^{I}_{ai}} - u^{I}_{ai}\frac \partial {\partial u^{=}_{ai}}$ is one of the harmonic covariant derivatives and
\begin{eqnarray}\label{cA2f+2b=q0}
&& \hspace{-1cm} {\cal A}_{I_1\dot{q}_2\dot{q}_3 I_4} ({k_1};k_2;k_3 ;{k_4})= {\cal A}_{I_1\dot{q}_2\dot{q}_3 I_4} ({k_1};k_2;k_3 ;{k_4}\vert 0)=  \qquad \nonumber \\ &  =& \; - \frac 1 {16s}\, \frac {\rho^{\#}_{(2)} \rho^{\#}_{(3)}} {\rho^{\#}_{1} +\rho^{\#}_{2}}
\; u^{I_1}_{a1}  v^{-\alpha}_{\dot{q}_22} (\sigma^a\tilde{\sigma}^b{\sigma}^c)_{\alpha\beta}v^{-\beta}_{\dot{q}_33}(k_{b1}+k_{b2})u^{I_4}_{c4}
- (2\leftrightarrow 3)
\; .\qquad
\end{eqnarray}
Just this part  of our BCFW amplitude has the structure close to the one of the matrix elements described in \cite{Schwarz:1982jn}.

However, the presence of the second term in (\ref{cA2f2b=cA+q}) brings us back to the problem of an unwanted dependence on a deformation vector $q^a$, which, in contradistinction to 4D case, is not fixed uniquely neither in our formalism, nor in generic discussion of
higher dimensional, $D>4$  BCFW relations presented in \cite{ArkaniHamed:2008yf,Cheung:2008dn}.

Coming back to  (\ref{qcA2f2b=10D}) where the deformation vector is considered as polarization vector of the first particle does not resolve the issue. As for $D>4$ the properties of the deformation vector do not fix it uniquely,
the apparent problem is how to treat the freedom in its  choice. If we assume that (\ref{qcA2f2b=10D})  is valid for an arbitrary consistent choice of the deformation vector, the (appropriately defined\footnote{\label{q=coset2} In the light of the statement in footnote \ref{q=coset}, this should be an appropriate covariant derivative on the coset $\frac {SO(D-2)}{SO(D-4)\otimes U(1)}$.  }) derivative of  (\ref{qcA2f2b=10D}) with respect to $q^I_{(1)}$ gives us an additional relation for the amplitudes. Probably in such a way one can reach an algorithmic formulation of the above mentioned covariantization procedure suggested in \cite{Cheung:2009dc}, but consistency of this is to be investigated.

Perhaps  this problem can be solved by exploiting the relation of the present approach with the analytic superamplitude formalism proposed recently in \cite{Bandos:2017zap}.
Indeed, as it was shown in conclusion of  \cite{Bandos:2017zap}, in its frame the freedom in choosing the deformation vector can be reduced to one complex number, like it is the case for  the deformations of Weyl spinors used in the study of 4D superamplitudes. To be more precise, the deformation  vector is expressed  there in terms of complex spinor helicity variables constructed form the spinor frame variables and internal harmonic variables parametrizing
$SO(D-2)/[SO(D-4)\otimes U(1)]$ coset, and thus it can be identified with a coordinate of such a coset. This brings us back to the problem of whether one should allow to differentiate the deformation vector in (\ref{qcA2f2b=10D}) to obtain a new relation for the amplitude, which have to be investigated.

This and, more generally, the use of interplay between  the present constrained and the analytic superamplitude formalisms for their mutual development  will be the subject of future work.

\bigskip

\section{Conclusion and outlook}

In this paper we have developed spinor helicity formalism for 11D supergravity (SUGRA) and  on-shell superfield formalism  for tree amplitudes of 11D SUGRA and   10D SYM. Another  superamplitude formalism for 10D SYM was proposed some years ago in \cite{CaronHuot:2010rj}. It was based on Clifford superfields and looks quite nonminimal and difficult to use. In contrast, the 10D spinor helicity formalism of \cite{CaronHuot:2010rj} coincide with ours; in this paper we clarify its structure and the nature of constrained spinor helicity variables.

The observation that the basic variables of the 10D spinor helicity formalism of \cite{CaronHuot:2010rj} coincide with spinor moving frame variables (Lorentz harmonics), allowed us to develop a more economic superfield description of 10D SYM amplitudes and also to obtain the 11D generalization of the spinor helicity and superamplitude formalisms.
Our superamplitudes are multiparticle generalizations of constrained superfields describing the linearized 11D SUGRA and 10D SYM which were found in  \cite{Galperin:1992pz}. They obey a number of equations, hence the name of constrained superamplitude formalism which we use for our approach.

We have  shown how the constrained on-shell superfields of \cite{Galperin:1992pz} can be obtained from quantization of 11D and 10D massless superparticle mechanics. Actually we have used the massless superparticle mechanics and its quantization  as a guide for the development of the superamplitude formalism.

The constrained $n$-particle  superamplitudes of 10D SYM (11D SUGRA) depend on the set of spinor helicity variables as well as on the  set of fermionic variables $\theta^-_{qi}$, $q=1,..,8$ ($q=1,..,16$), $i=1,..,n$. The spinor helicity variables include
the set of constrained spinors
$v_{\alpha q i}^-$, $\alpha=1,...,16$ ($\alpha=1,...,32$), spinor frame variables or  Lorentz harmonics, parametrizing celestial sphere ${\bb S}^8$ (${\bb S}^9$) and densities $\rho^\#_i$, $i=1,...,n$, which are allowed to be negative.
We have described the set of equations, which are imposed on constrained superamplitudes and restrict their dependence on $\theta^-_{qi}$.

We have made some stages towards calculation of superamplitudes and amplitudes in this formalism. In particular, we have found a gauge fixed expressions for the Lorentz harmonics which can be considered as a covariant counterpart of the light cone gauge and promises to be a very useful tool for the calculation of amplitudes and superamplitudes. We have also considered the consequences of the momentum conservation and described supermomentum, the fermionic superpartner of total momentum, in the spinor frame formalism.

We have obtained the supersymmetric Ward identities for 10D SYM and 11D SUGRA amplitudes and  used that to check our guess for 3-point 10D amplitude with two fermions. Such a calculation is simplified in the above  described  gauge fixed on the spinor frame variables.

We have also  discussed a natural candidate for the BCFW-type recurrent relations for our constrained  superamplitudes. Setting all the fermionic variables equal to zero, we reduce these candidate  BCFW-type  relations for superamplitudes to the relations for amplitudes.  As a check of completeness and consistency, we used the above 3-point amplitudes and the candidate BCFW relations to obtain a gauge fixed form of the 10D 4-fermion tree amplitude and also a covariant form of the  4 point amplitude with 2 fermionic and 2 bosonic legs.

This simple calculation however showed a problem indicating either incompleteness of the higher dimensional  BCFW deformation prescription or a need for improvement of our candidate  BCFW recurrent relations.
The above 4-point tree amplitudes calculated with their use suffer an  apparent dependence  on a deformation null-vector, which, in contradistinction to 4D case, is defined with a big degree of arbitrariness in D=10 and D=11.

One can observe that the deformation null-vector $q^a=q^I_{(1)}u_{1}^{aI}$ can be associated with breaking of the
$SO(D-2)_1$ gauge symmetry of the real spinor frame formalism down to its $SO(D-4)\times U(1)$ subgroup. Similar symmetry breaking  occurs inevitably in the analytic superfield approach of \cite{Bandos:2017zap}. It is related to the appearance of complex structure characteristic for the analytic superamplitude formalism and complex spinor frame variables used in it. As it was shown in \cite{Bandos:2017zap}, the BCFW deformation of complex spinor frame and fermionic variables of the analytic superfield formalism can be quite naturally  fixed uniquely up to a single complex number $z$, i.e. can reproduce the structure similar to the known from 4D case.
This suggests that, perhaps, a solution of the above discussed problem with candidate BCFW relations of the constrained superamplitude formalism can be found by elaborating its interrelation with the analytic superamplitude approach.

This issue is clearly the first in the que of problems to be addressed for further development of our approach to tree superamplitudes in D=10 and D=11, which will be the subject of forthcoming work. Upon solving it, it would be also natural to search for a generalization of our formalism for the case of loop superamplitudes\footnote{Despite of neither 10D SYM nor 11D supergravity is expected to be finite, a number of interesting problems can be addressed by studying their amplitudes: see for instance \cite{Green:1997as,Deser:2000xz} as well as recent \cite{Green:2016tfs} and refs. therein.}.

Another interesting direction of study
is to approach the constrained superamplitude formalism by quantizing  11D and 10D Green-Schwarz type counterparts of the so-called ambitwistor string
\footnote{Ambitwistor string
\cite{Mason:2013sva,Adamo:2014wea}
is known to reproduce the scattering equation of Cachazo, He and Yuan (CHY) \cite{Cachazo:2013hca,Adamo:2013tsa}. Originally the ambitwistor string was formulated as a Nevieu-Scwarz-Ramond (NSR) type model in d=10. However, as it was discussed in \cite{Bandos:2014lja}, at the classical level this model is equivalent to the so-called null-superstring  and thus (see \cite{Bandos:2006af}) to the twistor string \cite{Witten:2003nn,Berkovits:2004hg,Siegel:2004dj};
hence it should allow for  a consistent quantization in an arbitrary number of dimensions. The D=4 ambitwistor string models have been  developed in \cite{Geyer:2014fka,Lipstein:2015vxa,Bork:2017qyh,Farrow:2017eol}. See also \cite{Heslop:2016plj,Casali:2016atr,Casali:2017zkz,Adamo:2017sze} for related results.}. This   might lead us to a convenient superfield form of the  scattering equation approach \cite{Cachazo:2013hca,Adamo:2013tsa}.

\acknowledgments{
This work has been supported in part by the
Spanish Ministry of Economy, Industry and Competitiveness (MINECO) grant FPA 2015-66793-P, which is partially financed with FEDER/ERDF (European Regional Development Fund of the European
Union), by the Basque Government Grant IT-979-16, and the Basque Country University program UFI 11/55.

The author is thankful to the Theoretical Department of CERN (Geneva, Switzerland),
to the Galileo Galilei Institute for Theoretical Physics and the INFN (GGI, Florence, Italy),  to the organizers of the GGI program ''Supergravity: what next?'', and especially to Toine Van Proyeen,  as well as to the Simons Center for Geometry and Physics, Stony Brook University (New York, US) for the hospitality and partial support of his visits at certain stages of this work.

 The author is grateful to
Dima Sorokin and  Emeri Sokatchev  for useful discussions and suggestions,  and to Luis Alvarez-Gaume, Paolo Di Vecchia and Boris Piolin
for useful discussions on related topics.
}

\bigskip

\appendix

\setcounter{equation}0
\def\theequation{A.\arabic{equation}}

\section{Clifford superfield version of D=10 superamplitudes and fermionic  BCFW deformation }

\subsection{SUSY generator representation with Clifford algebra element}

In this Appendix we will show how the Clifford superfield form of the on-shell superfield formalism, a one-particle counterpart to the approach to 10D superamplitudes proposed by Carron-Huot and O'Connell in \cite{CaronHuot:2010rj}, and its 11D generalization, can be obtained from covariant quantization of the massless superparticle model in its spinor moving frame formulation.

As we have discussed in sec.
\ref{11Dsuperparticle},
 In the 'analytic basis' the 11D massless superparticle has only second class fermionic constraints (\ref{d+q})  which obey
\begin{eqnarray}\label{d+d+=PB}
\{ d^+_q, d^+_p\}_{_{P.B.}} =-4i\rho^\# \delta_{qp}\; . \end{eqnarray}
 If we pass to the Dirac brackets, the fermionic coordinate variable will obey
\begin{eqnarray}\label{t-t-=DB} {} \{ \theta^-_q , \theta^-_p\}_{_{D.B.}} = - \frac {i}{4\rho^{\#}}\delta_{qp}, \end{eqnarray}
so that after quantization
the algebra of fermionic operators $\hat{\theta}^-_q$ is
\begin{eqnarray}\label{t-t-=Cl}{} \{ \hat{\theta}^-_q , \hat{\theta}^-_p\}  = \frac {1}{4\rho^{\#}}\delta_{qp}\; . \end{eqnarray}
Thus they can be identified with the 16-dimensional  Clifford algebra generators
\begin{eqnarray}\label{th=G=}
\hat{\theta}^-_q= \frac {1} {\sqrt{2\rho^\# } }\; \mathfrak{C}_{q } \; , \qquad \{ \mathfrak{C}_{q } , \mathfrak{C}_{p }\}= 2\delta_{qp}\; .
\end{eqnarray}
The superfield formalism by Carron-Huot and O'Connell is constructed by considering wavefunctions  dependent on Clifford variable $\mathfrak{C}_{q }$ (we call this ''Clifford superfield approach'').

The realization of the D=10 supersymmetry generator on the state with light-like momentum $k_a$, used in \cite{CaronHuot:2010rj} can be written as  (see (\ref{Q=10D}))
\begin{eqnarray}\label{Q=rvG=10D}
Q_\alpha = \sqrt{\rho^{\#}} v_{\alpha q}^{\; -}\; \mathfrak{C}_{{q}}
\; . \qquad
\end{eqnarray}
Here
$v_{\alpha q}{}^-$ are the square roots of $k_a$ as defined by (\ref{k=pv-v-}), which implies $k_a = {1\over 8}\rho^{\#}v^-_{{q}}\tilde{\sigma}v^-_{{q}}$. It is easy to check that
\begin{eqnarray}\label{QQ=10D} {} \{ Q_\alpha , Q_\beta \} = 2  \rho^{\#} v_{\alpha q}{}^- v_{\beta q}{}^-= \sigma^a_{\alpha\beta}k_a\; . \qquad
\end{eqnarray}
The same equations with $ \alpha, \beta =1,...,32$, $\quad q,p=1,..,16$ and $\sigma^a\mapsto \Gamma^a$ with $a=0,1,...,10$ describe the realization of 11D
supersymmetry generator in terms of 16 dimensional Clifford algebra and the homogeneous coordinate of
${\bb S}^9$.

In D=10 case, one can realize these generators on fourier images of the fields
 $w^I=w^I(x^=, v_{\alpha q}^{\; -})$ and
$\psi_q=\psi_q(x^=, v_{\alpha q}^{\; -})$ (see
(\ref{Fab=kuwI}),
(\ref{chi10D=vpsi}),
(\ref{susy=8+8a}))
\begin{eqnarray}\label{Cliff=8+8}
\mathfrak{C}_{{q}} \psi_{\dot q }(\rho^{\#}) = \frac 1 { \sqrt{\rho^{\#}}}  \gamma^I_{q\dot{q}}  \; w^I(\rho^{\#}) \; , \qquad \mathfrak{C}_{{q}} w^I(\rho^{\#}) =   \gamma^I_{q\dot{q}}    \sqrt{\rho^{\#}}  \psi_{\dot q }(\rho^{\#})\; \; . \qquad
\end{eqnarray}

\bigskip

\label{susyAlg}

We can also  write the formal supersymmetry generator acting on the states of $n$ on-shell particles and on n-point amplitudes  \cite{CaronHuot:2010rj}:
\begin{eqnarray}\label{Q=10D} Q_\alpha = \sum\limits_{i=1}^n \sqrt{\rho^{\#}_{(i)}} v_{\alpha  q(i)}^{\; -}\; \mathfrak{C}_{ q{(i)}}
\; . \qquad
\end{eqnarray}
Here  $\mathfrak{C}_{q (i)}$
are $n$ sets of generators of 8d Clifford algebra (16d Clifford algebra in the case of D=11),
\begin{eqnarray}\label{GG=Ii}
{} \{ \mathfrak{C}_{{q}(i)} \, , \mathfrak{C}_{{p}(j)}  \} = 2   \delta_{(i)(j)} \delta_{{q}{p}} {\cal I}\; .  \qquad
\end{eqnarray}

It is easy to check that these supersymmetry generator obey the usual supersymmetry algebra (\ref{QQ=10D}),
\begin{eqnarray}\label{QQ=10D=P} {} \{ Q_\alpha , Q_\beta \} = 2  P\!\!\!/{}_{\alpha\beta}\; , \qquad
P\!\!\!/{}_{\alpha\beta}= \sigma^a_{\alpha\beta}P_a = \sum\limits_{i=1}^n {\rho^{\#}_{(i)}}  v_{\alpha q(i)}^{\; -} v_{\beta  q(i)}^{\; -}\;  \qquad
\end{eqnarray}
the right hand side of which contains the total momentum, the sum of momenta of all the scattered particle.
Notice that this vanishes due to the momentum conservation in scattering processes, so that
on the mass shell the 10D supersymmetry generator (\ref{Q=10D}) is just nilpotent,
\begin{eqnarray}\label{QQ=0}
P_a= \sum\limits_{i=1}^n k_{a(i)}=0\; \qquad \Rightarrow \qquad   \{ Q_\alpha , Q_\beta \} =0\; .
\end{eqnarray}

\subsection{Clifford superfield version of D=10  BCFW deformation }

\label{CliffSF}

The first generalization of the BCFW--type deformation of 10D on-shell superamplitudes was proposed in \cite{CaronHuot:2010rj}, where the amplitudes depending on the spinor helicity variables (which are, as we have seen,  essentially the spinor moving frame variables, $v_{\alpha q(i)}$)
 and Clifford algebra variables $\mathfrak{C}_{q}$ were considered. We call these
${\cal A}(\rho^{\#}_{(1)},  v_{\alpha q(1)}, \mathfrak{C}_{(1)}; \ldots, \rho^{\#}_{(n)},  v_{\alpha q(n)},  \mathfrak{C}_{(n)})$  Clifford superamplitudes.

The BCFW-type deformation of 10D spinor helicity variables from
\cite{CaronHuot:2010rj} can be reproduced from the deformation of spinor frame variables described in  (\ref{BCWF=vnM}) and  (\ref{BCWF=v1M}).
Here for completeness we present (in our notation) also the deformation of the Clifford algebra variables proposed in  \cite{CaronHuot:2010rj}.
It  reads
\begin{eqnarray}\label{BCFW=fn}
\widehat{ \mathfrak{C}_{{p}(n)}}= \mathfrak{C}_{{p}(n)}+ z \mathfrak{C}_{{q}(1)} {\bb M}_{{q}{p}} = e^{{\frac z2 \mathfrak{C}_{{q}(1)} {\bb M}_{{q}{p}} \mathfrak{C}_{{p}(n)}}}\mathfrak{C}_{{p}(n)}  e^{-{\frac z2 \mathfrak{C}_{{q}(1)} {\bb M}_{{q}{p}} \mathfrak{C}_{{p}(n)}}}\; , \qquad \\ \label{BCFW=f1}
\widehat{ \mathfrak{C}_{{q}(1)}}= \mathfrak{C}_{{q}(1)}- z  {\bb M}_{{q}{p}} \mathfrak{C}_{{p}(n)} = e^{{\frac z2 \mathfrak{C}_{{q}(1)} {\bb M}_{{q}{p}} \mathfrak{C}_{{p}(n)}}} \mathfrak{C}_{{q}(1)}  e^{{-\frac z2 \mathfrak{C}_{{q}(1)} {\bb M}_{{q}{p}} \mathfrak{C}_{{p}(n)}}}\; , \qquad \end{eqnarray}
where the matrix ${\bb M}_{{q}{p}} $ is nilpotent, (\ref{MMT=0}).


\begin{thebibliography}{99}
\renewcommand{\theequation}{R.\arabic{equation}}


\bibitem{Witten:2003nn}
  E.~Witten,
  ``Perturbative gauge theory as a string theory in twistor space,''
  Commun.\ Math.\ Phys.\  {\bf 252} (2004) 189
  [hep-th/0312171].




\bibitem{Penrose:1967wn}
  R.~Penrose,
  ``Twistor algebra,''
  J.\ Math.\ Phys.\  {\bf 8} (1967) 345.
  doi:10.1063/1.1705200

\bibitem{Penrose:1972ia}
  R.~Penrose and M.~A.~H.~MacCallum,
  ``Twistor theory: An Approach to the quantization of fields and space-time,''
  Phys.\ Rept.\  {\bf 6} (1972) 241.
  doi:10.1016/0370-1573(73)90008-2

\bibitem{Penrose:1986ca}
  R.~Penrose and W.~Rindler, ``Spinors And Space-time. Volume 1 , Two-Spinor Calculus and Relativistic Fields,'' Cambridge University Press 1984;
  ``Spinors And Space-time. Vol. 2: Spinor And Twistor Methods In Space-time Geometry,''  Cambridge University Press 1986.

\bibitem{Ferber:1977qx}
  A.~Ferber,
  ``Supertwistors and Conformal Supersymmetry,''
  Nucl.\ Phys.\ B {\bf 132} (1978) 55.
  doi:10.1016/0550-3213(78)90257-2


\bibitem{Shirafuji:1983zd}
  T.~Shirafuji,
  ``Lagrangian Mechanics of Massless Particles With Spin,''
  Prog.\ Theor.\ Phys.\  {\bf 70} (1983) 18.
  doi:10.1143/PTP.70.18


\bibitem{Witten:1978xx}
  E.~Witten,
  ``An Interpretation of Classical Yang-Mills Theory,''
  Phys.\ Lett.\ B {\bf 77} (1978) 394.
  doi:10.1016/0370-2693(78)90585-3

\bibitem{Bern:2011qn}
  Z.~Bern, J.~J.~Carrasco, L.~J.~Dixon, H.~Johansson and R.~Roiban,
  ``Amplitudes and Ultraviolet Behavior of N = 8 Supergravity,''
  Fortsch.\ Phys.\  {\bf 59} (2011) 561
  doi:10.1002/prop.201100037
  [arXiv:1103.1848 [hep-th]].

\bibitem{Drummond:2008vq}
  J.~M.~Drummond, J.~Henn, G.~P.~Korchemsky and E.~Sokatchev,
  ``Dual superconformal symmetry of scattering amplitudes in N=4 super-Yang-Mills theory,''
  Nucl.\ Phys.\ B {\bf 828} (2010) 317
  doi:10.1016/j.nuclphysb.2009.11.022
  [arXiv:0807.1095 [hep-th]].

\bibitem{Drummond:2009fd}
  J.~M.~Drummond, J.~M.~Henn and J.~Plefka,
  ``Yangian symmetry of scattering amplitudes in N=4 super Yang-Mills theory,''
  JHEP {\bf 0905} (2009) 046
  doi:10.1088/1126-6708/2009/05/046
  [arXiv:0902.2987 [hep-th]].


\bibitem{Eden:2011ku}
  B.~Eden, P.~Heslop, G.~P.~Korchemsky and E.~Sokatchev,
  Nucl.\ Phys.\ B {\bf 869} (2013) 378
  doi:10.1016/j.nuclphysb.2012.12.014
  [arXiv:1103.4353 [hep-th]].

\bibitem{Kallosh:2012yy}
  R.~Kallosh and T.~Ortin,
  ``New E77 invariants and amplitudes,''
  JHEP {\bf 1209} (2012) 137
  doi:10.1007/JHEP09(2012)137
  [arXiv:1205.4437 [hep-th]].

\bibitem{Elvang:2015rqa}
  H. Elvang and Y.t. Huang, Scattering Amplitudes in Gauge Theory and Gravity. Cambrodge: CUP, 2015.

\bibitem{ArkaniHamed:2017}
  N. Arkani-Hamed, J.L. Bourjaily, F. Cachazo, A.B. Goncharov, A. Postnikov and  J. Trnka,  Grassmannian Geometry of Scattering Amplitudes.  Cambridge: CUP, 2015, 194pp.




\bibitem{Britto:2005fq}
  R.~Britto, F.~Cachazo, B.~Feng and E.~Witten,
  ``Direct proof of tree-level recursion relation in Yang-Mills theory,''
  Phys.\ Rev.\ Lett.\  {\bf 94} (2005) 181602
  doi:10.1103/PhysRevLett.94.181602
  [hep-th/0501052].




\bibitem{Britto:2004ap}
  R.~Britto, F.~Cachazo and B.~Feng,
  ``New recursion relations for tree amplitudes of gluons,''
  Nucl.\ Phys.\ B {\bf 715} (2005) 499
  doi:10.1016/j.nuclphysb.2005.02.030
  [hep-th/0412308].

\bibitem{Bianchi:2008pu}
  M.~Bianchi, H.~Elvang and D.~Z.~Freedman,
  JHEP {\bf 0809} (2008) 063
  doi:10.1088/1126-6708/2008/09/063
  [arXiv:0805.0757 [hep-th]].


\bibitem{ArkaniHamed:2008gz}
  N.~Arkani-Hamed, F.~Cachazo and J.~Kaplan,
  ``What is the Simplest Quantum Field Theory?,''
  JHEP {\bf 1009} (2010) 016
  doi:10.1007/JHEP09(2010)016
  [arXiv:0808.1446 [hep-th]].

\bibitem{Brandhuber:2008pf}
  A.~Brandhuber, P.~Heslop and G.~Travaglini,
  ``A Note on dual superconformal symmetry of the N=4 super Yang-Mills S-matrix,''
  Phys.\ Rev.\ D {\bf 78} (2008) 125005
  doi:10.1103/PhysRevD.78.125005
  [arXiv:0807.4097 [hep-th]].


\bibitem{Heslop:2016plj}
  P.~Heslop and A.~E.~Lipstein,
  ``On-shell diagrams for $ \mathcal{N} $ = 8 supergravity amplitudes,''
  JHEP {\bf 1606} (2016) 069
  doi:10.1007/JHEP06(2016)069
  [arXiv:1604.03046 [hep-th]].

\bibitem{Herrmann:2016qea}
  E.~Herrmann and J.~Trnka,
  ``Gravity On-shell Diagrams,''
  JHEP {\bf 1611} (2016) 136
  doi:10.1007/JHEP11(2016)136
  [arXiv:1604.03479 [hep-th]].


\bibitem{Hodges:2009hk}
  A.~Hodges,
  ``Eliminating spurious poles from gauge-theoretic amplitudes,''
  JHEP {\bf 1305} (2013) 135
  doi:10.1007/JHEP05(2013)135
  [arXiv:0905.1473 [hep-th]].



\bibitem{Mason:2009qx}
  L.~J.~Mason and D.~Skinner,
  ``Dual Superconformal Invariance, Momentum Twistors and Grassmannians,''
  JHEP {\bf 0911} (2009) 045
  doi:10.1088/1126-6708/2009/11/045
  [arXiv:0909.0250 [hep-th]].


\bibitem{CaronHuot:2010rj}
  S.~Caron-Huot and D.~O'Connell,
  ``Spinor Helicity and Dual Conformal Symmetry in Ten Dimensions,''
  JHEP {\bf 1108} (2011) 014
  [arXiv:1010.5487 [hep-th]].

\bibitem{Boels:2012ie}
  R.~H.~Boels and D.~O'Connell,
  ``Simple superamplitudes in higher dimensions,''
  JHEP {\bf 1206} (2012) 163
  [arXiv:1201.2653 [hep-th]].




\bibitem{Boels:2012zr}
  R.~H.~Boels,
  ``Maximal R-symmetry violating amplitudes in type IIB superstring theory,''
  Phys.\ Rev.\ Lett.\  {\bf 109} (2012) 081602
  [arXiv:1204.4208 [hep-th]].




\bibitem{Wang:2015jna}
  Y.~Wang and X.~Yin,
  ``Constraining Higher Derivative Supergravity with Scattering Amplitudes,''
  arXiv:1502.03810 [hep-th].

\bibitem{Wang:2015aua}
  Y.~Wang and X.~Yin,
  ``Supervertices and Non-renormalization Conditions in Maximal Supergravity Theories,''
  arXiv:1505.05861 [hep-th].


\bibitem{Bandos:2016tsm}
  I.~Bandos, ``Britto-Cachazo-Feng-Witten-{}Type recurrent relations for tree amplitudes of $D=$ 11 supergravity,''
  Phys.\ Rev.\ Lett.\  {\bf 118} (2017) no.3,  031601
  [arXiv:1605.00036 [hep-th]].


\bibitem{Galperin:1991gk}
  A.~S.~Galperin, P.~S.~Howe and K.~S.~Stelle,
  ``The Superparticle and the Lorentz group,''
  Nucl.\ Phys.\ B {\bf 368} (1992) 248
  [hep-th/9201020].

\bibitem{Delduc:1991ir}
  F.~Delduc, A.~Galperin and E.~Sokatchev,
  ``Lorentz harmonic (super)fields and (super)particles,''
  Nucl.\ Phys.\ B {\bf 368} (1992) 143.

\bibitem{Bandos:1996ju}
  I.~A.~Bandos and A.~Y.~Nurmagambetov,
  ``Generalized action principle and extrinsic geometry for N=1 superparticle,''
  Class.\ Quant.\ Grav.\  {\bf 14} (1997) 1597
  [hep-th/9610098].

\bibitem{Uvarov:2015rxa}
  D.~V.~Uvarov,
  Class.\ Quant.\ Grav.\  {\bf 33}, no. 13, 135010 (2016)
  doi:10.1088/0264-9381/33/13/135010
  [arXiv:1506.01881 [hep-th]].

\bibitem{Galperin:1992pz}
  A.~S.~Galperin, P.~S.~Howe and P.~K.~Townsend,
  ``Twistor transform for superfields,''
  Nucl.\ Phys.\ B {\bf 402} (1993) 531.



\bibitem{Bandos:2006nr}
  I.~A.~Bandos, J.~A.~de Azcarraga and D.~P.~Sorokin,
  ``On D=11 supertwistors, superparticle quantization and a hidden SO(16) symmetry of supergravity,''  in: "Quantum, Super and Twistors, Proc. XXII Max Born Symposium, Wroclaw (Poland) 2006", Eds: J. Kowalski-Glikman and Ludwik Turko,  Wroclaw University Press  2008, pp. 25-32 [hep-th/0612252].

\bibitem{Bandos:2007mi}
  I.~A.~Bandos,
  ``Spinor moving frame, M0-brane covariant BRST quantization and intrinsic complexity of the pure spinor approach,''
  Phys.\ Lett.\ B {\bf 659} (2008) 388
  [arXiv:0707.2336 [hep-th]].

\bibitem{Bandos:2007wm}
  I.~A.~Bandos,
  ``D=11 massless superparticle covariant quantization, pure spinor BRST charge and hidden symmetries,''
  Nucl.\ Phys.\ B {\bf 796} (2008) 360
  [arXiv:0710.4342 [hep-th]].


\bibitem{BZ-str}
 I.~A.~Bandos and A.~A.~Zheltukhin,
 {\it Spinor Cartan moving n hedron, Lorentz harmonic formulations of superstrings, and kappa symmetry},
  JETP Lett.\  {\bf 54} (1991) 421--424;
\\ I.~A.~Bandos and A.~A.~Zheltukhin,
{\it Green-Schwarz   superstrings in spinor moving frame formalism},
  Phys.\ Lett.\  {\bf B288}, 77-83 (1992).
\bibitem{Bandos:1992ze}
  I.~A.~Bandos and A.~A.~Zheltukhin,
  ``Twistor-like approach in the Green-Schwarz D=10 superstring theory,''
  Phys.\ Part.\ Nucl.\  {\bf 25} (1994) 453-477 [{\it D = 10 superstring:
Lagrangian and Hamiltonian mechanics in twistor-like Lorentz harmonic formulation}, Preprint IC-92-422, ICTP, Trieste, 1992,
81pp.].

\bibitem{BZ-M2}
I.~A.~Bandos and A.~A.~Zheltukhin, ``Generalization of Newman-Penrose dyads in
connection with the action integral for supermembranes in an eleven-dimensional
space'',   JETP Lett.\ {\bf 55} (1992) 81 [Pisma Zh.\ Eksp.\ Teor.\ Fiz.\ {\bf 55}
(1992) 81 ].
\\
I.~A.~Bandos and A.~A.~Zheltukhin, ``Eleven-dimensional supermembrane in a spinor
moving repere formalism'', Int.\ J.\ Mod.\ Phys.\ A {\bf 8}, 1081 (1993);

\bibitem{BZ-p}
I.~A.~Bandos and A.~A.~Zheltukhin, ``N=1 super-p-branes in twistor - like Lorentz
harmonic formulation'', Class.\ Quant.\ Grav.\ {\bf 12}, 609 (1995)
[arXiv:hep-th/9405113].

\bibitem{Sorokin:1989zi}
  D.~P.~Sorokin, V.~I.~Tkach and D.~V.~Volkov,
  ``Superparticles, Twistors and Siegel Symmetry,''
  Mod.\ Phys.\ Lett.\ A {\bf 4} (1989) 901.
  doi:10.1142/S0217732389001064

\bibitem{Delduc:1992fk}
  F.~Delduc, A.~Galperin, P.~S.~Howe and E.~Sokatchev,
  ``A Twistor formulation of the heterotic D = 10 superstring with manifest (8,0) world sheet supersymmetry,''
  Phys.\ Rev.\ D {\bf 47} (1993) 578
  doi:10.1103/PhysRevD.47.578
  [hep-th/9207050].

\bibitem{Delduc:1992fc}
  F.~Delduc, E.~Ivanov and E.~Sokatchev,
  ``Twistor like superstrings with D = 3, D = 4, D = 6 target superspace and N=(1,0), N=(2,0), N=(4,0) world sheet supersymmetry,''
  Nucl.\ Phys.\ B {\bf 384} (1992) 334
  doi:10.1016/0550-3213(92)90470-V
  [hep-th/9204071].

\bibitem{Sorokin:1999jx}
  D.~P.~Sorokin,
  ``Superbranes and superembeddings,''
  Phys.\ Rept.\  {\bf 329} (2000) 1
  doi:10.1016/S0370-1573(99)00104-0
  [hep-th/9906142].



\bibitem{Bandos:1995zw}
  I.~A.~Bandos, D.~P.~Sorokin, M.~Tonin, P.~Pasti and D.~V.~Volkov,
  ``Superstrings and supermembranes in the doubly supersymmetric geometrical approach,''
  Nucl.\ Phys.\ B {\bf 446} (1995) 79
  [hep-th/9501113].

\bibitem{Howe:1996yn}
  P.~S.~Howe and E.~Sezgin,
  Phys.\ Lett.\ B {\bf 394} (1997) 62
  doi:10.1016/S0370-2693(96)01672-3
  [hep-th/9611008].

\bibitem{Bandos:1997ui}
  I.~A.~Bandos, K.~Lechner, A.~Nurmagambetov, P.~Pasti, D.~P.~Sorokin and M.~Tonin,
  ``Covariant action for the superfive-brane of M theory,''
  Phys.\ Rev.\ Lett.\  {\bf 78} (1997) 4332
  doi:10.1103/PhysRevLett.78.4332
  [hep-th/9701149].

\bibitem{Aganagic:1997zq}
  M.~Aganagic, J.~Park, C.~Popescu and J.~H.~Schwarz,
  ``World volume action of the M theory five-brane,''
  Nucl.\ Phys.\ B {\bf 496} (1997) 191
  doi:10.1016/S0550-3213(97)00227-7
  [hep-th/9701166].



\bibitem{Bandos:2017zap}
  I.~Bandos,
  ``An analytic superfield formalism for tree superamplitudes in D=10 and D=11,''
  JHEP {\bf 1805} (2018) 103
  doi:10.1007/JHEP05(2018)103
  [arXiv:1705.09550 [hep-th]].





\bibitem{Mandelstam:1982cb}
  S.~Mandelstam,
  ``Light Cone Superspace and the Ultraviolet Finiteness of the N=4 Model,''
  Nucl.\ Phys.\ B {\bf 213} (1983) 149.
  doi:10.1016/0550-3213(83)90179-7

\bibitem{Brink:1982pd}
  L.~Brink, O.~Lindgren and B.~E.~W.~Nilsson,
  ``N=4 Yang-Mills Theory on the Light Cone,''
  Nucl.\ Phys.\ B {\bf 212} (1983) 401.
  doi:10.1016/0550-3213(83)90678-8

\bibitem{Brink:1982wv}
  L.~Brink, O.~Lindgren and B.~E.~W.~Nilsson,
  ``The Ultraviolet Finiteness of the N=4 Yang-Mills Theory,''
  Phys.\ Lett.\  {\bf 123B} (1983) 323.
  doi:10.1016/0370-2693(83)91210-8

\bibitem{Green:1983hw}
  M.~B.~Green, J.~H.~Schwarz and L.~Brink,
  ``Superfield Theory of Type II Superstrings,''
  Nucl.\ Phys.\ B {\bf 219} (1983) 437.
  doi:10.1016/0550-3213(83)90651-X

\bibitem{Howe:1983sr}
  P.~S.~Howe, K.~S.~Stelle and P.~K.~Townsend,
  ``Miraculous Ultraviolet Cancellations in Supersymmetry Made Manifest,''
  Nucl.\ Phys.\ B {\bf 236} (1984) 125.
  doi:10.1016/0550-3213(84)90528-5

\bibitem{Bossard:2009sy}
  G.~Bossard, P.~S.~Howe and K.~S.~Stelle,
  ``The Ultra-violet question in maximally supersymmetric field theories,''
  Gen.\ Rel.\ Grav.\  {\bf 41} (2009) 919
  doi:10.1007/s10714-009-0775-0
  [arXiv:0901.4661 [hep-th]].

\bibitem{Howe:1980th}
  P.~S.~Howe and U.~Lindstrom,
  ``Higher Order Invariants in Extended Supergravity,''
  Nucl.\ Phys.\ B {\bf 181} (1981) 487.
  doi:10.1016/0550-3213(81)90537-X

\bibitem{Kallosh:1980fi}
  R.~E.~Kallosh,
  ``Counterterms in extended supergravities,''
  Phys.\ Lett.\  {\bf 99B} (1981) 122.
  doi:10.1016/0370-2693(81)90964-3


\bibitem{Berkovits:2000fe}
  N.~Berkovits,
  ``Super Poincare covariant quantization of the superstring,''
  JHEP {\bf 0004} (2000) 018
  doi:10.1088/1126-6708/2000/04/018
  [hep-th/0001035].

\bibitem{Berkovits:2004px}
  N.~Berkovits,
  ``Multiloop amplitudes and vanishing theorems using the pure spinor formalism for the superstring,''
  JHEP {\bf 0409} (2004) 047
  doi:10.1088/1126-6708/2004/09/047
  [hep-th/0406055].

\bibitem{Berkovits:2006vi}
  N.~Berkovits and N.~Nekrasov,
  ``Multiloop superstring amplitudes from non-minimal pure spinor formalism,''
  JHEP {\bf 0612} (2006) 029
  doi:10.1088/1126-6708/2006/12/029
  [hep-th/0609012].

\bibitem{Berkovits:2017ldz}
  N.~Berkovits and H.~Gomez,
  ``An Introduction to Pure Spinor Superstring Theory,''
  Math.\ Phys.\ Stud.\  (2017) 221
  doi:10.1007/978-3-319-65427-0\_6
  [arXiv:1711.09966 [hep-th]].

\bibitem{Bjornsson:2010wu}
  J.~Bjornsson,
  ``Multi-loop amplitudes in maximally supersymmetric pure spinor field theory,''
  JHEP {\bf 1101} (2011) 002
  doi:10.1007/JHEP01(2011)002
  [arXiv:1009.5906 [hep-th]].




\bibitem{Bjornsson:2010wm}
  J.~Bjornsson and M.~B.~Green,
  ``5 loops in 24/5 dimensions,''
  JHEP {\bf 1008} (2010) 132
  doi:10.1007/JHEP08(2010)132
  [arXiv:1004.2692 [hep-th]].
\bibitem{Mafra:2014gja}
  C.~R.~Mafra and O.~Schlotterer,
  ``Towards one-loop SYM amplitudes from the pure spinor BRST cohomology,''
  Fortsch.\ Phys.\  {\bf 63} (2015) no.2,  105
  doi:10.1002/prop.201400076
  [arXiv:1410.0668 [hep-th]].


\bibitem{Mafra:2015mja}
  C.~R.~Mafra and O.~Schlotterer,
  ``Two-loop five-point amplitudes of super Yang-Mills and supergravity in pure spinor superspace,''
  JHEP {\bf 1510} (2015) 124
  doi:10.1007/JHEP10(2015)124
  [arXiv:1505.02746 [hep-th]].

\bibitem{Cederwall:2010tn}
  M.~Cederwall,
  ``D=11 supergravity with manifest supersymmetry,''
  Mod.\ Phys.\ Lett.\ A {\bf 25} (2010) 3201
  doi:10.1142/S0217732310034407
  [arXiv:1001.0112 [hep-th]].




\bibitem{Cederwall:2012es}
  M.~Cederwall and A.~Karlsson,
  ``Loop amplitudes in maximal supergravity with manifest supersymmetry,''
  JHEP {\bf 1303} (2013) 114
  doi:10.1007/JHEP03(2013)114
  [arXiv:1212.5175 [hep-th]].

\bibitem{Berkovits:2018gbq}
  N.~Berkovits and M.~Guillen,
  ``Equations of motion from Cederwall's pure spinor superspace actions,''
  arXiv:1804.06979 [hep-th].

\bibitem{Karlsson:2014xva}
  A.~Karlsson,
  ``Ultraviolet divergences in maximal supergravity from a pure spinor point of view,''
  JHEP {\bf 1504} (2015) 165
  doi:10.1007/JHEP04(2015)165
  [arXiv:1412.5983 [hep-th]].


\bibitem{Berkovits:2004bw}
  N.~Berkovits and S.~A.~Cherkis,
  ``Higher-dimensional twistor transforms using pure spinors,''
  JHEP {\bf 0412} (2004) 049
  doi:10.1088/1126-6708/2004/12/049
  [hep-th/0409243].

\bibitem{Berkovits:2009by}
  N.~Berkovits,
``Ten-Dimensional Super-Twistors and Super-Yang-Mills,''
  JHEP {\bf 1004} (2010) 067
  doi:10.1007/JHEP04(2010)067
  [arXiv:0910.1684 [hep-th]].

\bibitem{Berkovits:2014aia}
  N.~Berkovits,
  ``Twistor Origin of the Superstring,''
  JHEP {\bf 1503} (2015) 122
  doi:10.1007/JHEP03(2015)122
  [arXiv:1409.2510 [hep-th]].







\bibitem{Parke:1986gb}
  S.~J.~Parke and T.~R.~Taylor,
  ``An Amplitude for $n$ Gluon Scattering,''
  Phys.\ Rev.\ Lett.\  {\bf 56} (1986) 2459.
  doi:10.1103/PhysRevLett.56.2459


\bibitem{Sokatchev:1985tc}
  E.~Sokatchev,
  ``Light Cone Harmonic Superspace and Its Applications,''
  Phys.\ Lett.\ B {\bf 169} (1986) 209.
  doi:10.1016/0370-2693(86)90652-0

\bibitem{Sokatchev:1987nk}
  E.~Sokatchev,
  ``Harmonic Superparticle,''
  Class.\ Quant.\ Grav.\  {\bf 4} (1987) 237.
  doi:10.1088/0264-9381/4/2/007



\bibitem{Siegel:1983hh}
  W.~Siegel,
  ``Hidden Local Supersymmetry in the Supersymmetric Particle Action,''
  Phys.\ Lett.\ B {\bf 128} (1983) 397.

\bibitem{de Azcarraga:1982dw}
  J.~A.~de Azcarraga and J.~Lukierski,
  ``Supersymmetric Particles with Internal Symmetries and Central Charges,''
  Phys.\ Lett.\ B {\bf 113} (1982) 170.





\bibitem{Galperin:1984av}
  A.~Galperin, E.~Ivanov, S.~Kalitsyn, V.~Ogievetsky and E.~Sokatchev,
  ``Unconstrained N=2 Matter, Yang-Mills and Supergravity Theories in Harmonic Superspace,''
  Class.\ Quant.\ Grav.\  {\bf 1} (1984) 469
   Erratum: [Class.\ Quant.\ Grav.\  {\bf 2} (1985) 127].
  doi:10.1088/0264-9381/1/5/004

\bibitem{Galperin:1984bu}
  A.~Galperin, E.~Ivanov, S.~Kalitsyn, V.~Ogievetsky and E.~Sokatchev,
  ``Unconstrained Off-Shell N=3 Supersymmetric Yang-Mills Theory,''
  Class.\ Quant.\ Grav.\  {\bf 2} (1985) 155.
  doi:10.1088/0264-9381/2/2/009

\bibitem{Galperin:2001uw}
  A.~S.~Galperin, E.~A.~Ivanov, V.~I.~Ogievetsky and E.~S.~Sokatchev,
  ``Harmonic superspace,''
  Cambridge, UK: Univ. Pr. (2001) 306 p
  doi:10.1017/CBO9780511535109

\bibitem{Bandos:1990ji}
  I.~A.~Bandos,
  ``Superparticle in Lorentz harmonic superspace,''
  Sov.\ J.\ Nucl.\ Phys.\  {\bf 51} (1990) 906


\bibitem{Eden:1966dnq}
  R.~J.~Eden, P.~V.~Landshoff, D.~I.~Olive and J.~C.~Polkinghorne,
  ``The analytic S-matrix,'' Cambridge at the University Press, 1966. 287pp.

\bibitem{Dirac:1963}
P.A.M. Dirac, ''Lectures on Quantum mechanics'', Yeshiva Univ., New York, 1964. 84pp.

\bibitem{Berezin:1976eg}
  F.~A.~Berezin and M.~S.~Marinov,
 ``Particle Spin Dynamics as the Grassmann Variant of Classical Mechanics,''
  Annals Phys.\  {\bf 104} (1977) 336.
  doi:10.1016/0003-4916(77)90335-9

\bibitem{Brink:1976uf}
  L.~Brink, P.~Di Vecchia and P.~S.~Howe,
  ``A Lagrangian Formulation of the Classical and Quantum Dynamics of Spinning Particles,''
  Nucl.\ Phys.\ B {\bf 118} (1977) 76.
  doi:10.1016/0550-3213(77)90364-9


\bibitem{Green:1999by}
  M.~B.~Green, M.~Gutperle and H.~H.~Kwon,
  ``Light cone quantum mechanics of the eleven-dimensional superparticle,''
  JHEP {\bf 9908} (1999) 012
  [hep-th/9907155].


\bibitem{Schweber:1961zz}
  S.~S.~Schweber,
  ``An Introduction to Relativistic Quantum Field Theory,'' Row, Peterson, Evanston, Ill, 1961, 905pp.



\bibitem{Tonin:1991ii}
  M.~Tonin,
  ``World sheet supersymmetric formulations of Green-Schwarz superstrings,''
  Phys.\ Lett.\ B {\bf 266} (1991) 312.
  doi:10.1016/0370-2693(91)91046-X


\bibitem{Tonin:1991ia}
  M.~Tonin,
  ``Kappa symmetry as world sheet supersymmetry in D = 10 heterotic superstring,''
  Int.\ J.\ Mod.\ Phys.\ A {\bf 7} (1992) 6013.
  doi:10.1142/S0217751X92002726


\bibitem{Zupnik:1989bw}
  B.~M.~Zupnik and D.~G.~Pak,
  ``Differential and Integral Forms in Supergauge Theories and Supergravity,''
  Class.\ Quant.\ Grav.\  {\bf 6} (1989) 723.
  doi:10.1088/0264-9381/6/5/014

\bibitem{ArkaniHamed:2008yf}
  N.~Arkani-Hamed and J.~Kaplan,
  ``On Tree Amplitudes in Gauge Theory and Gravity,''
  JHEP {\bf 0804} (2008) 076
  doi:10.1088/1126-6708/2008/04/076
  [arXiv:0801.2385 [hep-th]].

\bibitem{Cheung:2008dn}
  C.~Cheung,
 ``On-Shell Recursion Relations for Generic Theories,''
  JHEP {\bf 1003} (2010) 098
  doi:10.1007/JHEP03(2010)098
  [arXiv:0808.0504 [hep-th]].

\bibitem{Cheung:2009dc}
  C.~Cheung and D.~O'Connell,
  ``Amplitudes and Spinor-Helicity in Six Dimensions,''
  JHEP {\bf 0907} (2009) 075
  doi:10.1088/1126-6708/2009/07/075
  [arXiv:0902.0981 [hep-th]].

\bibitem{Green:1997as}
  M.~B.~Green, M.~Gutperle and P.~Vanhove,
  ``One loop in eleven-dimensions,''
  Phys.\ Lett.\ B {\bf 409} (1997) 177
  doi:10.1016/S0370-2693(97)00931-3
  [hep-th/9706175].

\bibitem{Deser:2000xz}
  S.~Deser and D.~Seminara,
  ``Tree amplitudes and two loop counterterms in D = 11 supergravity,''
  Phys.\ Rev.\ D {\bf 62} (2000) 084010
  doi:10.1103/PhysRevD.62.084010
  [hep-th/0002241].


\bibitem{Green:2016tfs}
  M.~B.~Green and A.~Rudra,
  ``Type I/heterotic duality and M-theory amplitudes,''
  arXiv:1604.00324 [hep-th].

\bibitem{Green:1981xx}
  M.~B.~Green and J.~H.~Schwarz,
  ``Supersymmetrical Dual String Theory. 2. Vertices and Trees,''
  Nucl.\ Phys.\ B {\bf 198} (1982) 252.
  doi:10.1016/0550-3213(82)90556-9

\bibitem{Green:1981ya}
  M.~B.~Green and J.~H.~Schwarz,
  ``Supersymmetrical Dual String Theory. 3. Loops and Renormalization,''
  Nucl.\ Phys.\ B {\bf 198} (1982) 441.
  doi:10.1016/0550-3213(82)90334-0
\bibitem{Schwarz:1982jn}
  J.~H.~Schwarz,
  ``Superstring Theory,''
  Phys.\ Rept.\  {\bf 89} (1982) 223.
  doi:10.1016/0370-1573(82)90087-4


\bibitem{Gross:1986iv}
  D.~J.~Gross and E.~Witten,
  ``Superstring Modifications of Einstein's Equations,''
  Nucl.\ Phys.\ B {\bf 277} (1986) 1.
  doi:10.1016/0550-3213(86)90429-3

\bibitem{Sannan:1986tz}
  S.~Sannan,
  ``Gravity as the Limit of the Type {II} Superstring Theory,''
  Phys.\ Rev.\ D {\bf 34} (1986) 1749.
  doi:10.1103/PhysRevD.34.1749

\bibitem{Galperin:1985va}
  A.~Galperin, E.~Ivanov, V.~Ogievetsky and E.~Sokatchev,
  ``Harmonic Supergraphs. Feynman Rules and Examples,''
  Class.\ Quant.\ Grav.\  {\bf 2} (1985) 617.
  doi:10.1088/0264-9381/2/5/005



\bibitem{Mason:2013sva}
  L.~Mason and D.~Skinner,
  ``Ambitwistor strings and the scattering equations,''
  arXiv:1311.2564 [hep-th].


\bibitem{Adamo:2014wea}
  T.~Adamo, E.~Casali and D.~Skinner,
  ``A Worldsheet Theory for Supergravity,''
  JHEP {\bf 1502} (2015) 116
  [arXiv:1409.5656 [hep-th]].


\bibitem{Cachazo:2013hca}
  F.~Cachazo, S.~He and E.~Y.~Yuan,
  ``Scattering of Massless Particles in Arbitrary Dimension,''
  arXiv:1307.2199 [hep-th];
  ``Scattering of Massless Particles: Scalars, Gluons and Gravitons,''
  JHEP {\bf 1407} (2014) 033
  [arXiv:1309.0885 [hep-th]].


\bibitem{Adamo:2013tsa}
  T.~Adamo, E.~Casali and D.~Skinner,
  ``Ambitwistor strings and the scattering equations at one loop,''
  JHEP {\bf 1404} (2014) 104
  [arXiv:1312.3828 [hep-th]].






\bibitem{Bandos:2014lja}
  I.~Bandos,
  ``Twistor/ambitwistor strings and null-superstrings in spacetime of D=4, 10 and 11 dimensions,''
  JHEP {\bf 1409} (2014) 086
  [arXiv:1404.1299 [hep-th]].


\bibitem{Bandos:2006af}
  I.~A.~Bandos, J.~A.~de Azcarraga and C.~Miquel-Espanya,
  ``Superspace formulations of the (super)twistor string,''
  JHEP {\bf 0607} (2006) 005
  [hep-th/0604037].


\bibitem{Berkovits:2004hg}
  N.~Berkovits,
  ``An Alternative string theory in twistor space for N=4 superYang-Mills,''
  Phys.\ Rev.\ Lett.\  {\bf 93} (2004) 011601
  [hep-th/0402045].

\bibitem{Siegel:2004dj}
  W.~Siegel,
  ``Untwisting the twistor superstring,''
  hep-th/0404255.

\bibitem{Geyer:2014fka}
  Y.~Geyer, A.~E.~Lipstein and L.~J.~Mason,
  ``Ambitwistor Strings in Four Dimensions,''
  Phys.\ Rev.\ Lett.\  {\bf 113} (2014) 8,  081602
  [arXiv:1404.6219 [hep-th]].

\bibitem{Lipstein:2015vxa}
  A.~Lipstein and V.~Schomerus,
  ``Towards a Worldsheet Description of N=8 Supergravity,''
  arXiv:1507.02936 [hep-th].


\bibitem{Bork:2017qyh}
  L.~V.~Bork and A.~I.~Onishchenko,
  ``Four dimensional ambitwistor strings and form factors of local and Wilson line operators,''
  arXiv:1704.04758 [hep-th].

\bibitem{Farrow:2017eol}
  J.~A.~Farrow and A.~E.~Lipstein,
  ``From 4d Ambitwistor Strings to On Shell Diagrams and Back,''
  JHEP {\bf 1707} (2017) 114
  doi:10.1007/JHEP07(2017)114
  [arXiv:1705.07087 [hep-th]].

\bibitem{Casali:2016atr}
  E.~Casali and P.~Tourkine,
  ``On the null origin of the ambitwistor string,''
  JHEP {\bf 1611} (2016) 036
  doi:10.1007/JHEP11(2016)036
  [arXiv:1606.05636 [hep-th]].

\bibitem{Casali:2017zkz}
  E.~Casali, Y.~Herfray and P.~Tourkine,
  ``The complex null string, Galilean conformal algebra and scattering equations,''
  JHEP {\bf 1710} (2017) 164
  doi:10.1007/JHEP10(2017)164
  [arXiv:1707.09900 [hep-th]].

\bibitem{Adamo:2017sze}
  T.~Adamo, E.~Casali, L.~Mason and S.~Nekovar,
  ``Amplitudes on plane waves from ambitwistor strings,''
  JHEP {\bf 1711} (2017) 160
  doi:10.1007/JHEP11(2017)160
  [arXiv:1708.09249 [hep-th]].

\end{thebibliography}
\end{document}